\documentclass[prd,nofootinbib,twocolumn,superscriptaddress,preprintnumbers,balancelastpage,nofootinbib]{revtex4-1}

\usepackage{float}
\usepackage{graphicx}  
\usepackage{dcolumn}   
\usepackage{bm}        
\usepackage{amssymb, amsmath}
\usepackage{slashed}   

\usepackage{hyperref}
\hypersetup{colorlinks=true, citecolor=blue, urlcolor=blue, linkcolor=blue}
\usepackage[usenames,dvipsnames,svgnames,table]{xcolor}
\def\be{\begin{equation}}
\def\ee{\end{equation}}
\def\bea{\begin{eqnarray}}
\def\eea{\end{eqnarray}}
\newcommand{\aprime}{A^\prime}

\usepackage{calrsfs}
\DeclareMathAlphabet{\pazocal}{OMS}{zplm}{m}{n}

\newcommand{\Eq}[1]{Eq.~(\ref{#1})}

\newcommand{\GeV}{{\rm GeV}}
\newcommand{\MeV}{{\rm MeV}}

\newcommand{\Nevt}{N_{\mathrm{sig}}}
\newcommand{\mAp}{m_{A^\prime}}

\def\beq{\begin{equation}}
\def\eeq{\end{equation}}

\def\be{\begin{eqnarray}}
\def\ee{\end{eqnarray}}

\def\bi{\begin{itemize}}
\def\ei{\end{itemize}}

\begin{document}
\widetext

\newcommand\UCSC{University of California, Santa Cruz, and Santa Cruz Institute for Particle Physics, Santa Cruz, CA 95064, USA}
\newcommand\KICP{ University of Chicago, Kavli Institute for Cosmological Physics, Chicago, IL, USA}
\newcommand\FNAL{Fermi National Accelerator Laboratory, Batavia, IL, USA}
\newcommand\carleton{Ottawa-Carleton Institute for Physics, Department of Physics, Carleton University, Ottawa, ON, Canada}

\preprint{FERMILAB-PUB-20-501-T}

\title{Characterizing Dark Matter Signals with Missing Momentum Experiments}

\author{Nikita Blinov}
\affiliation{\FNAL}
\affiliation{\KICP}

\author{Gordan Krnjaic}
\affiliation{\FNAL}
\affiliation{\KICP}

\author{Douglas Tuckler}
\affiliation{\FNAL}
\affiliation{\UCSC}
\affiliation{\carleton}

\date{\today}

\begin{abstract}
\noindent
Fixed target missing-momentum experiments such as LDMX and M$^3$ are powerful probes of light dark matter and
other light, weakly coupled particles beyond the Standard Model (SM). Such experiments involve $\sim$ 10 GeV 
beam particles whose energy and momentum are individually measured before and after passing through a suitably thin target. 
If new states are radiatively produced in the target, the recoiling beam particle loses a large fraction of its initial momentum, 
and no SM particles are observed in a downstream veto detector. We explore how such experiments can 
use kinematic variables and experimental parameters, such as beam energy and polarization,
to measure properties of the radiated particles and discriminate between models if a signal is discovered. In particular, the transverse momentum of 
recoiling particles is shown to be a powerful tool to measure the masses of new radiated states, offering significantly 
better discriminating ability compared to the recoil energy alone. We further illustrate how variations in beam energy, polarization, and lepton
flavor (i.e., electron or muon) can be used to disentangle the possible the Lorentz structure of the new interactions. 
\end{abstract}

\pacs{}
\maketitle




\section{Introduction}

Over the past decade the experimental dark matter (DM) search effort has greatly expanded in scope
to explore the sub-GeV mass range. This push towards lower masses has been driven by several
 complementary strategies, including new direct-detection techniques (e.g., electron ionization)~\cite{
 Essig:2011nj,Graham:2012su,Essig:2013vha,Hochberg:2015fth,Hochberg:2015pha,Essig:2016crl,Knapen:2016cue,Essig:2018tss,Knapen:2017ekk,Essig:2016crl,Knapen:2016cue,Essig:2018tss,Knapen:2017ekk} and low-energy accelerator searches~\cite{deNiverville:2011it,deNiverville:2016rqh,Izaguirre:2015yja,Izaguirre:2017bqb,Akesson:2019iul,Battaglieri:2017qen,Kahn:2018cqs,Berlin:2018bsc,Izaguirre:2017bqb,Izaguirre:2014bca,Kahn:2014sra,Izaguirre:2014cza,Izaguirre:2014dua,Izaguirre:2013uxa,Berlin:2020uwy,Buonocore:2019esg,Tsai:2019mtm,deNiverville:2018dbu,Aguilar-Arevalo:2018wea,Berlin:2018pwi,Aguilar-Arevalo:2017mqx} -- see Refs.~\cite{Essig:2013lka,Alexander:2016aln,Battaglieri:2017aum,Strategy:2019vxc} for reviews.

A particularly promising accelerator-based strategy involves the fixed-target missing-momentum (MM) concept~\cite{Izaguirre:2014bca}. In this setup, a low-current ${\pazocal O}(1-10)$ GeV lepton beam is passed through a thin target, which is surrounded on both sides by tracking material and positioned upstream of a veto detector; the energy and momentum of {\it individual} beam particles are measured on both sides of the target. If DM (or any other invisible or long-lived particle) is produced in the target, the beam particle loses a large fraction of its energy and momentum, and no other visible particles are observed in the veto detector.



\begin{figure*}
        \includegraphics[width=0.99\textwidth]{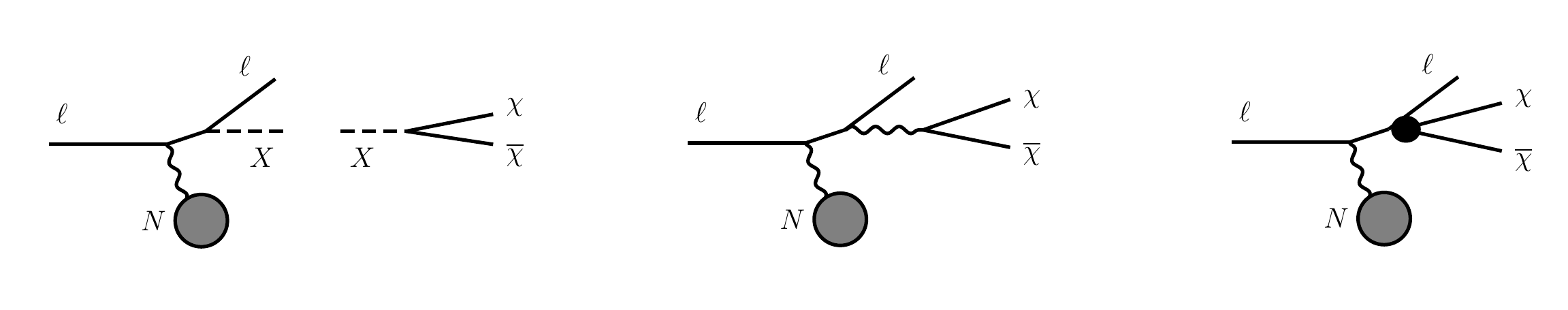} \\
    \centering
      \caption{Schematic diagrams depicting the signal topologies outlined in Sec.~\ref{models}. {\bf Left:}
      two-step DM production via $2 \to 3$ scattering in which a massive mediator particle $X$ is produced 
      on shell and subsequently decays invisibly via $X \to \bar \chi \chi$. {\bf Middle:} Direct DM production through virtual
      light mediator exchange corresponding to the $(\bar e  e)(\bar\chi\chi)/q^2$ ``millicharge'' interaction in Eq.~(\ref{L24}).
      {\bf Right: } Direct DM production through a contact $(\bar e  e)(\bar\chi\chi)/\Lambda^2$ interaction from Eq.~(\ref{L24})
      represented by the black circle. 
      }
    \label{cartoon}
\end{figure*}



\begin{figure}
\hspace{-0.5cm}
        \includegraphics[width=0.4\textwidth]{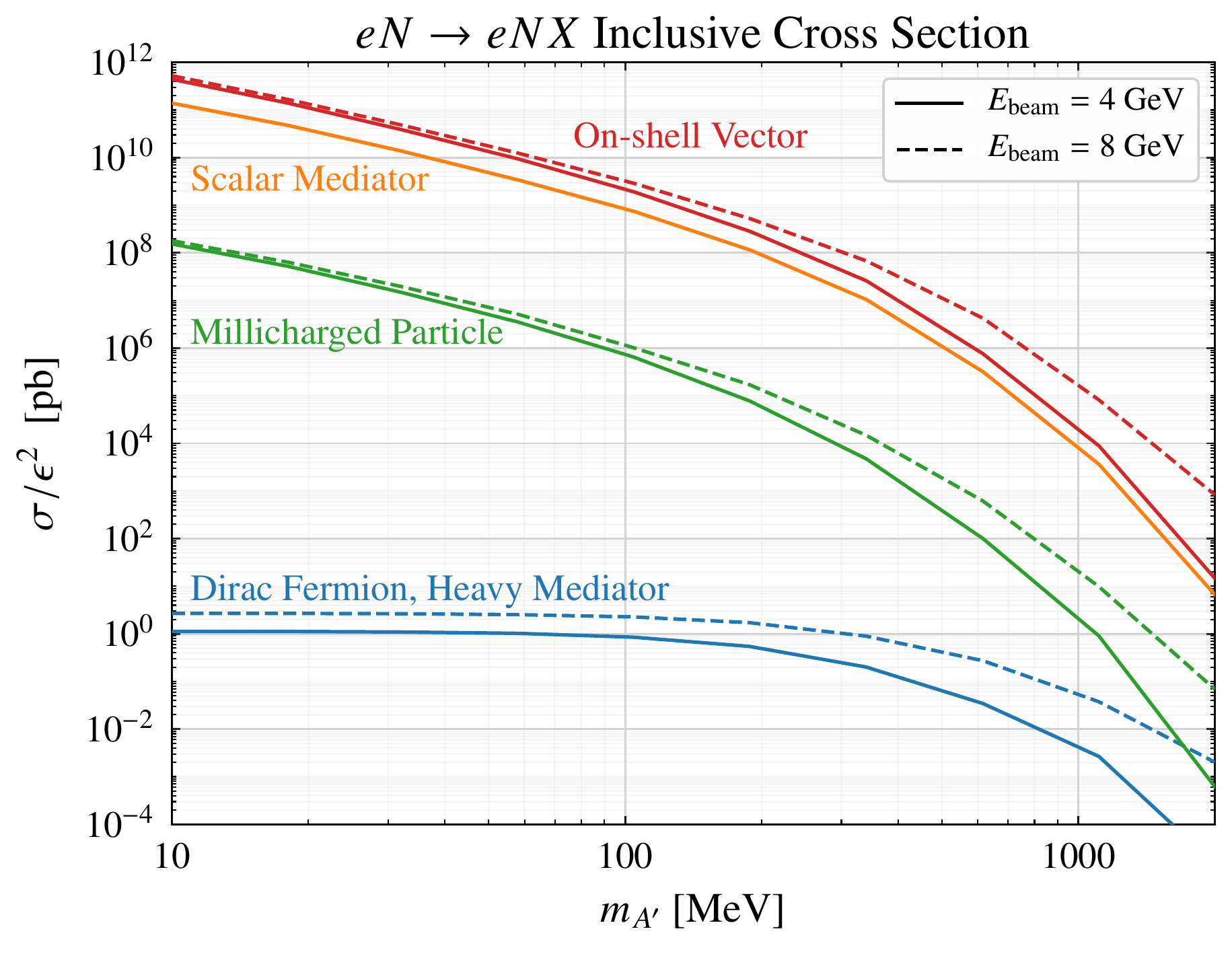} \\
    \centering
      \caption{
      Fixed target production cross sections for a representative subset of scenarios 
      in Eqs.~(\ref{LS1})-(\ref{L24}) for 4 (solid lines) and 8 (dashed lines) GeV electron beams on a tungsten target. 
      The horizontal axis represents the mediator mass $m_{A^\prime}$ 
      for the radiative production of $A^\prime$ vector mediators through the interaction in \Eq{LS1} (top) and $2 m_\chi$  for the DM pair production
      through the interactions in  \Eq{L24}. The heavy mediator case corresponds to $\mAp = 10\;\GeV$ and exhibits 
      a systematic increase in the cross section as the beam energy is increased. This behavior 
    reflects the higher-dimensionality of the operator mediating this interaction (first line of \Eq{L24}).}
    \label{fig:cross-sections}
\end{figure}

This technique has several appealing features:
\bi
\item{\bf  Experimental Control:} Unlike direct detection searches, whose sensitivity is subject to
to astrophysical uncertainties and environmental backgrounds, the MM signal strength is
fully calculable and irreducible background events occur at a rate of $\sim 10^{-15}$ per incident
beam particle~\cite{Akesson:2019iul}. 
\item{\bf  Coverage Breadth:} Since DM production at accelerators is relativistic, the signal strength 
 is largely insensitive to the Lorentz structure of the underlying interaction. Consequently, the same MM search simultaneously 
 covers a wide range of DM spin and coupling varieties. 
\item{\bf Parametric Enhancement:}  Unlike beam-dump DM searches whose signal involves DM production 
followed by its scattering in a downstream detector, the MM setup only requires DM production. Thus, the signal rate depends only on the production rate without the added penalty of a small scattering probability.  
\ei
The electron beam MM strategy is currently being developed by the LDMX 
collaboration~\cite{Akesson:2018vlm,Akesson:2019iul} and additional studies are underway to explore a future
muon beam MM experiment (M$^3$) at Fermilab~\cite{Kahn:2018cqs}.
Collectively, these efforts have the potential to test nearly every model of 
sub-GeV freeze-out for which the MM signal strength is parametrically related to the DM annihilation rate
in the early universe~\cite{Izaguirre:2015yja,Berlin:2018bsc,Berlin:2020uwy}.\footnote{This conclusion only fails to hold if early universe annihilation is 
on resonance, corresponding to a tuned region of parameter space~\cite{Feng:2017drg,Berlin:2020uwy}.}
In this work we explore the optimistic scenario in which one of these experiments convincingly discovers a new physics signal.



\begin{figure*}
    \centering
    \includegraphics[width=0.3275\textwidth]{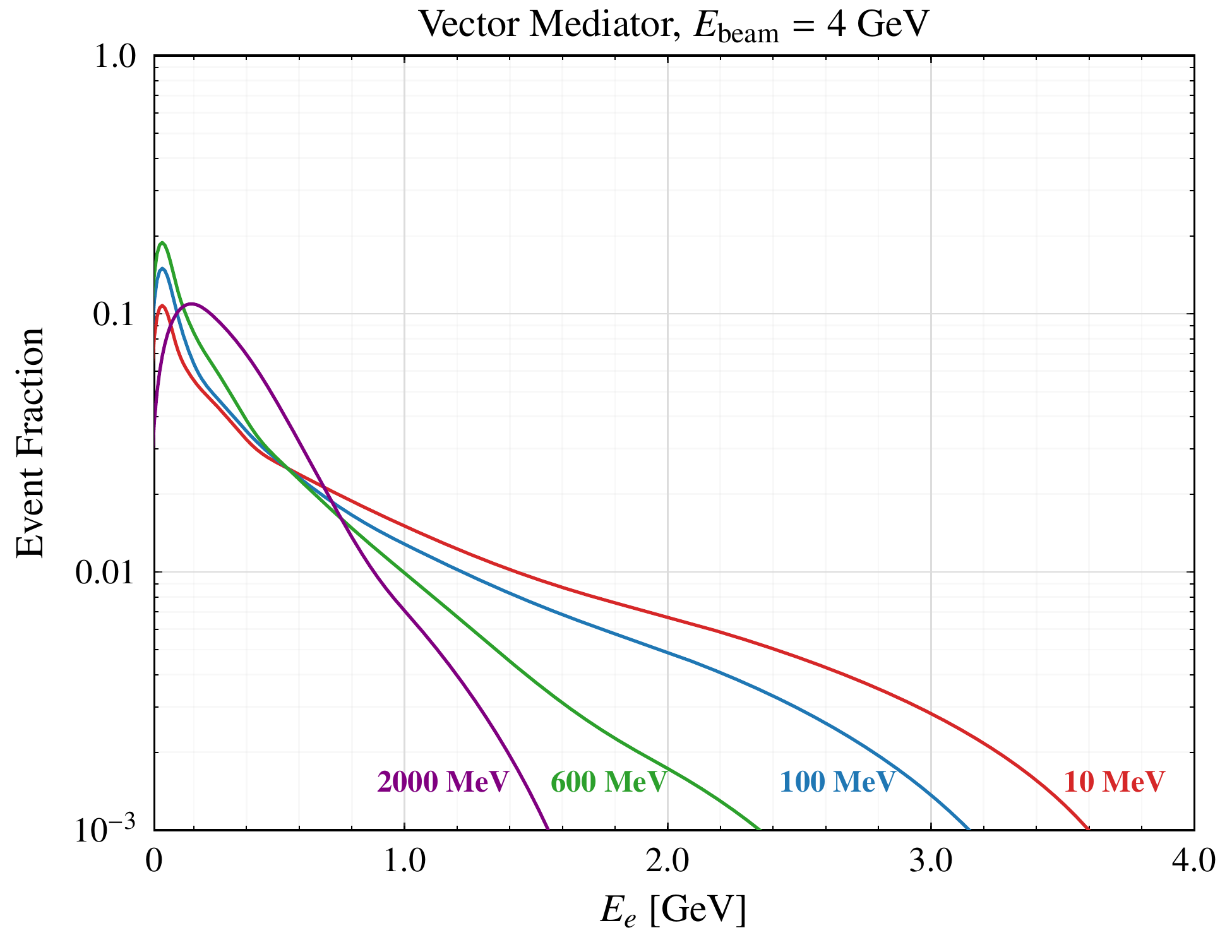}
    \includegraphics[width=0.3275\textwidth]{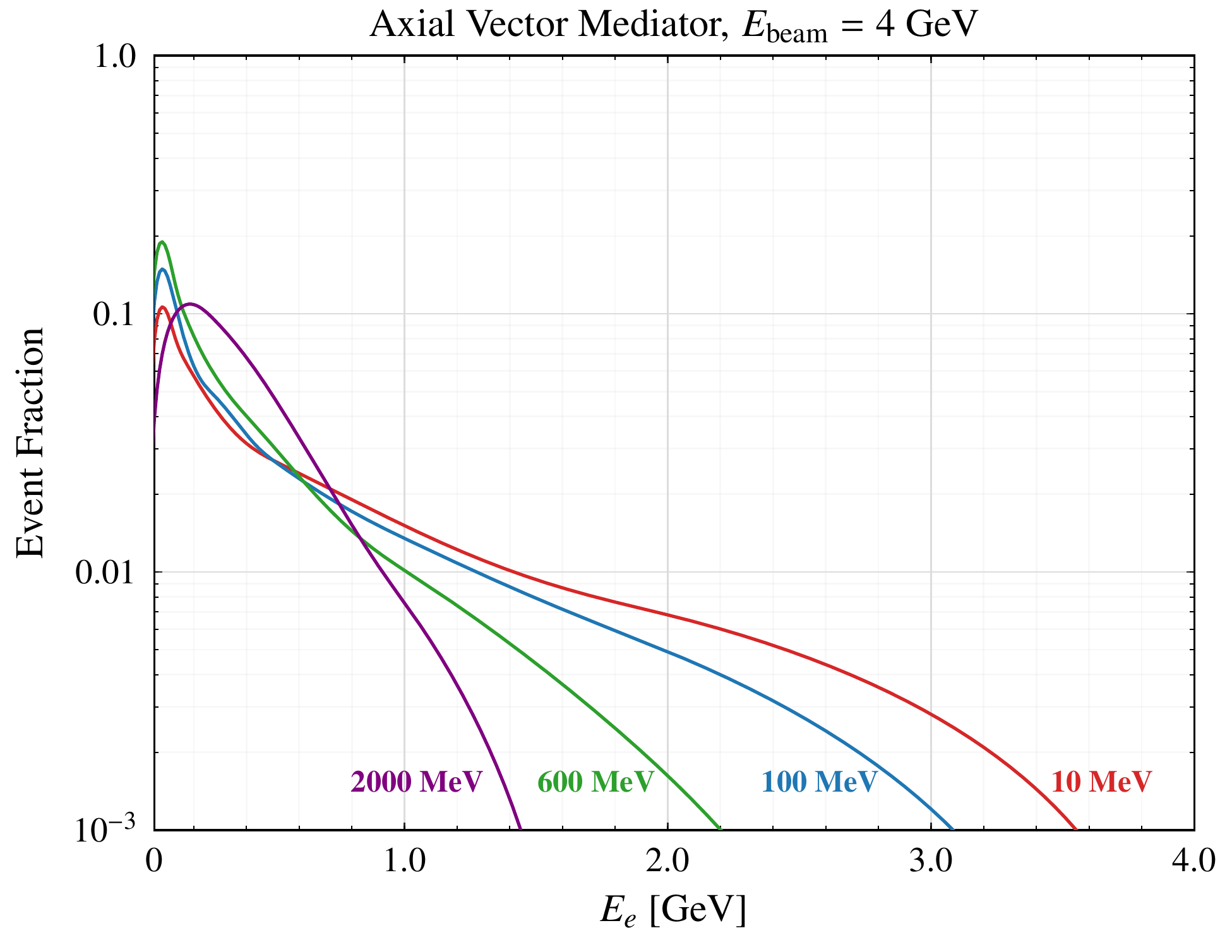}
    \includegraphics[width=0.3275\textwidth]{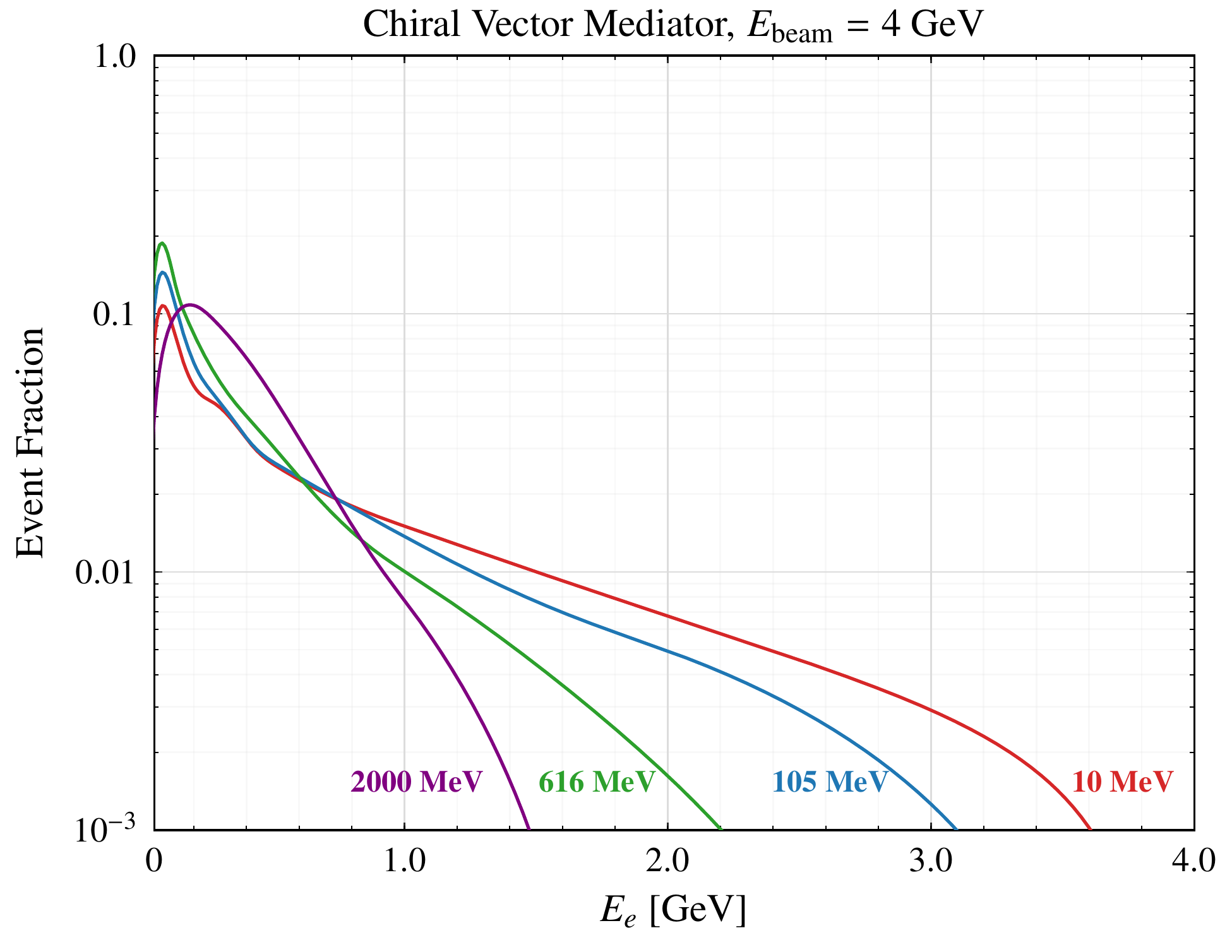}\\
        \medskip
   \includegraphics[width=0.3275\textwidth]{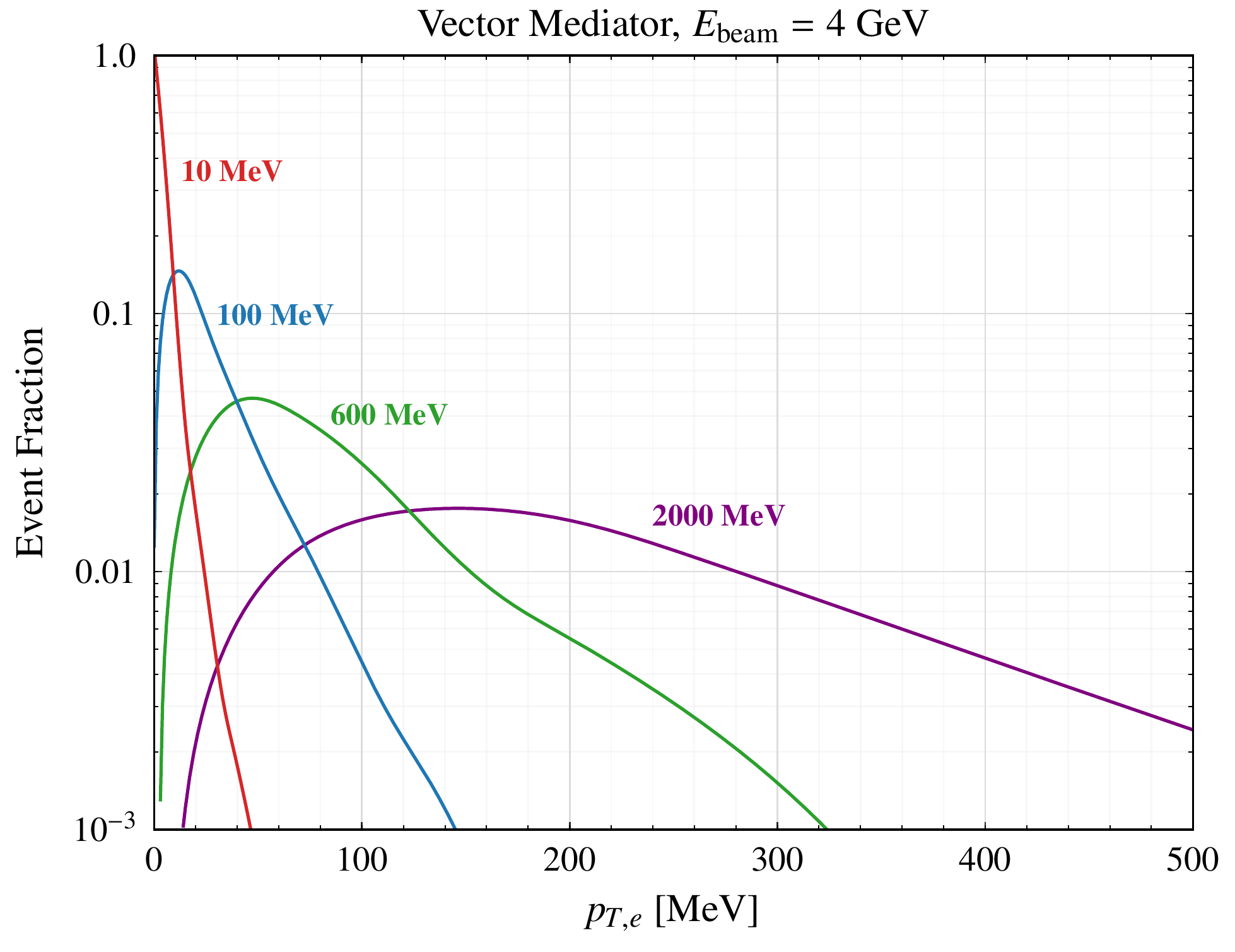}
    \includegraphics[width=0.3275\textwidth]{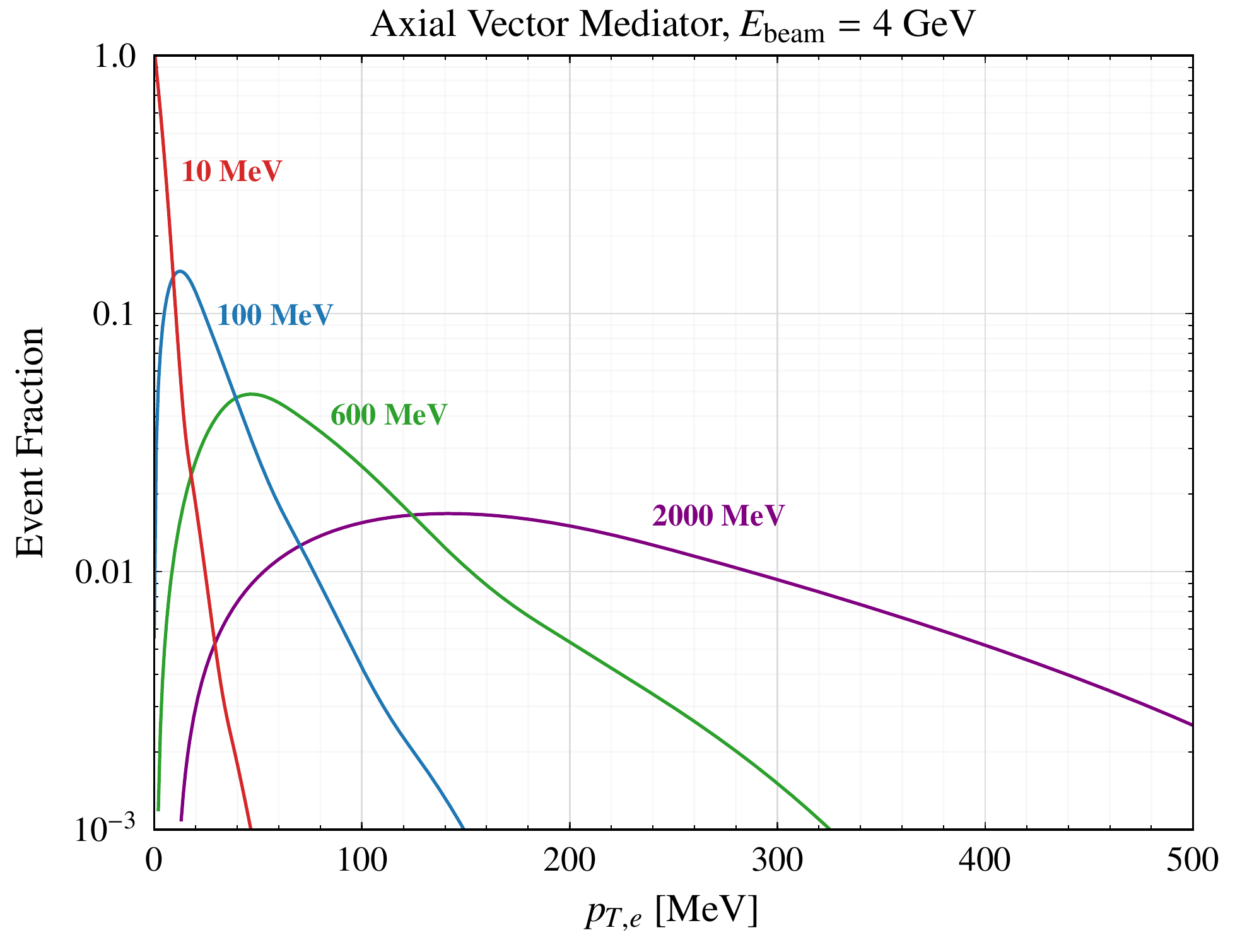}
    \includegraphics[width=0.3275\textwidth]{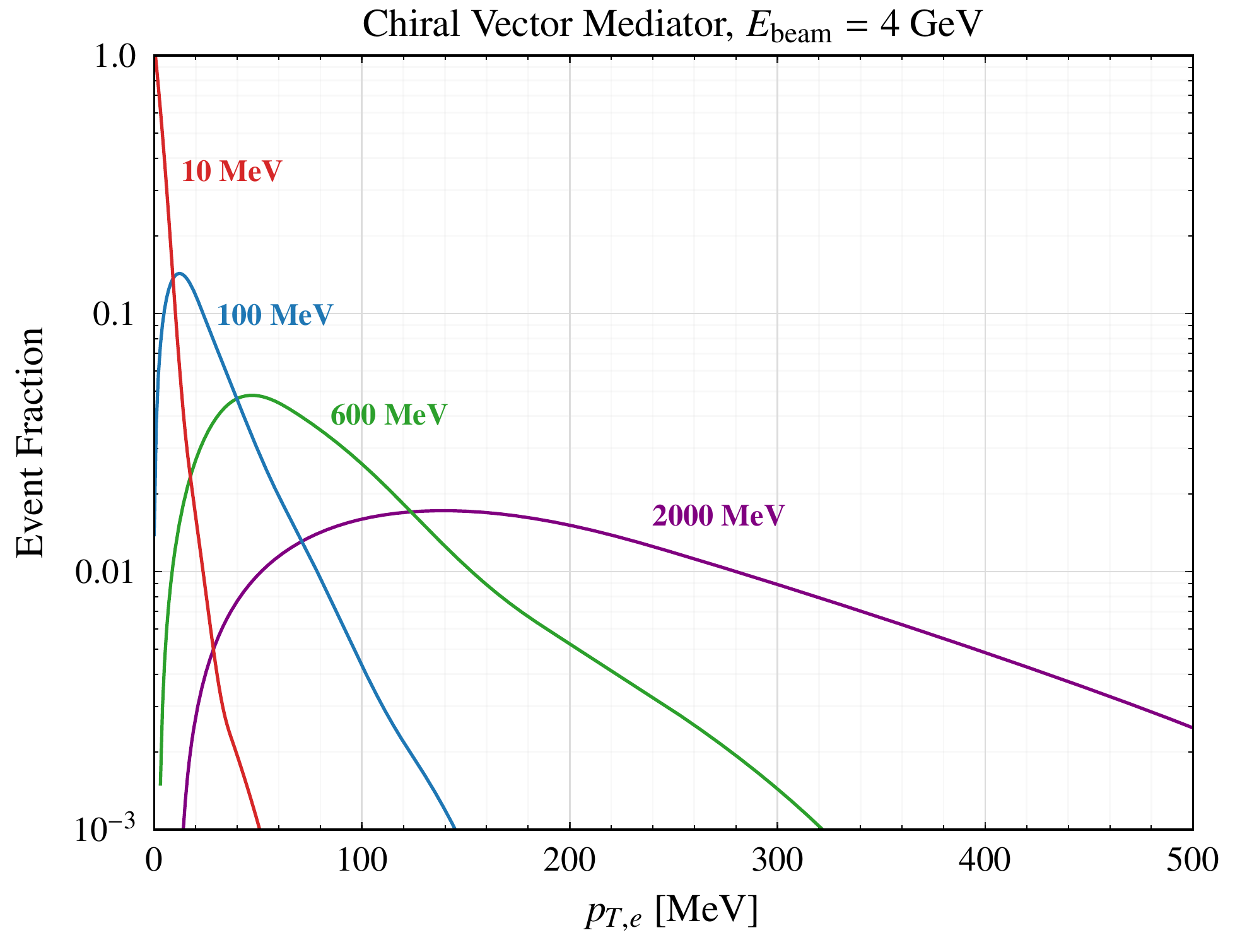}\\

      \caption{ Kinematic distributions for the recoiling electron in on shell production of a massive mediator particle in $e N \to e N X$ fixed target reactions. 
        The models considered in the left, center, and right panels correspond to the vector, axial vector, and chiral vector interactions
 in Eq.~(\ref{LS1}), respectively. The {\bf top row} shows the outgoing electron's recoil energy for the three 
 interaction types for various choices of $m_\chi$ and the {\bf bottom row}
 shows the corresponding electron $p_{T,e}$ distributions. In all cases, the incident electron
 beam energy of 4 GeV and a tungsten target are chosen to match projections for LDMX Phase 1~\cite{Akesson:2018vlm}.	
      } 
      
    \label{vectorhistograms}
\end{figure*}



\begin{figure*}
    \centering
    \includegraphics[width=0.3275\textwidth]{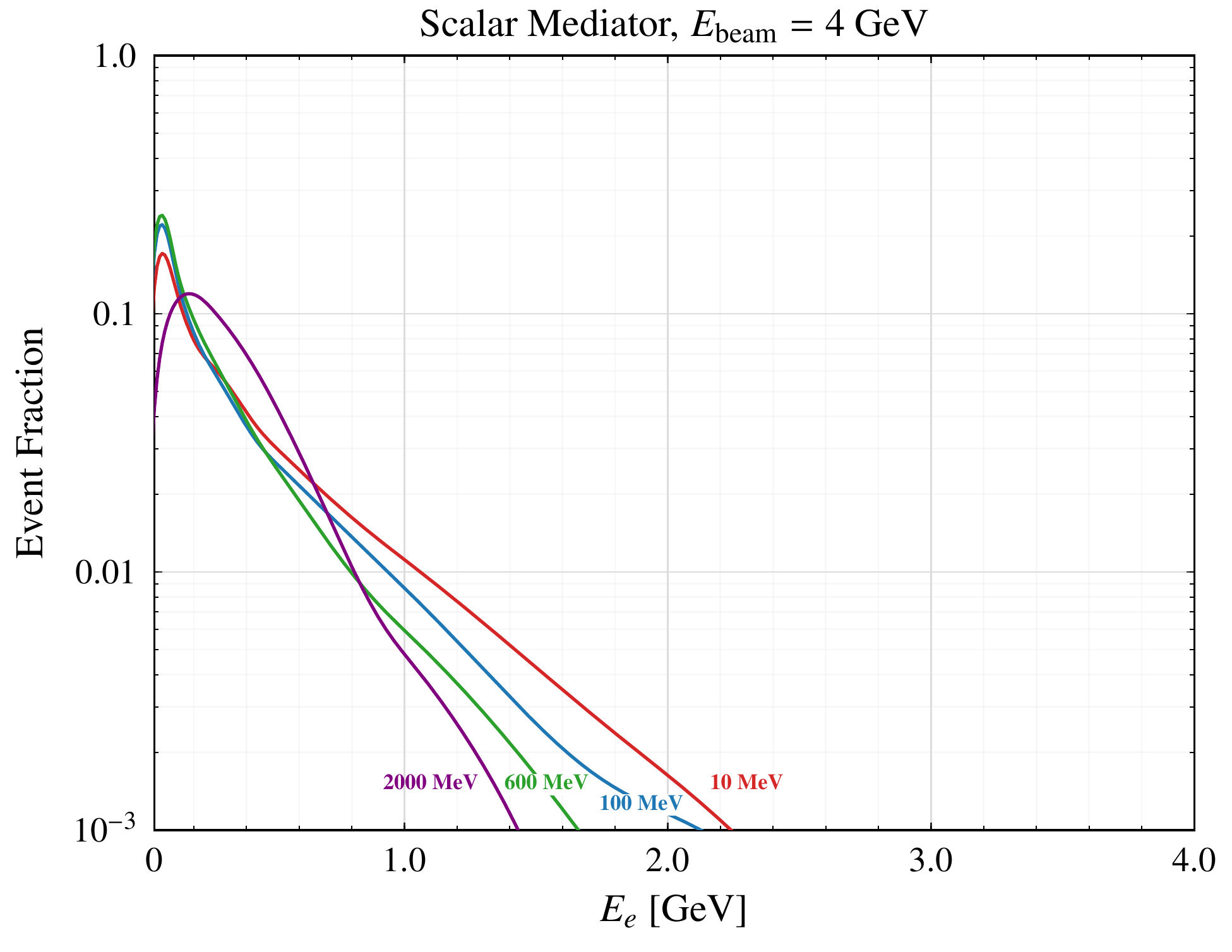}~~
        \includegraphics[width=0.3275\textwidth]{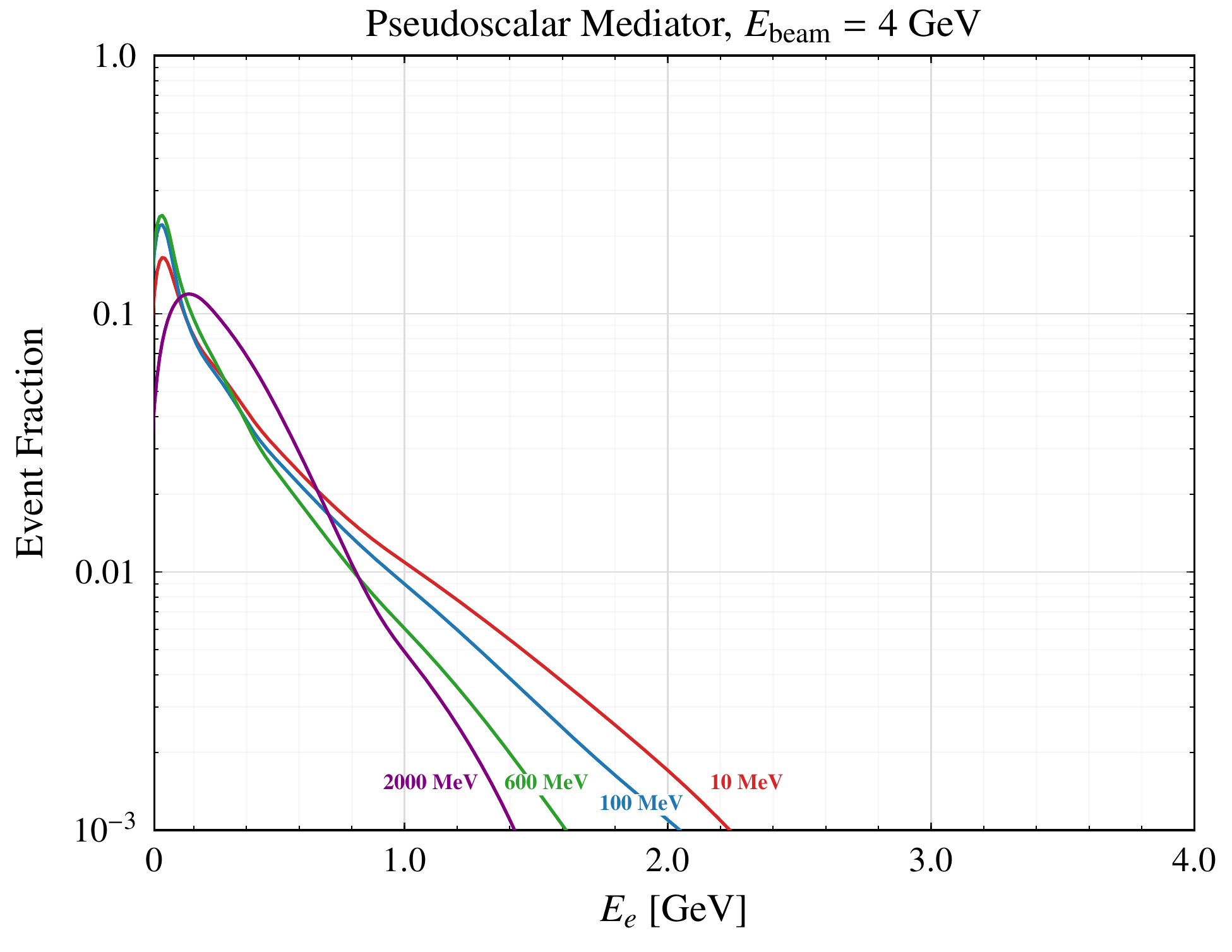} \\ 
        \medskip
    \includegraphics[width=0.3275\textwidth]{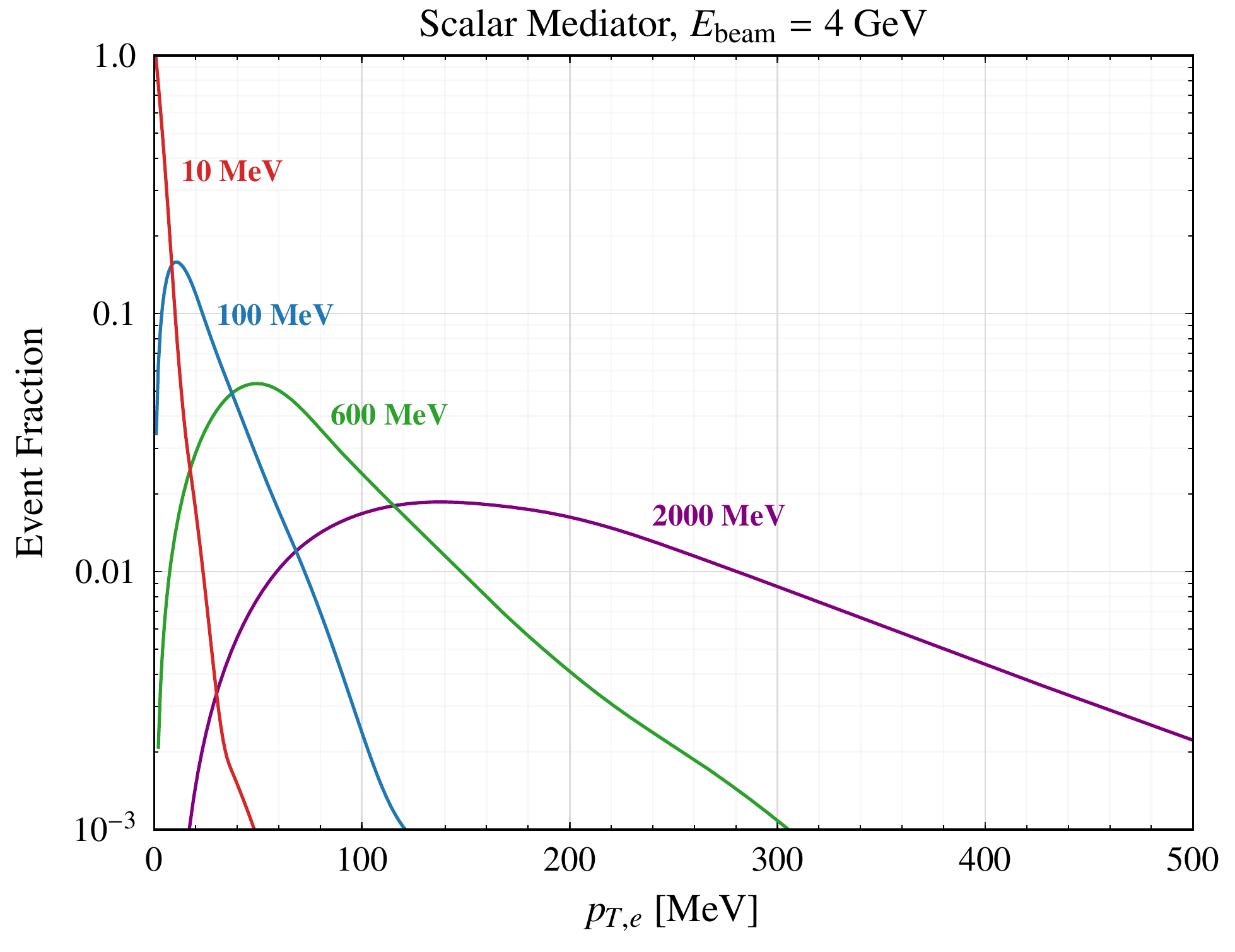}~~
        \includegraphics[width=0.3275\textwidth]{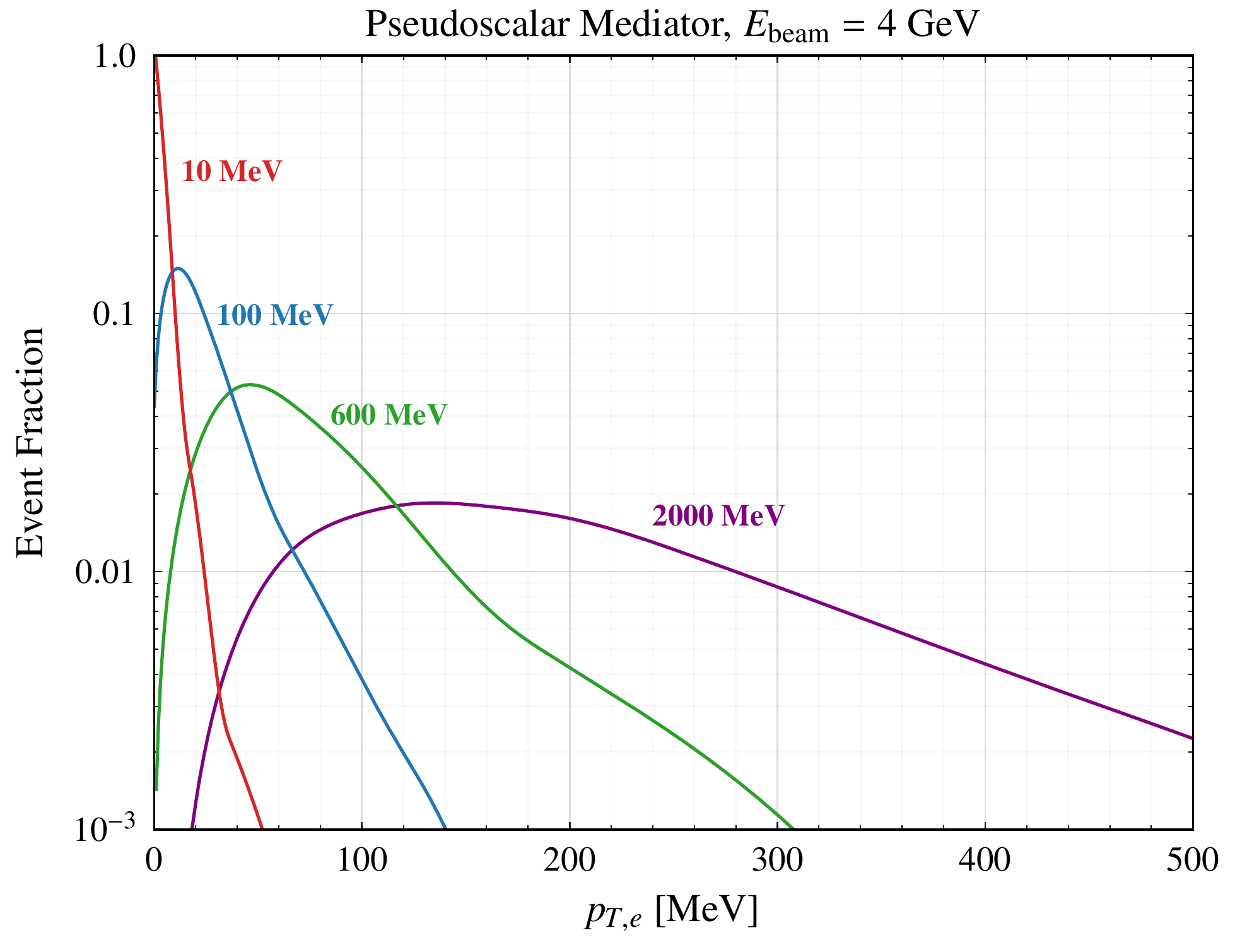}
      \caption{ Kinematic distributions for the recoiling electron in radiative scalar
      and pseudoscalar emission in $e N \to e N s/a$ fixed target reactions that utilize the interactions 
 in Eq.~(\ref{LS0}). The {\bf top row} shows the outgoing electron's recoil energy  for the two 
 interaction types for various choices of $m_\chi$ and the {\bf bottom row}
 shows the corresponding electron $p_{T,e}$ distributions.
 In all cases, the incident electron
 beam energy of 4 GeV and a tungsten target are chosen to match projections for LDMX Phase 1~\cite{Akesson:2018vlm}.
 }
    \label{scalarhistograms}
\end{figure*}




\begin{figure*}
    \centering
    \includegraphics[width=0.3275\textwidth]{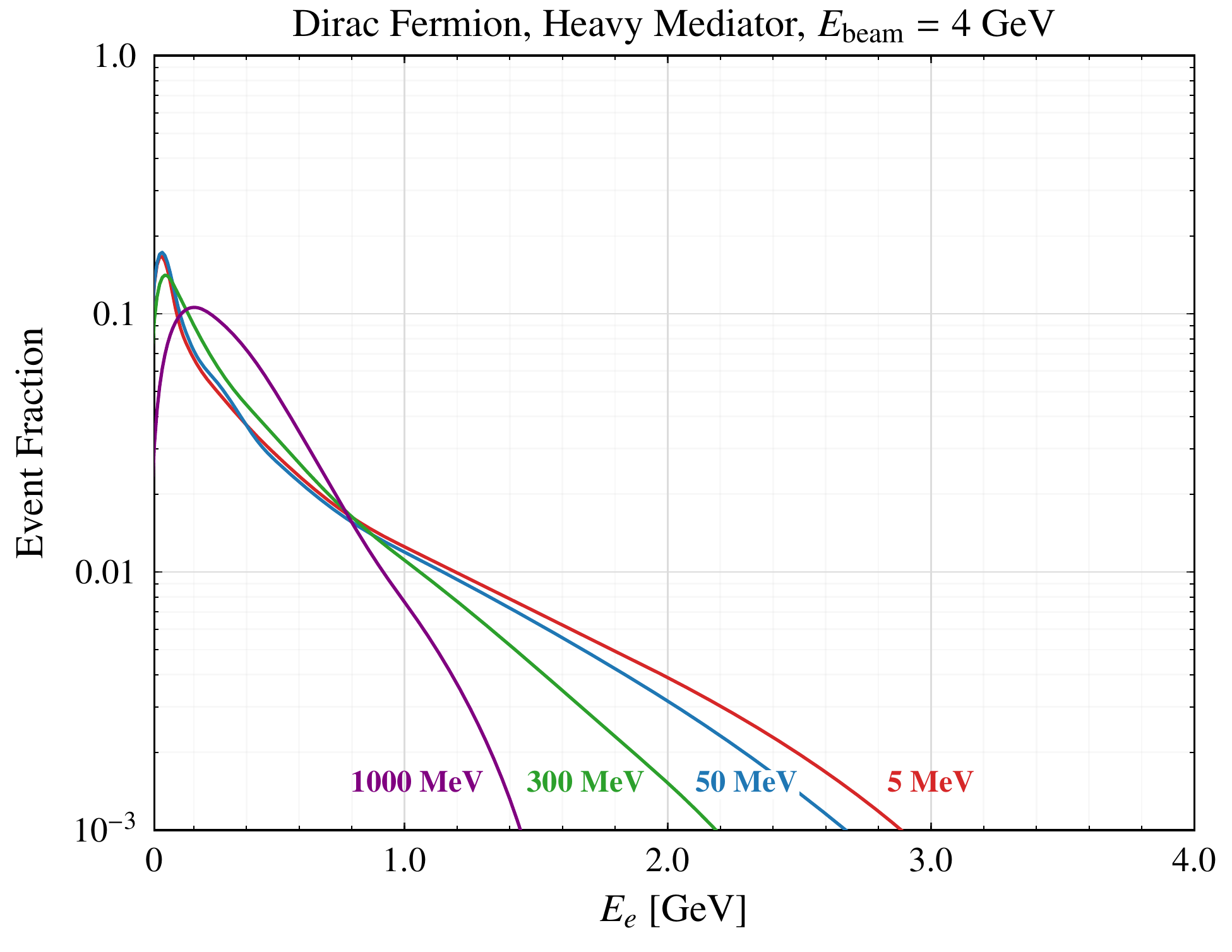}~~
        \includegraphics[width=0.3275\textwidth]{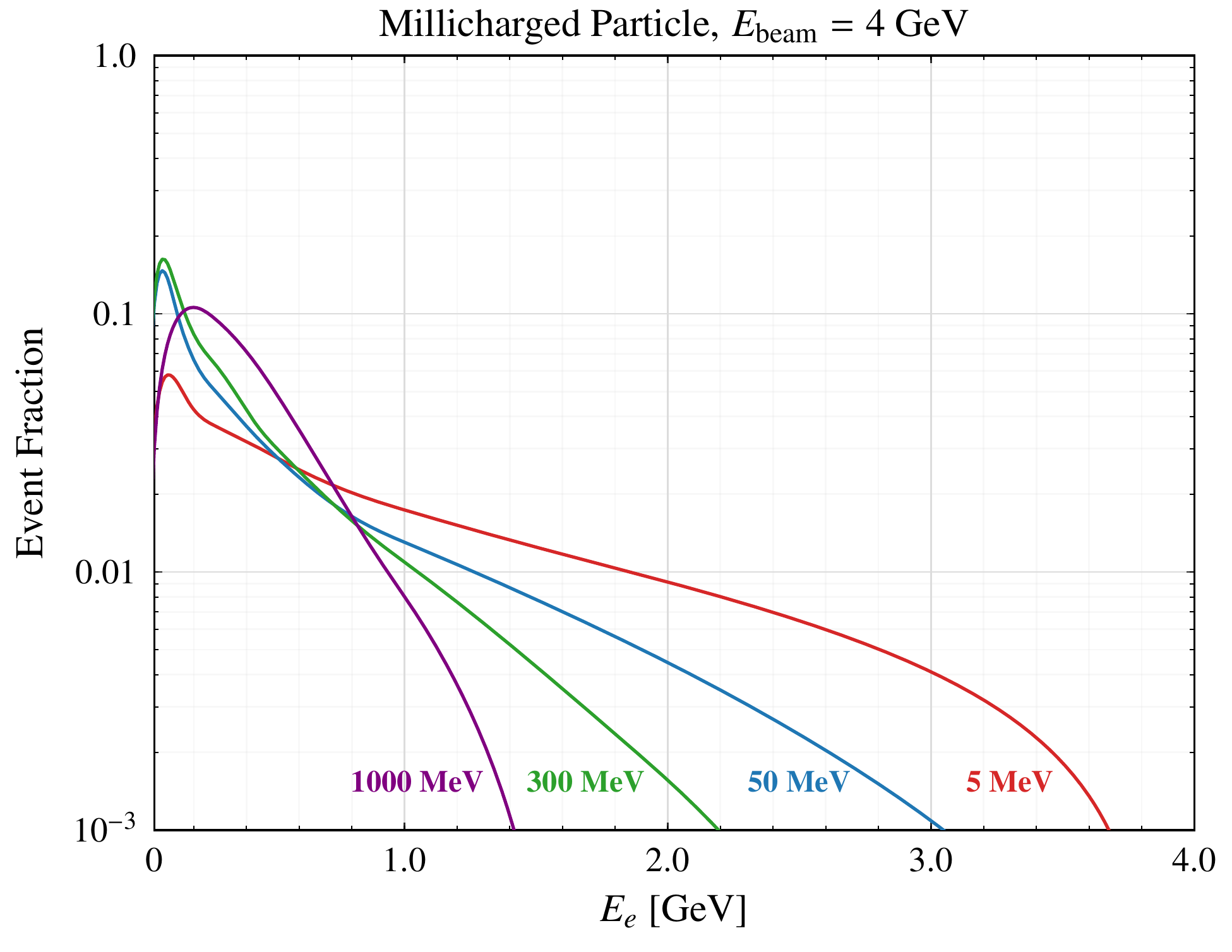} \\ 
        \medskip
    \includegraphics[width=0.3275\textwidth]{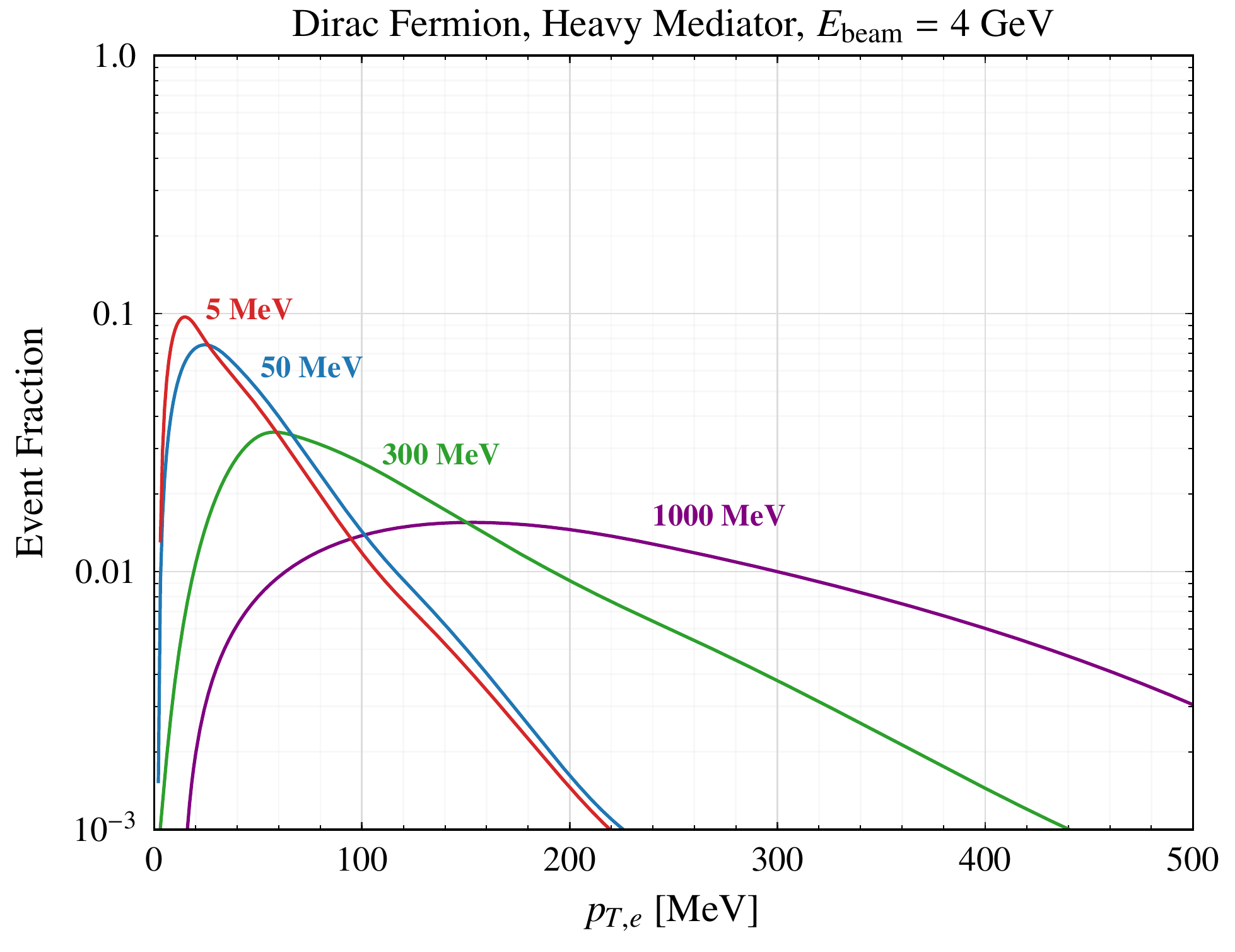}~~
        \includegraphics[width=0.3275\textwidth]{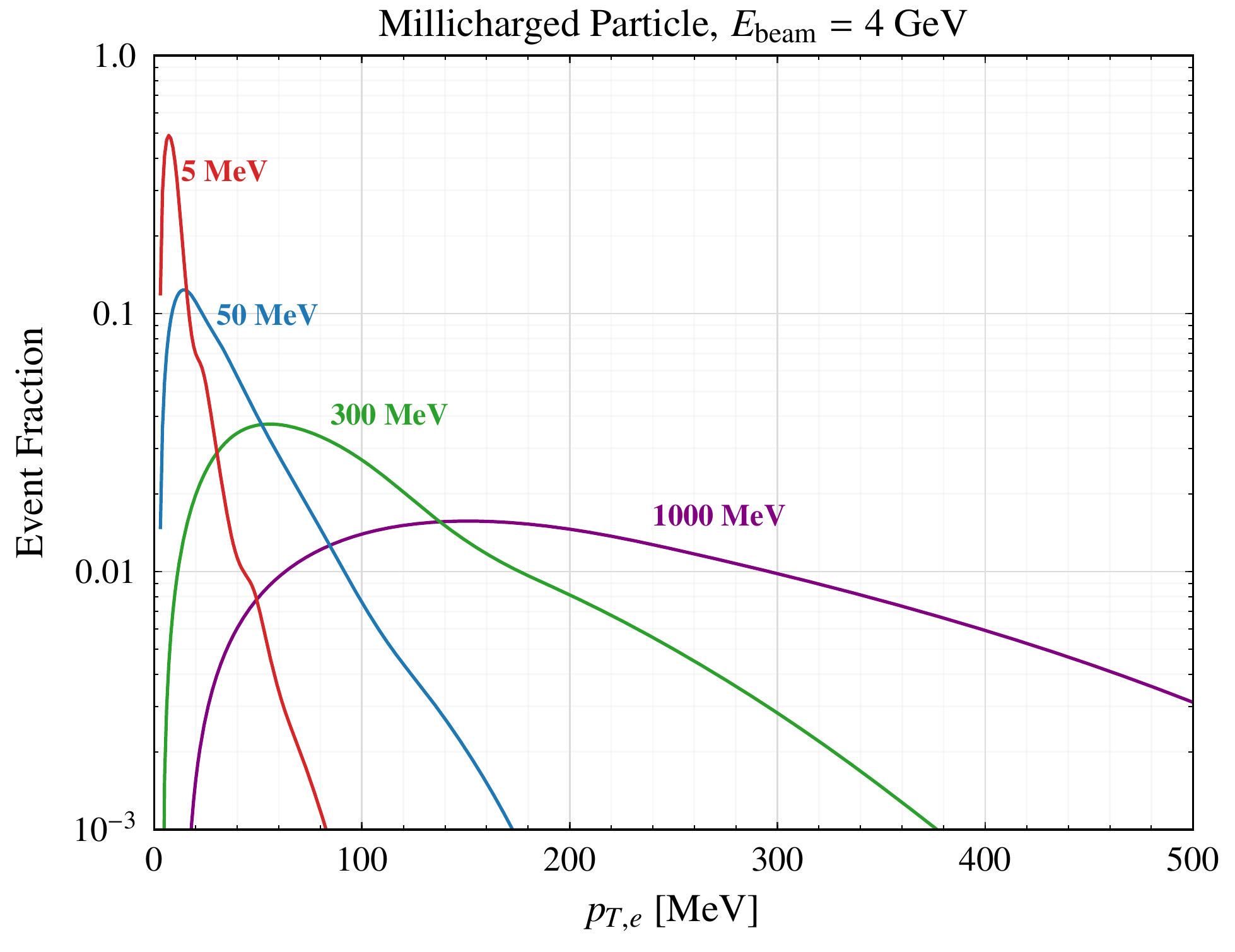}
      \caption{ Kinematic distributions for the recoiling electron in direct pair-production of Dirac
      fermions in $e N \to e N \chi \bar \chi$ fixed target reactions. The models considered in the left and 
      right columns correspond to the contact (heavy mediator) and ``millicharge" (light mediator) interactions
 in Eq.~(\ref{L24}). The {\bf top row} shows the outgoing electron's recoil energy  for the two 
 interaction types for various choices of $m_\chi$ and the {\bf bottom row}
 shows the corresponding electron $p_{T,e}$ distributions.  
  In all cases, the incident electron
 beam energy of 4 GeV and a tungsten target are chosen to match projections for LDMX Phase 1~\cite{Akesson:2018vlm}.
}
    \label{directhistograms}
\end{figure*}

The existing MM literature has largely focused on the kinematics of signals arising from the radiative production of
massive spin-1 bosons with vector current interactions (e.g. kinetically mixed dark photons)~\cite{Izaguirre:2014bca,Akesson:2018vlm,Akesson:2019iul}. 
In this paper we generalize these studies to explore the 
kinematics of recoiling beam leptons in fixed target interactions
\be
\ell N \to \ell N +X\dots ,
\ee
where $\ell = e$ or $\mu$, and $X$ represents a broad range of possible final state particles, including single boson 
emission in $2\to 3$ scattering and direct DM pair production in $2\to 4$ processes as depicted in Fig. \ref{cartoon}.
Furthermore, we develop novel strategies and statistical tests to distinguish these different model hypotheses 
and quantify the signal samples required for model discrimination. 

This paper is organized as follows: in Section~\ref{models} we introduce representative phenomenological interactions for MM signals and describe the details of our numerical simulations; in Section~\ref{kinematic} we evaluate the model discrimination potential of final state kinematic variables; in Section~\ref{polarization} we study how varying the initial state beam energy and polarization yields new observables; in Section~\ref{muons} we explore the additional clues that muon beams 
can offer; we offer some concluding remarks in Section~\ref{conclusion}.




\section{Models and Simulations}
\label{models}

Throughout this paper our goal is to distinguish between representative signal models categorized according to 
how they couple to leptons or the mass of new particles involved.
In this section, we define
these interactions by their fixed-target production mode as radiative $2 \to 3$ reaction $\ell N \to \ell N X$, where 
$X$ is an invisibly decaying particle, or a $2 \to 4$
reaction that pair-produces the DM $\chi$ directly through an off-shell mediator.\footnote{For our purposes, $X$ and $\chi$ need only be ``invisible" on 
detector length scales of order a few meters; if they decay promptly into visible or semi-visible final states, additional model-dependent
signals may be more useful for model discrimination. However, exploiting these features is beyond the scope of this paper.}

\subsection{Mediator Production: \texorpdfstring{$2\to 3$}{2 to 3} Processes}
\label{mediators}
The nominal LDMX/M$^3$ signal process involves renormalizable 
interactions through which a single ``mediator" particle $X$ is emitted
in $\ell N \to \ell N X$ reactions inside the target, where $\ell = e$ or $\mu$ is the 
lepton beam particle and $N$ is a target nucleus.

The renormalizable (mass dimension $\le$ 4) interactions with new spin-1 
states  are
\be
\label{LS1}
{\cal L}_{S=1} = 
e \epsilon \times \begin{cases}
A^\prime_\mu \bar \ell \gamma^\mu \ell &  \text{vector} \\
Z^\prime_\mu \bar \ell \gamma^\mu P_{L,R} \ell &  \text{chiral vector} \\
V_\mu \bar \ell \gamma^\mu \gamma^5 \ell &  \text{axial vector} ,
\end{cases}
\ee
where we have adopted an arbitrary overall normalization $e\epsilon$ to match
the interaction of a kinetically mixed dark photon; $e = \sqrt{4\pi \alpha_{\rm EM}}$ is
the electric charge and $\epsilon$ parametrizes the strength of this interaction relative to electromagnetism. 
Here $P_{L,R} \equiv (1\mp\gamma^5)/2$ is a left (right) projector and we consider various mass choices
for each possible scenario $m_{A^\prime},m_{Z^\prime}, m_{V}$. 
Constraints on light new vector boson interactions with anomaly-free couplings to Standard Model (SM) particles
are summarized in Refs.~\cite{Ilten:2018crw,Bauer:2018onh}; bounds on chiral and axial vector 
interactions are more model dependent, but are typically very strong because such interactions require additional SM-charged field content for
anomaly cancellation~\cite{Kahn:2016vjr,Dror_2017dark}. 

We also consider renormalizable spin-0 interactions
\be
\label{LS0}
{\cal L}_{S=0}= 
e \epsilon \times \begin{cases}
s \bar  \ell\ell &  \text{scalar} \\
i a \bar \ell \gamma^5  \ell &  \text{pseudo-scalar} ,\\
\end{cases}
\ee
for scalar and pseudoscalar interactions 
with corresponding masses $m_s$ and $m_a \lesssim$ GeV.
Note that unlike the vector interactions in \Eq{LS1}, which can be gauge-invariant at high energies,
 the Yukawa couplings in \Eq{LS0} are not invariant under the $SU(2)_L \times U(1)_Y$ gauge group, so 
 they must arise from higher dimension operators proportional to the source of
 electroweak symmetry breaking. We therefore expect $\epsilon \propto v/F$, where $v = 246$ GeV is the Higgs
 vacuum expectation value and $F$ is the mass scale 
 of the heavy particle whose quantum numbers restore SM gauge gauge invariance.
As such, the bounds on these interactions depend on the details of the ultraviolet completion; 
they are summarized in Refs.~\cite{Essig:2010gu,Dolan:2014ska,Krnjaic:2015mbs,Batell:2017kty,Egana_Ugrinovic_2020,Liu:2020qgx} 
for many representative examples.

\medskip
\subsection{DM Pair Production: \texorpdfstring{$2\to 4$}{2 to 4} Processes}
\label{contact}

If the mediators in Sec \ref{mediators} are too heavy for direct production or 
if their decays to DM pairs are kinematically forbidden, the DM can still be produced directly 
via $\ell N \to \ell N \bar \chi \chi$ reactions with virtual mediator exchange. 
Here we consider two representative limiting cases in which the amplitudes for DM pair production 
are proportional to
\be
\label{L24}
{\cal M}_{\rm pair} \propto 
e \epsilon \times \begin{cases}
\dfrac{1}{\Lambda^{2}} \big(\bar \ell \gamma^\mu  \ell\big)(\bar \chi \gamma^\mu \chi)  &  \text{heavy mediator} \\
&\vspace{-0.3cm}  \\
\dfrac{1}{q^2} \big(\bar \ell \gamma^\mu \ell\big)(\bar \chi \gamma_\mu \chi)  &  \text{``millicharge"} \\
\end{cases}~~~~~
\ee
where $\Lambda \gtrsim$ GeV represents the mass scale of a vector mediator that has been integrated
out to generate a contact operator and $q$ is the momentum imparted to the $\bar \chi \chi$
system in the limit $\Lambda < 2m_\chi$ where the mediator cannot decay to DM particles. 
Note that in our numerical studies the heavy and light mediator cases are modeled using 
renormalizable interactions from Sec.~\ref{mediators} by taking the mediator mass $m_{A'} \gg E_{\rm beam}$ or $m_{A'} \ll 2m_\chi$, 
respectively.
These two classes of processes are represented schematically by the middle and right diagrams in Fig. \ref{cartoon}. 
We note that the list of operators in \Eq{L24} is not exhaustive and can include different Lorentz structures (e.g., additional
$\gamma^5$ insertions from integrating out the axial vector in \Eq{LS1}).

\subsection{Simulation Details}

For the numerical studies in the remainder of this paper, 
we generate signal samples for electron and muon beams based on Eqs. (\ref{LS1}) - (\ref{L24})
using version 2.6.4 of \texttt{MadGraph 5 aMC@NLO}~\cite{Alwall:2014hca} including 
elastic and inelastic atomic and nuclear form factors for the target~\cite{Kim:1973he,Tsai:1973py}. 
The inclusive signal production
cross sections for these scenarios are presented in Fig. \ref{fig:cross-sections}
as functions of either the mass of the radiated particle for the vector mediator
in Eq. (\ref{LS1}) or as a function of  $2m_\chi$  in Eq. (\ref{L24}) where appropriate.

Although our qualitative conclusions below are largely independent of any particular
experimental setup, for electron beam studies, our simulation is designed to match the anticipated
 LDMX phase 1 design with a 4 GeV electron beam impinging on a thin tungsten target 
 \cite{Akesson:2018vlm}.  Similarly, our muon beam simulation is motivated 
 by the M$^3$ concept, which has been studied for a 15 GeV 
 beam energy and also with a tungsten target \cite{Kahn:2018cqs}.
 For our signal samples we select events with $E_{e} \leq 1.2\;\GeV$ and 
 $E_\mu \leq 9\;\GeV$ for LDMX and M$^3$, respectively.




\section{Kinematic Variables}
\label{kinematic}

\begin{figure*}
  \centering
   \includegraphics[width=0.35\textwidth]{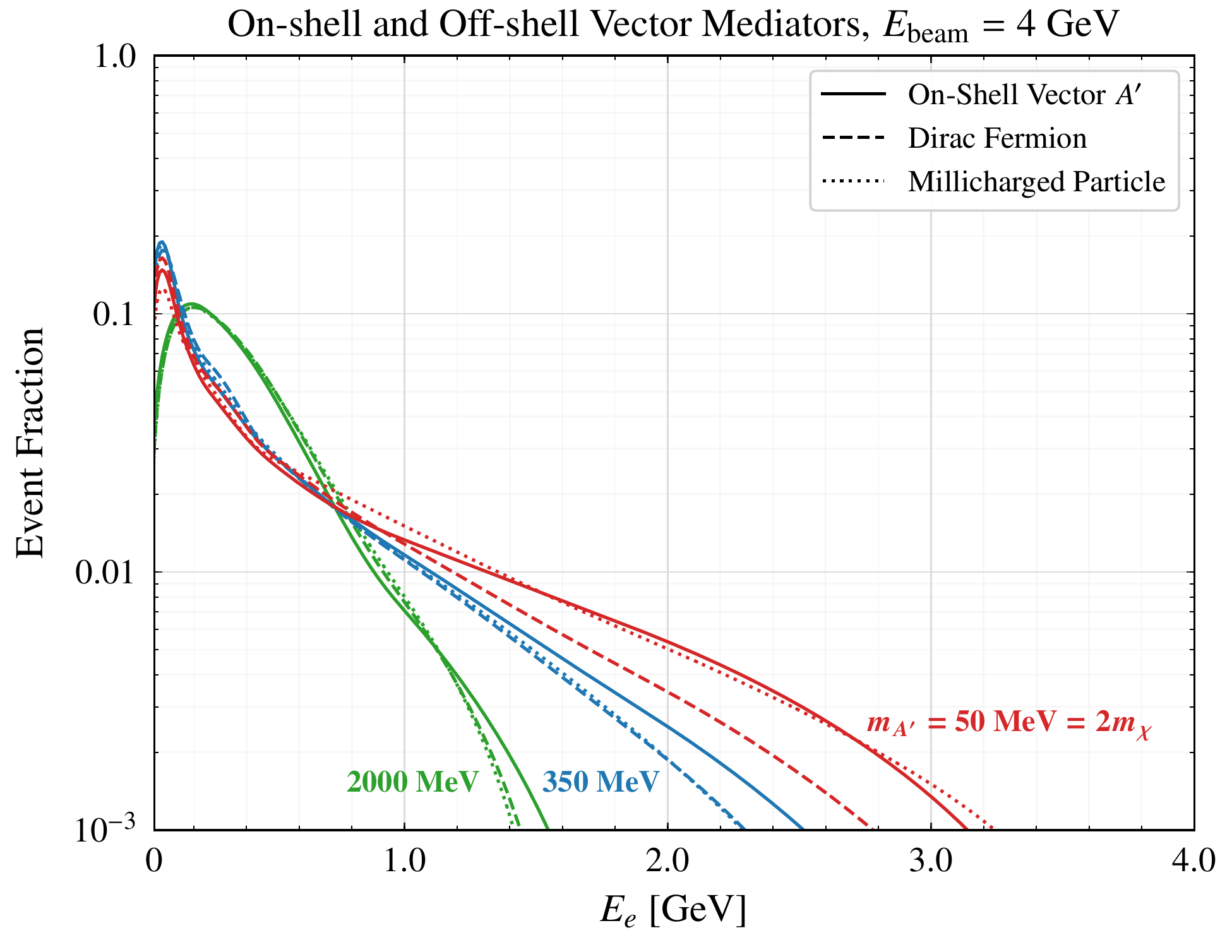}~~~
    \includegraphics[width=0.35\textwidth]{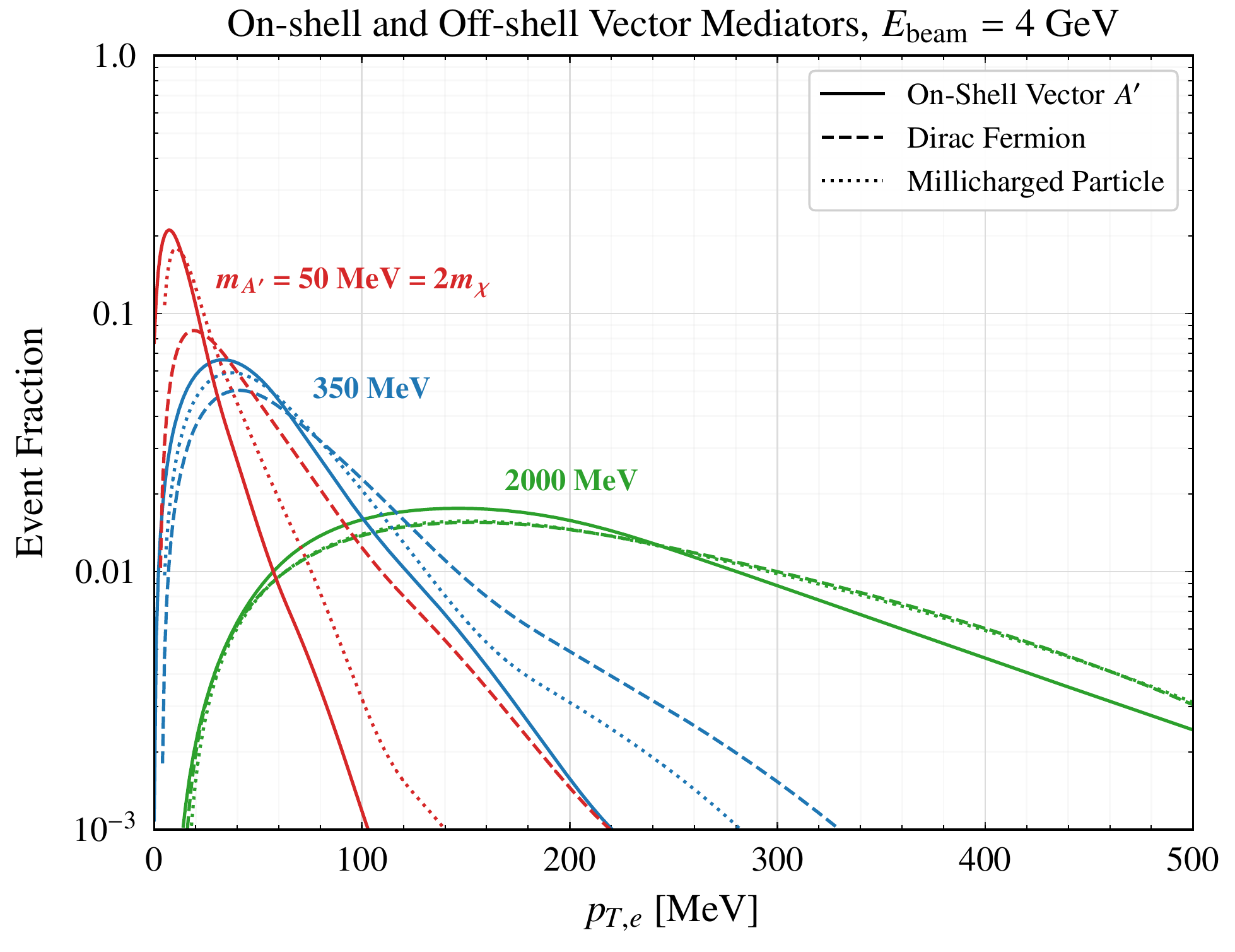}
    \caption{Kinematic distributions for on-shell vector mediator (solid curves), off shell vector mediator with $m_{A^\prime} = 10$ GeV (dashed curves), and off shell massless vector mediator (dotted curves) for different masses. All distributions are for a 4 GeV electron beam colliding with a tungsten target. 
}
    \label{fig:onvsoffshell}
\end{figure*}

This section generalizes earlier studies of dark photon production in 
missing momentum experiments~\cite{Izaguirre:2014bca,Akesson:2018vlm,Akesson:2019iul} by using
kinematic variables -- $E_{\rm beam}$ (beam energy), $E_{e}$ (electron recoil energy), and $p_{T,e}$ (recoil electron transverse momentum) -- as tools for distinguishing various new physics scenarios. 
 The differential distributions of $E_e$ and $p_{T,e}$  are shown in Figs. \ref{vectorhistograms}  and \ref{scalarhistograms} for the on-shell emission of spin-1 and spin-0 particles (Sec. \ref{mediators}), 
 and in Fig.~\ref{directhistograms} for direct DM production (Sec. \ref{contact}) via heavy and light mediators; in Fig.~\ref{fig:onvsoffshell} we 
 directly compare the distributions of these models. We discuss these results in more detail below.  
 
\subsection{On-Shell Mediators: Mass Measurement}
On shell mediator emission arises in $2\to 3$ processes as shown in the left panel of Fig.~\ref{cartoon}.
The corresponding kinematic distributions shown in Fig.~\ref{vectorhistograms} and \ref{scalarhistograms} 
are sensitive to the mass of the emitted particle. It is therefore interesting to investigate the ability 
of missing momentum experiments to distinguish different mass hypotheses. 
We quantify this discriminating power using a simple likelihood ratio test as described below.

We start by representing a histogram of experimental events (either in $E_e$, $p_{T,e}$ or both) as a vector 
$\vec d = (d_1, \dots , d_N )$.
We treat each bin as an independent counting experiment with Poisson statistics; the $d_i$ are then drawn from the Poisson distribution $f_P(d_i,\nu_i)$ 
\be
f_P(d,\nu) = \frac{\nu^d}{d!} e^{-\nu},
\ee
where $\nu$ are the predicted means in each bin which define a model hypothesis. 
The joint probability distribution or likelihood for $\vec d$ to be observed given the means $\vec \nu$ is~\cite{Cowan:1998ji,Cowan:2013pha}
\be
L(\vec{d}\; | \; \vec{\nu}) = \prod_{i=1}^N f_P(d_i,\nu_i).
\ee
Models that describe the data better have a larger likelihood. 
For computational simplicity we work with the logarithm of the likelihood
\be
\ln L (\vec  d | \vec \nu) = 
\sum_{i=1}^N \left( d_i \ln \nu_i - \nu_i - \ln d_i ! \right),~~~
\label{eq:simple_log_lkl}
\ee
and define a test statistic (TS) $\lambda$ to compare two different models $A$ and $B$ for 
a given data set
\be
\lambda = - 2 \left[   \ln L(\vec d\,|\, \vec \nu_{A}) -   \ln L(\vec d\,|\, \vec \nu_{B})  \right].
\label{eq:ts}
\ee
This test statistic is negative when model $A$ is preferred over $B$ and positive otherwise (in the limit of 
large statistics in all bins $\lambda$ simply becomes a difference of $\chi^2$ values for the two models). 
We use our Monte Carlo event samples to generate many mock experiments and study the resulting 
distribution of $\lambda$ for various combinations of hypotheses. 
If there are enough signal events, one can reject the $A$ hypothesis with a given confidence level CL compared to $B$
if the probability of obtaining $\lambda < 0$ is $1 - \mathrm{CL}$. In other words, we can find the number of events $\Nevt$ 
such that 
\beq
p = \int_{-\infty}^{0} d \lambda f(\lambda; \Nevt) = 1 - \mathrm{CL},
\label{eq:p_value_eq_direct_comparison}
\eeq
where $f(\lambda; \Nevt)$ is the distribution of the test statistic for a given number of 
signal events.



\begin{figure*}
        \includegraphics[width=0.43\textwidth]{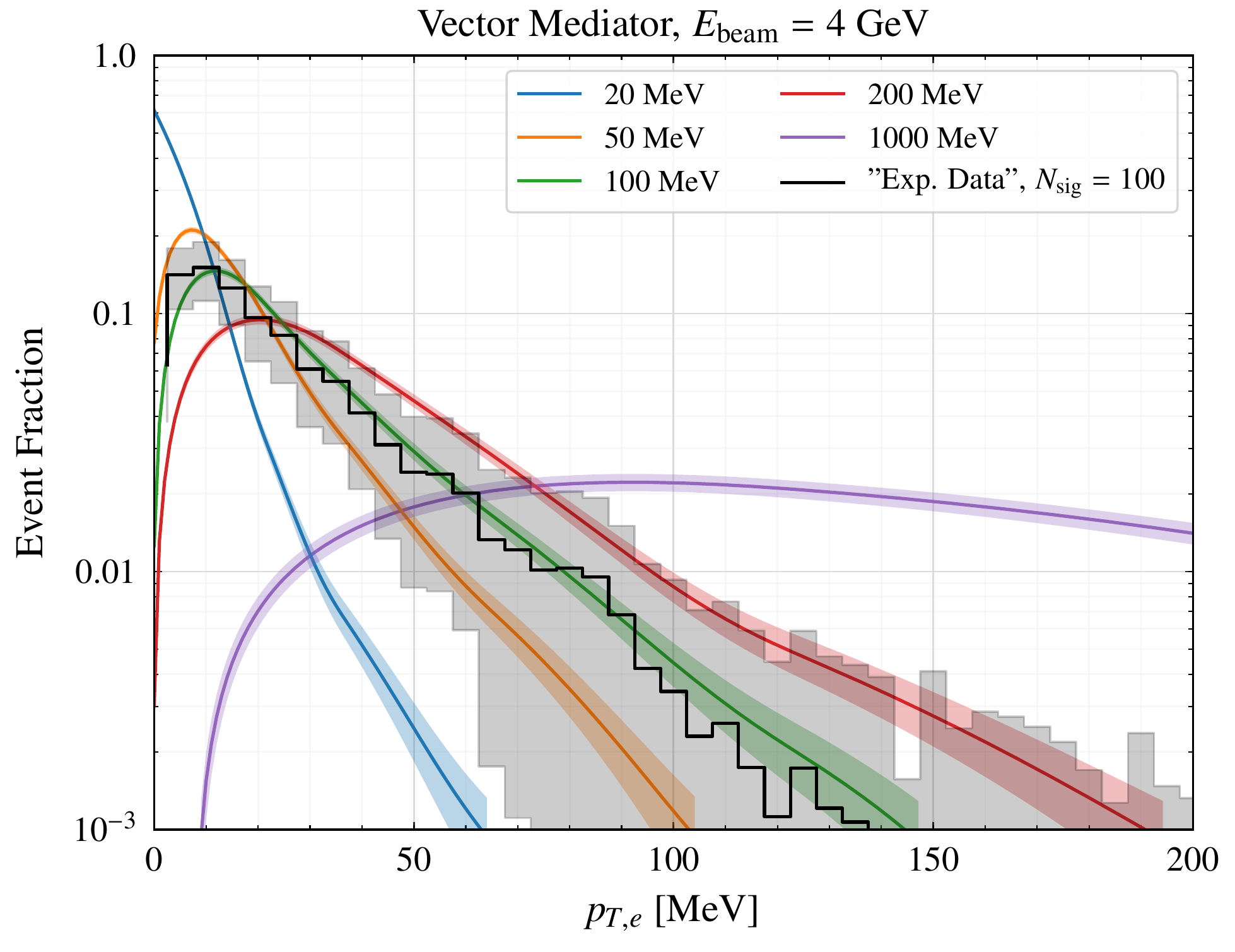}~~
        \includegraphics[width=0.47\textwidth]{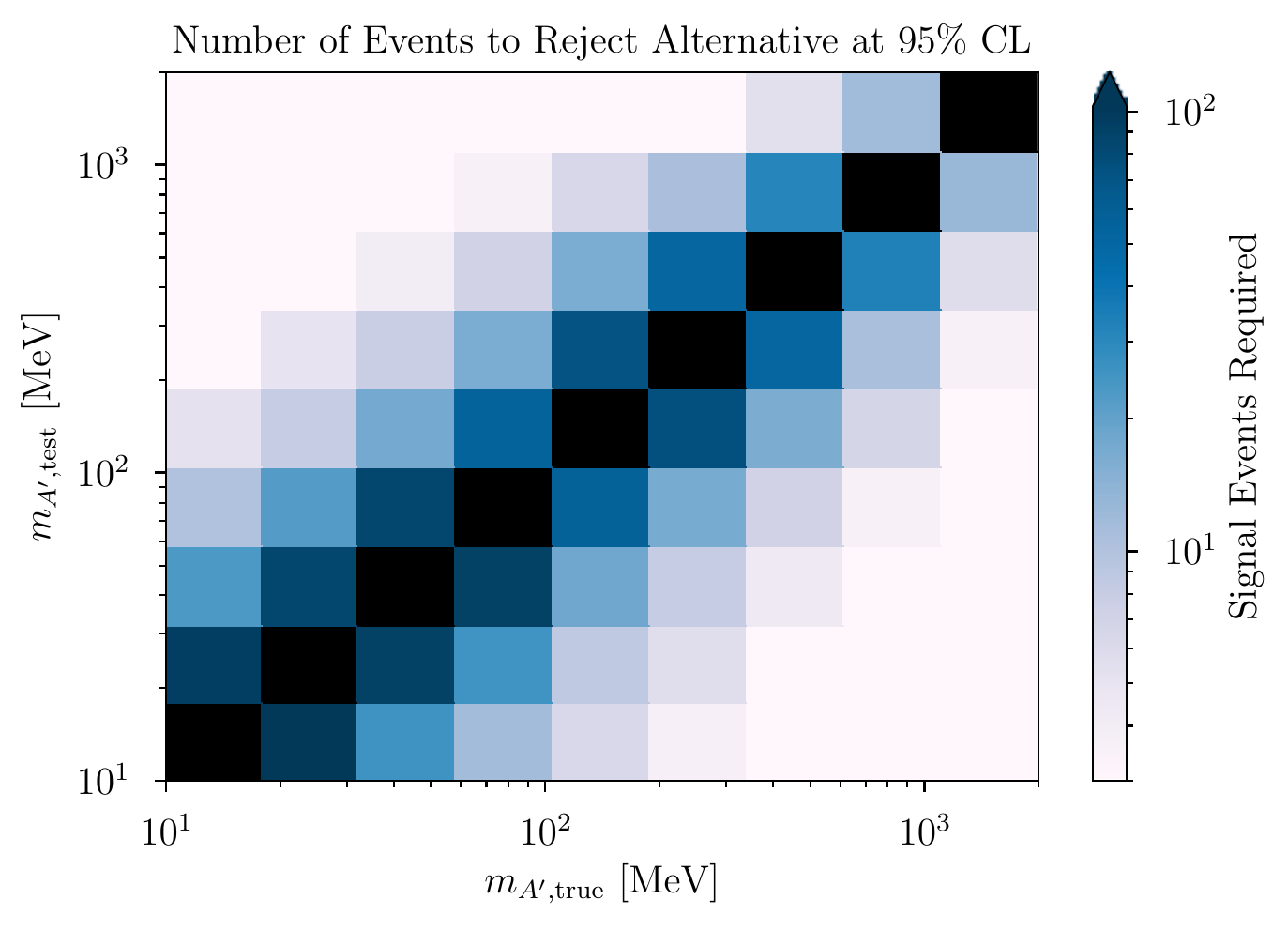} ~\\
	\includegraphics[width=0.43\textwidth]{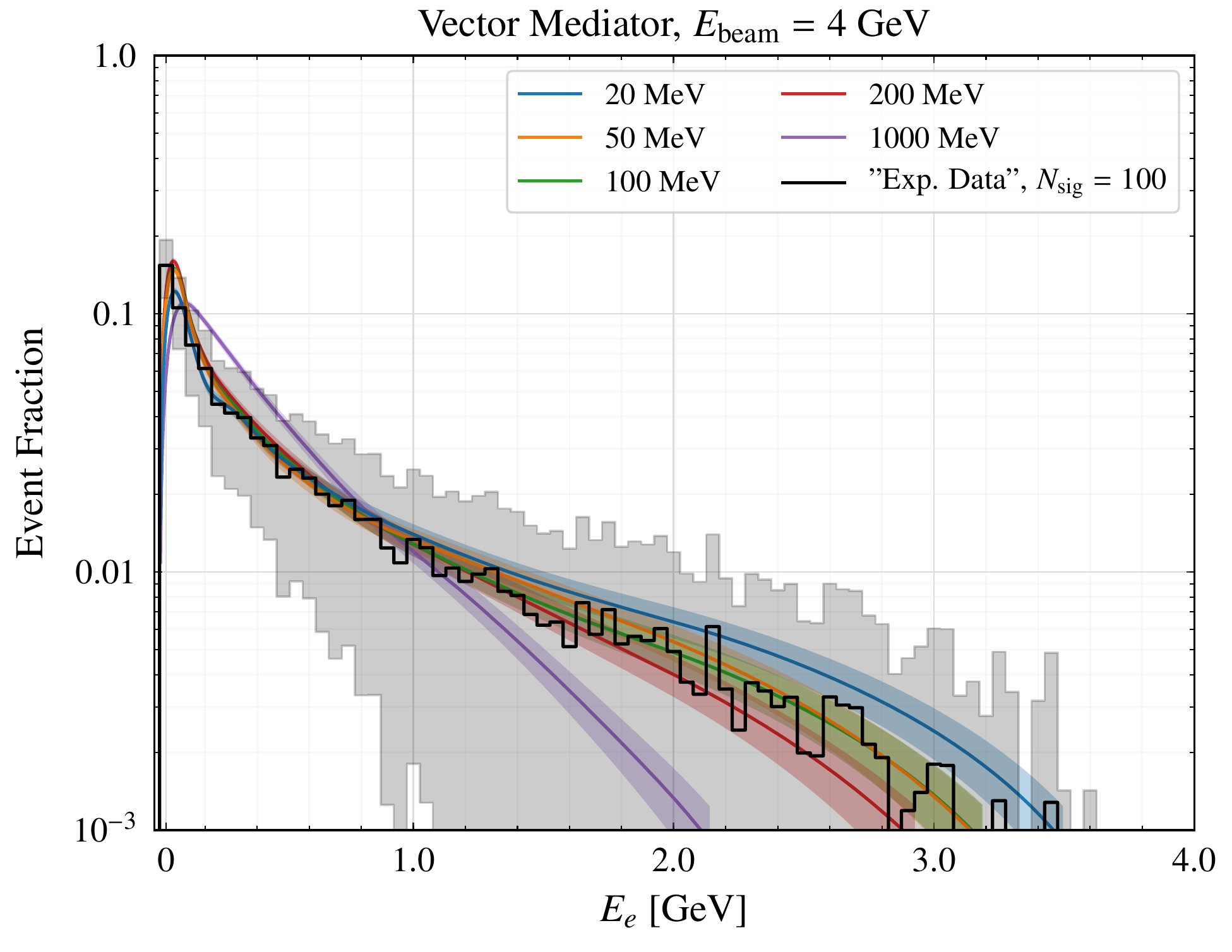}~~
  \includegraphics[width=0.47\textwidth]{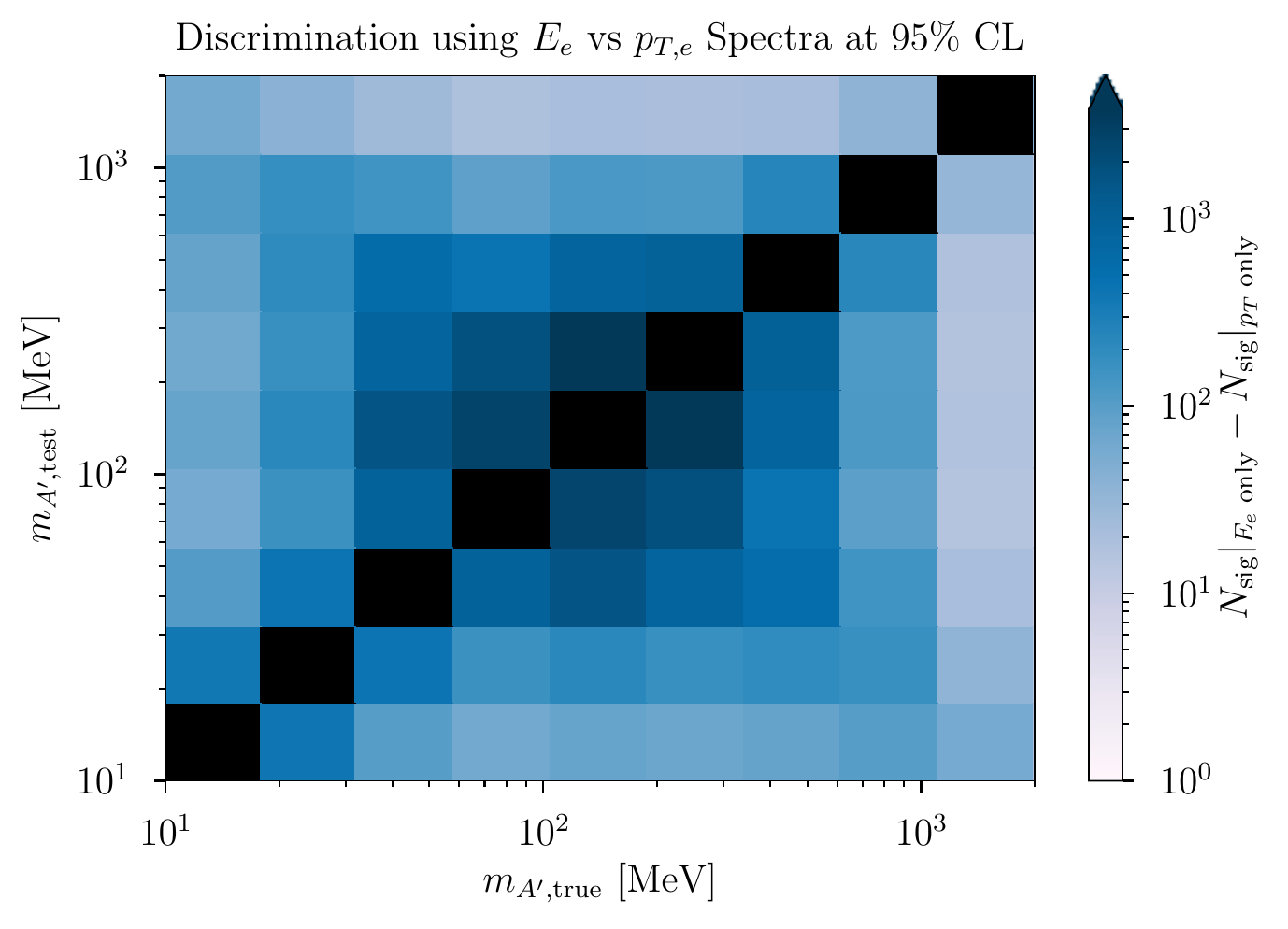}
      \caption{ 
      Mass discrimination sensitivity for vector mediators using electron transverse  
      momentum $(p_{T,e})$ and electron recoil energy $(E_e)$ distributions. {\bf Top left:} $p_{T,e}$ distributions
      for representative test hypotheses for different mass values (colored curves) plotted alongside  
  a mock dataset with $N_\text{sig} = 100$ signal events drawn from a distribution of $m_{\aprime} = 100$ MeV; the gray shaded band surrounding the black histogram represents the statistical uncertainty of the mock dataset.
{\bf Top right:} 
Two dimensional histogram of the number of 
signal events required to distinguish a given test mass hypothesis on the vertical axis from 
the true mass on the horizontal axis (the one used to generate the mock dataset) at 95\% confidence level assuming only statistical errors. 
Darker colors correspond to more signal events needed. This result makes use of the $p_{T,e}$ spectral information only.
{\bf Bottom left:} same as top left, but showing the electron recoil energy $(E_e)$ instead of transverse momentum $(p_{T,e})$.
{\bf Bottom right:} comparison of recoil energy-only versus transverse momentum-only 
analyses to discriminate between different mass hypotheses. The density plot shows the difference between the 
number of signal events required to distinguish a test model from the ``true'' scenario using only $E_{e}$ and 
$p_{T,e}$ information. This difference is strictly positive, implying that transverse momentum enables superior model 
discrimination with a much smaller number of signal events. This improvement is particularly pronounced 
for models with similar masses where the recoil energy spectra are nearly identical.
 }
    \label{fig:2D}
\end{figure*}

Note that due to the finite size of any Monte Carlo sample, a sufficiently fine binning of the events (or binning in multiple variables) 
will result in small statistics and significant fluctuations in individual bins. 
We address this issue by using a modification of \Eq{eq:simple_log_lkl} based on Ref.~\cite{Barlow:1993dm} as described in Appendix~\ref{sec:stat}. 
We also do not account for detailed experimental effects such as energy and momentum smearing, so 
our results should be viewed as an idealized best-case scenario. However, 
our choices of $p_{T,\ell}$ and $E_\ell$ binning are motivated by the detailed 
simulation-based detector studies of Refs.~\cite{Akesson:2018vlm,Akesson:2019iul}. 
In particular, we use $5\;\MeV$ $p_T$ bins and $1\%$ $E_{\ell}$ bins. While the 
latter is almost certainly an overly optimistic choice, we will show that 
$p_T$ spectra still offer superior model discrimination ability compared to recoil energy.

We show an application of this TS in Fig.~\ref{fig:2D}. 
The upper left panel shows a histogram of a mock dataset with 100 events drawn from 
a high-statistics MC $p_{T,e}$ distribution for a 100 MeV vector mediator emitted on shell as in Fig \ref{cartoon} (left). 
This distribution is compared to expected distributions for a set of different masses. It 
is clear that some hypotheses fall outside of the gray  $1\sigma$ statistical uncertainty band 
of the mock data and are therefore disfavored, while others are indistinguishable 
within errors. Equations~(\ref{eq:ts}) and~(\ref{eq:p_value_eq_direct_comparison}) 
allow us to quantify this observation. The result is shown in the upper right panel of Fig.~\ref{fig:2D}, 
which illustrates the number of signal events needed to distinguish a test mass hypothesis on the vertical axis from 
the ``true'' model (i.e., the one that was used to generate the mock data set) on the horizontal axis. We see that 
very different masses can already be distinguished with the background-free ``discovery threshold'' number of signal events $\Nevt = 3$, 
while comparable masses will require ${\pazocal O}(10 - 100)$ events to disentangle. 


\subsection{Missing Momentum vs. Missing Energy}

Armed with the statistical test introduced above, we can now 
compare the model discrimination power of 
missing momentum experiments like LDMX and M$^3$ against other 
fixed-target lepton beam techniques that only measure
the missing energy of the beam (e.g., the NA64 experiment \cite{NA64:2019imj,Gninenko:2019qiv}).

As in the previous section, we use simulated data to estimate the 
number of signal events required to distinguish a test hypothesis from the ``true'' model 
using recoil energy alone at a given confidence level, yielding a two-dimensional histogram similar to the 
upper right panel of Fig.~\ref{fig:2D}. We compare this to $p_{T,e}$-only discrimination 
by subtracting the two histograms. The resulting histogram is shown in the bottom right panel of Fig.~\ref{fig:2D}.
This difference is strictly positive, implying that transverse momentum enables superior model 
discrimination with a much smaller sample of signal events. 

For illustration, in the left column of Fig. \ref{fig:2D} we also show various $p_{T,e}$ (top) and $E_e$ (bottom) test  hypothesis templates (colored curves) plotted alongside $N_\text{sig} = 100$ events of simulated data drawn from
a vector mediator sample with $m_{A^\prime} = 100$ MeV (black curve). The gray band surrounding the black curve is the statistical uncertainty of the mock data. Visually it is clear that, relative to the $E_e$ distributions, the $p_{T,e}$ distributions span a greater variety of shapes for the same model parameters and thereby offer more discriminating power. For nearly the entire mass range shown in this figure, recoil energy distributions require ${\pazocal O}(100)$ events to discriminate between widely separated mass hypotheses (e.g., MeV and GeV), whereas transverse momentum can already exclude closely spaced choices (e.g., 10 and 100 MeV) requiring only a few events, near the discovery threshold for a zero-background environment.

\subsection{Off-Shell Mediators}

Ref.~\cite{Berlin:2020uwy} has shown that missing momentum 
experiments are sensitive to a wide range of mediator masses in direct DM production, including off-shell 
production through heavy mediators. In this section, we instead focus on the kinematic features of these signals and 
extend the discussion to the emission of invisible particles through a massless off-shell mediator (e.g., production of 
millicharged particles via the photon).

In Fig.~\ref{directhistograms} we show the distributions of electron recoil energies and transverse momenta 
for direct DM or millicharged particle production for a few different masses. 
These distributions look qualitatively similar to on-shell mediator emission. However, 
in the off-shell case, the invariant mass of the emitted particles is not fixed, so one naively expects broader distributions 
in $p_{T,e}$ compared to the on-shell case. This is explicitly illustrated in right plot of Fig.~\ref{fig:onvsoffshell} in the heavy mediator regime. 
A massless mediator leads to an enhancement of the amplitude at invariant mass close to the kinematic minimum of $2m_\chi$, 
which compensates for the expected broadening. Consequently, the transverse momentum distributions in the 
millicharge and on-shell cases look very similar. 

Figure~\ref{fig:onvsoffshell} illustrates another important point: the kinematic distributions 
are somewhat degenerate in ``theory space." That is, a given distribution can be interpreted as coming from an on-shell mediator 
of a certain mass, or from off-shell mediators and DM of different masses. Thus, it is essential 
to have a complementary array of experiments that can probe these interactions at different $\sqrt{s}$.
For example, it is clear from Fig.~\ref{fig:onvsoffshell} that a MM experiment with a few signal events will not be able 
to distinguish between the production of an on-shell mediator with $\mAp = 350\;\MeV$ or 
direct DM or millicharge production with $2m_\chi = 350\;\MeV$ via an off-shell mediator. 
A $B$-factory experiment like BaBar, Belle II or BESIII with $\sqrt{s} = 10.58\;\GeV$, on the other hand, can 
potentially observe very different signals in monophoton searches depending on the nature of the mediator: 
a sharp peak in the photon energy spectrum if the mediator is massive and $\mAp < \sqrt{s}$, or a 
broader excess if the mediator is massless~\cite{Essig:2013vha,Bevan:2014iga, Lees:2017lec, Kou:2018nap, Liang:2019zkb}.

However, we note that if the beam energy in a MM experiment is varied over a sufficiently broad range of values, the 
energy dependence of the new-physics cross section can be extracted, in principle, as shown in
Fig. \ref{fig:cross-sections}. Indeed, the interaction for a sufficiently heavy mediator arises from a higher-dimension operator,  so the signal rate grows more prominently with energy. Consequently, the signal's beam energy dependence can be used to distinguish this class of models from on-shell mediator production or millicharge-like production through a virtual light mediator.




\section{Beam Polarization}
\label{polarization}

In this section we quantify the discriminating power of the incident 
beam energy and polarization. While the nominal LDMX setup
is not designed to support polarization, electron beam polarimetry
for ${\pazocal O}$(few-10 GeV) beams is well established and
 there is no {\it a priori} impediment to the measurements we consider here~\cite{Benesch:2014bas,Zhao:2017xej}.
  Although detailed studies are ultimately needed to firmly establish the feasibility of a polarized source
  that can satisfy the other LDMX beam requirements (e.g., $\sim$ 100 pA currents required
to avoid pileup-related backgrounds), such efforts are beyond the scope of the present work.


\begin{figure}
\includegraphics[width=0.45\textwidth]{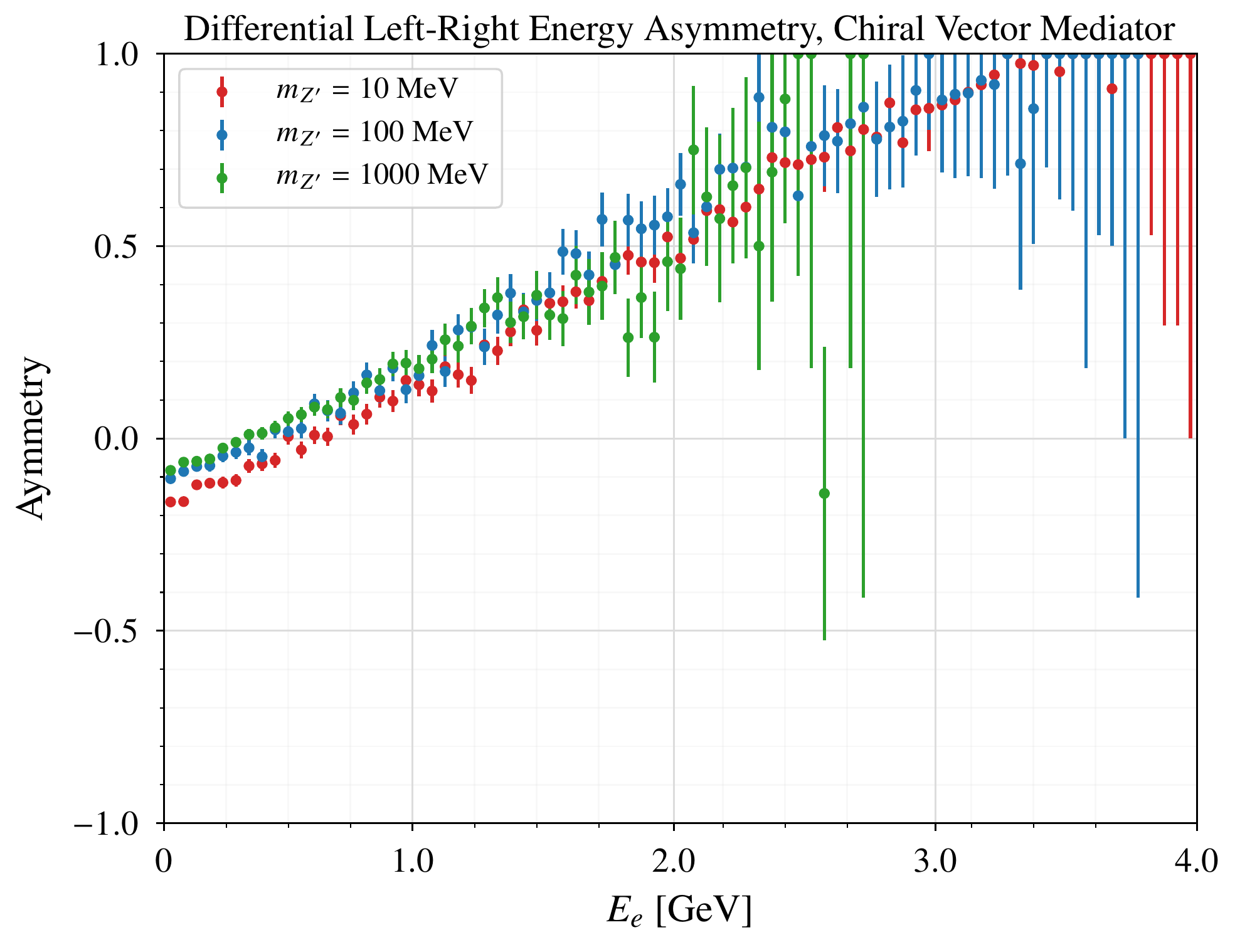}
\caption{
  Differential left/right asymmetry $A_{\rm LR}(E_e)$ from \Eq{asymm} for the (left-handed) chiral vector model plotted against
      electron recoil energy $E_e$ for a beam energy of 4 GeV. Also shown
      are the MC statistical uncertainties for each energy bin.
       }
\label{energyasymm}
\end{figure}
We define the differential left/right energy asymmetry in terms of a new observable

\be
\label{asymm}
{ A}_{\rm LR}(E_e) \equiv 
\left(   \dfrac{       d\sigma_{L} }{dE_e}      -  \dfrac{   d\sigma_{R} }{dE_e}                     \right)
\left(
\dfrac{  d\sigma_{L} }{dE_e}      + \dfrac{   d\sigma_{R} }{dE_e}    
\right)^{-1},
\ee
where $\sigma_{L/R}$ are signal production cross sections for left/right polarized beam particles. For the vector and axial vector interactions it can been shown that the differential asymmetry vanishes since the amplitudes for left- and right-handed electron beam are independent of the polarization. 

On the other hand, for chiral vector interactions involving a $P_L$ operator, the cross section for a right-handed polarized electron beam vanishes when $m_e \ll E_\text{beam}$. This is can be seen in Fig.~\ref{energyasymm} where we plot the differential left/right polarization asymmetry in signal events as a function of electron recoil energy $E_e$ for different chiral vector mediator masses, and we only show statistical 
error bars. We see that there is noticeable asymmetry for all masses, making the chiral vector scenario easily distinguishable from the vector and axial vector scenarios, which do not exhibit an asymmetry. 

Note that all potential backgrounds from QED processes (such as photonuclear hadron production) 
involve vector Lorentz structures, so no asymmetry is expected even if such events cannot 
be vetoed; electroweak ``invisible" backgrounds from direct neutrino production via off-shell
$Z$ exchange $eN\to eN\bar \nu \nu$ will contribute to the asymmetry, but the event rates 
for these processes are negligible for a Phase 1 LDMX run~\cite{Akesson:2018vlm,Akesson:2019iul}.

Independently of how the asymmetry in \Eq{asymm} is binned, we can also define an inclusive total asymmetry  observable 
\be
\label{Atot}
A_{\rm tot} \equiv \frac{   N_L - N_R }{N_L+N_R},
\ee
where $N_{L/R}$ is the number of signal evens observed using a left/right beam polarized
beam, assuming equal luminosities for the two data samples. The uncertainty on the total asymmetry is 
\be
\label{deltaALR}
\delta {A_{\rm tot}} =  2  \sqrt{ \frac{  N_L N_R      }{  (N_L+N_R)^3 }    },
\ee
where we have used standard error propagation and Poisson uncertainties for the
signal events $\delta N_{L/R} = \sqrt{N_{L/R}}$; to claim $n \sigma$  evidence of a polarization asymmetry, we require $A_{\rm tot} > n \delta A_{\rm tot}.$  Note that, while the total asymmetry defined in \Eq{Atot} is related to $A_{\rm LR}(E_e)$ from \Eq{asymm}, the former is not the integral of the latter. Furthermore, in the limit of a purely chiral interaction $(\propto P_L $ or $P_R$), signal events will overwhelmingly be produced with only one beam polarization, so an asymmetry can be identified with a small number of signal events, as soon as the number of signal events exceeds the Poisson error on the total event count.



\begin{figure*}
  \centering
    \includegraphics[width=0.34\textwidth]{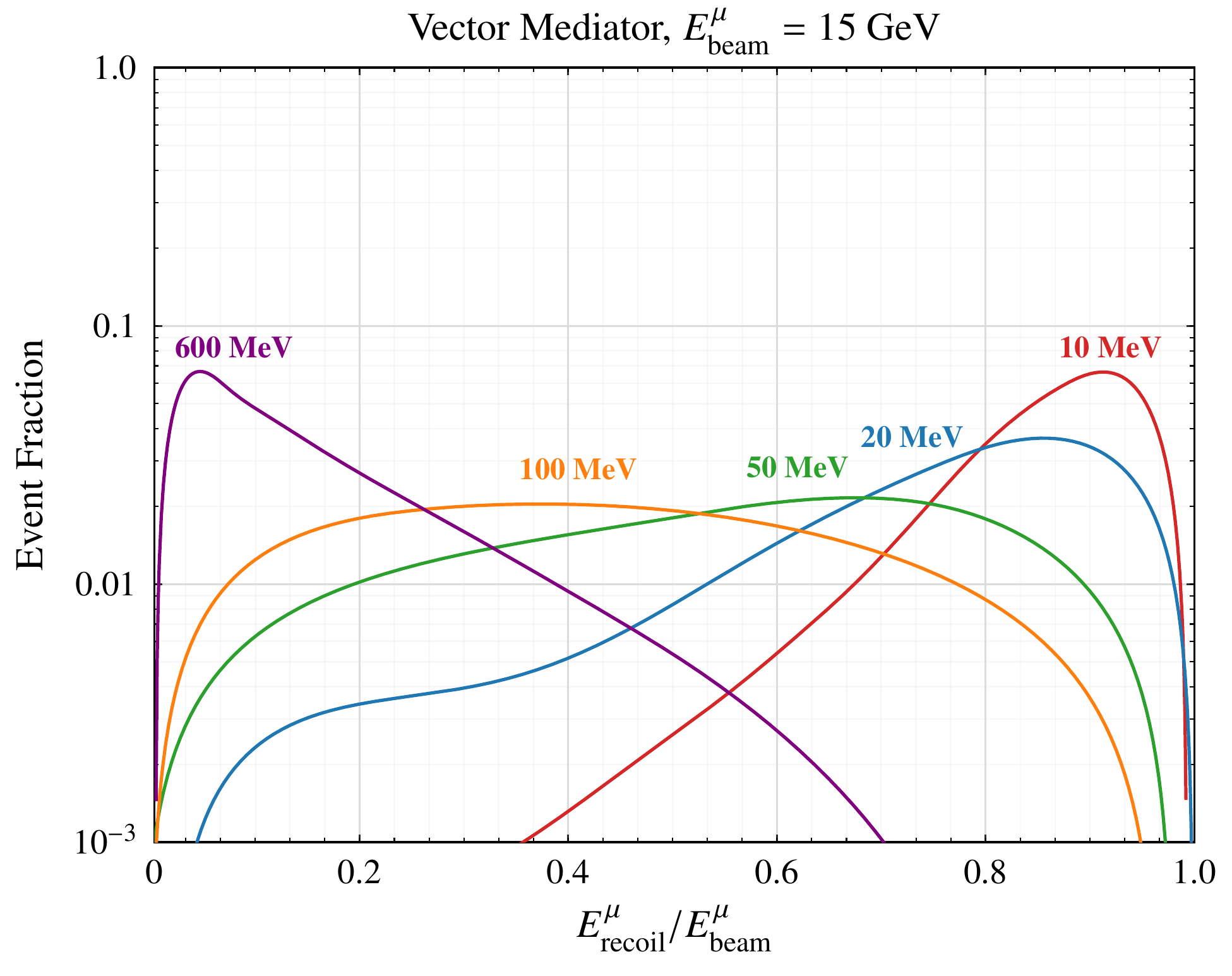}~~
        \includegraphics[width=0.34\textwidth]{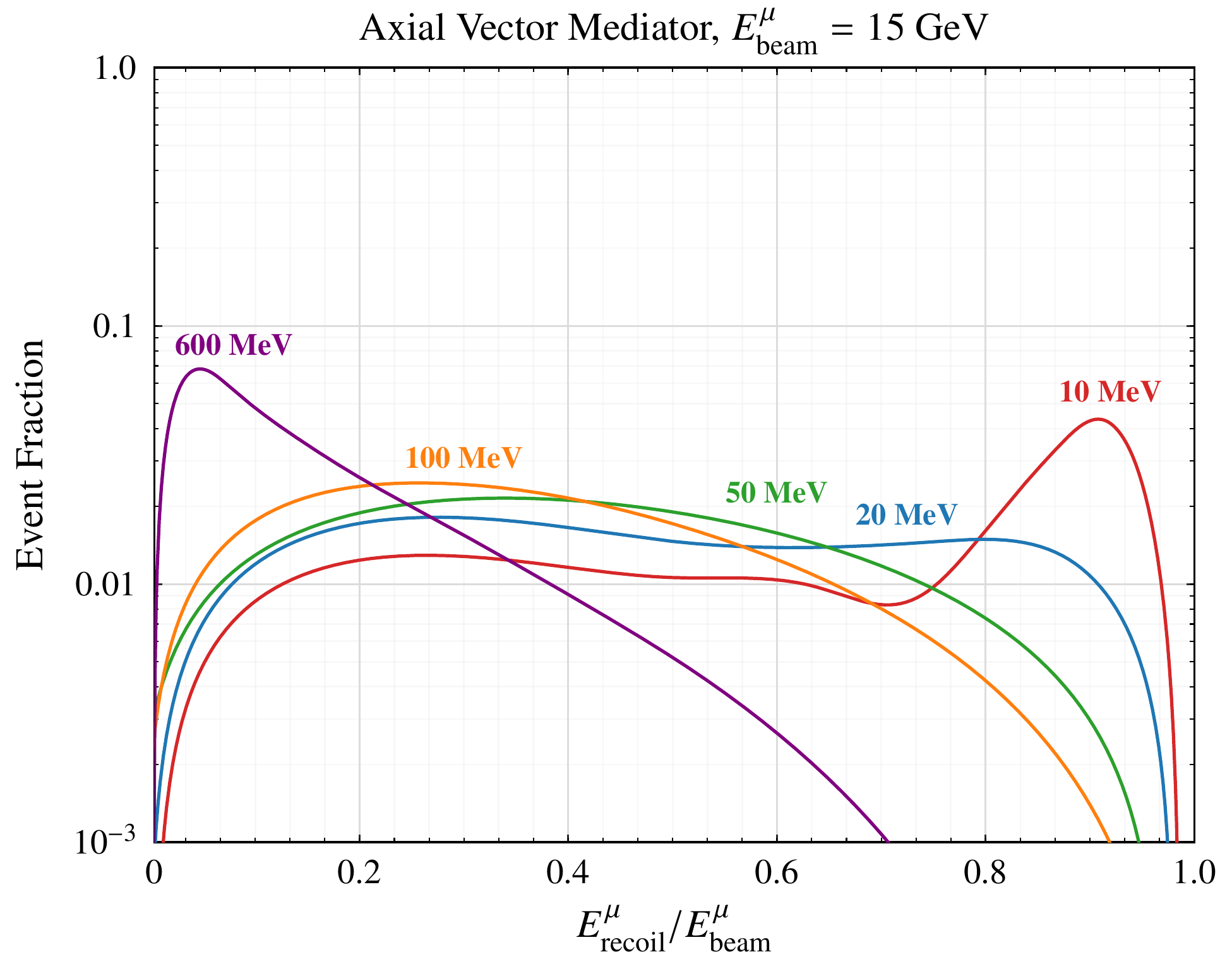} \\ 
        \medskip
    \includegraphics[width=0.34\textwidth]{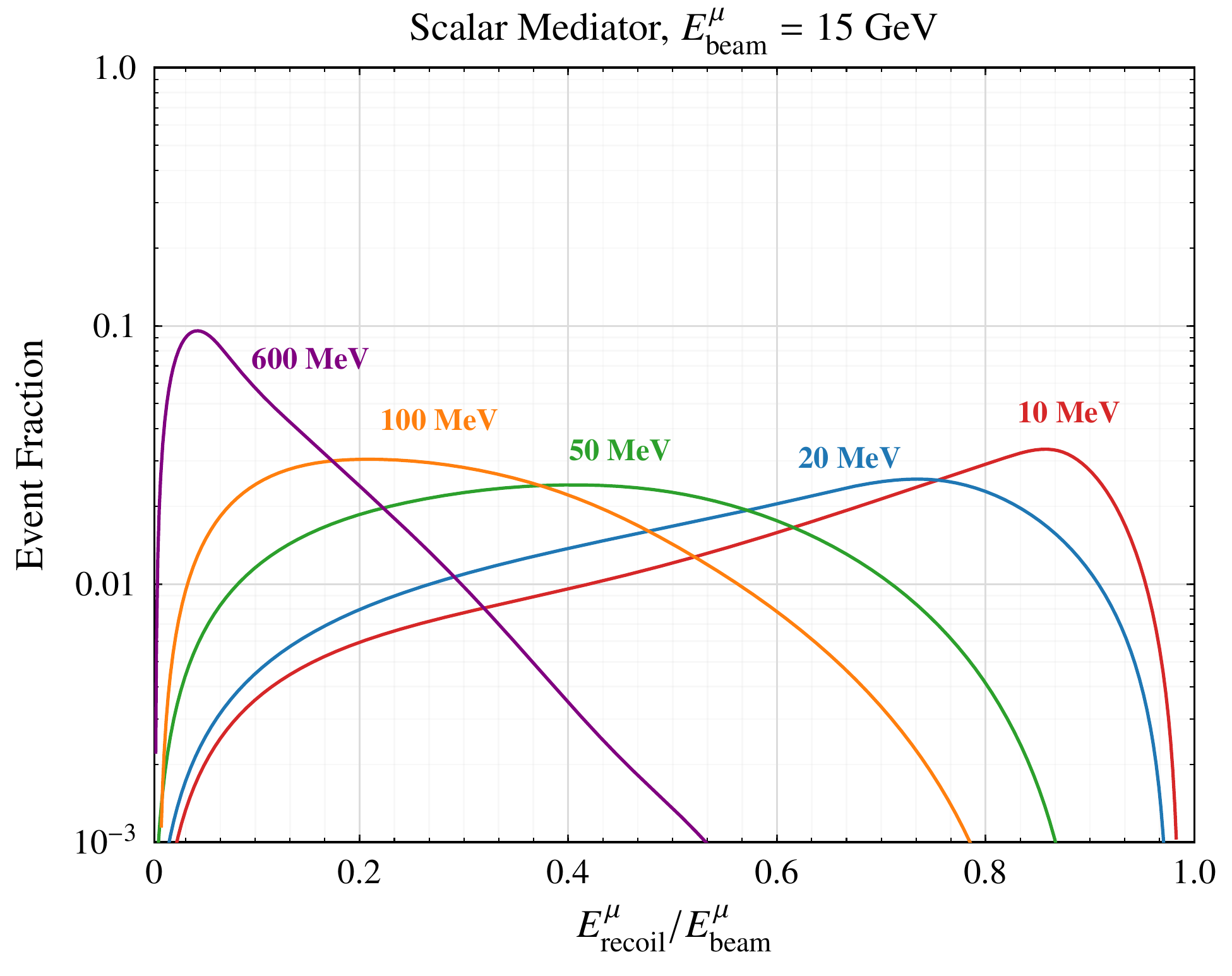}~~
        \includegraphics[width=0.34\textwidth]{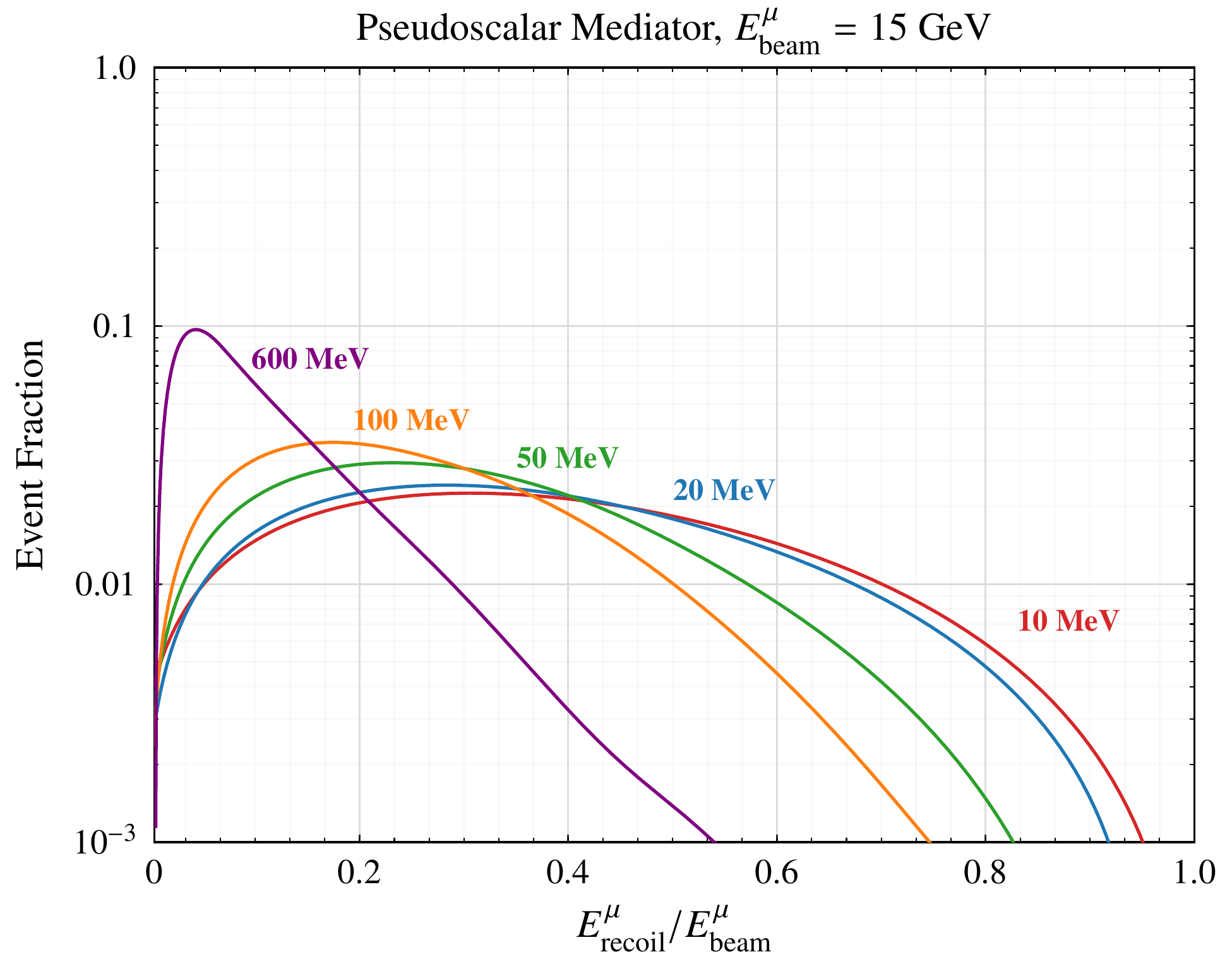}

      \caption{Energy distributions for the recoiling muon in on-shell production of a massive mediator particle in $\mu N \to \mu N X$ fixed target reactions. The {\bf top row} shows the outgoing muon's recoil energy for vector ({\it left}) and axial vector ({\it right}) interactions and  the {\bf bottom row} for scalar ({\it left}) and pseudoscalar ({\it right}) interactions. In all cases, the incident muon
 beam energy of 15 GeV is chosen to match projections for M$^3$ Phase 1 \cite{Kahn:2018cqs}.
}
    \label{muon-distributions}
\end{figure*}

\section{Electron vs. Muon Beams}\label{muons}

In the previous section we have seen that the observation of a polarization asymmetry in the electron energy distribution indicates that DM interacts with the SM via a chiral vector mediator, while non-observation of an asymmetry indicates that the mediator has either vector or axial vector interactions (similar conclusions hold 
for scalar mediators). Distinguishing between vector and axial vector interactions would require modifying the beam in a different way. In particular, the difference between the differential energy distributions of the vector and axial vector mediator scenarios is proportional to mass of the beam particle $m_\ell^2$. This can be easily seen by considering the $\ell^- \gamma \to \ell^- A'$ 
process for the vector and axial couplings of $A'$;\footnote{This sub-process forms the 
basis of the equivalent photon approximation for computing the kinematics of the full collision $\ell N \rightarrow \ell N + A'$. The $\gamma$ here is 
therefore the virtual photon sourced by the target nucleus.} in the limit of $m_\ell/\sqrt{s} \ll 1$ the squared amplitudes differ by 
\beq
|\mathcal{M}_V|^2 - |\mathcal{M}_A|^2 \propto \frac{4m_\ell^2 (\mAp^4 - 8 \mAp^2 t +t^2)}{\mAp^2 s (\mAp^2 - s - t)},
\eeq
where $s$ is the Mandelstam variable for the $2\to 2$ process (rather than the $\ell N \to \ell N +A'$ collision).
This expression illustrates two important points: the difference between the axial and vector interactions is amplified 
for heavier beam particles, and the axial interaction tends to produce more 
events at lower $\ell^-$ recoil energies (this can be confirmed by expressing $t$ in terms of the 
recoil $\ell^-$ energy and $p_T$). These differences disappear at larger $\mAp$ or beam 
energies, both of which increase $s$.
This intuition is borne out in the full MC simulation of a muon missing momentum experiment from Ref.~\cite{Kahn:2018cqs} 
which features a $15\;\GeV$ muon beam colliding with a tungsten target.
We show the recoiling muon energy distributions in Fig.~\ref{muon-distributions}
for various $A'$ masses in the vector (upper left panel) and axial models (upper right panel) .
We see that for $\mAp \lesssim m_\mu$, the distributions are visibly different at low recoil energies as expected. 
An analogous behavior is also shown in the recoil energy distributions for 
scalar and pseudo-scalar interactions in the lower row of Fig.~\ref{muon-distributions}.
We therefore conclude that experiments utilizing muon beam and electron beams are 
complimentary for probing both the flavor and Lorentz structure of BSM interactions.

 Although the muon beam missing momentum concept, as 
  demonstrated in \cite{Kahn:2018cqs} for M$^3$, is 
similar to an LDMX-style electron beam experiment, there 
are some important differences. For example, LDMX anticipates a
 monochromatic electron beam, utilizes a thin target $(\ll$ radiation length),
 and requires $\sim 10^{14} - 10^{16}$ electrons on target for phases 1 and 
 2, respectively. By contrast, muon beams are typically prepared
  from boosted pion decays with a broader spread of beam energies and lower
  luminosities; as a result, M$^3$ specifically is designed to run with $\sim 10^{13}$ total muons 
  delivered to the target. 
  The optimal recoil energy cut to suppress SM backgrounds was 
  also found to be different. 

\begin{figure}
    \includegraphics[width=0.47\textwidth]{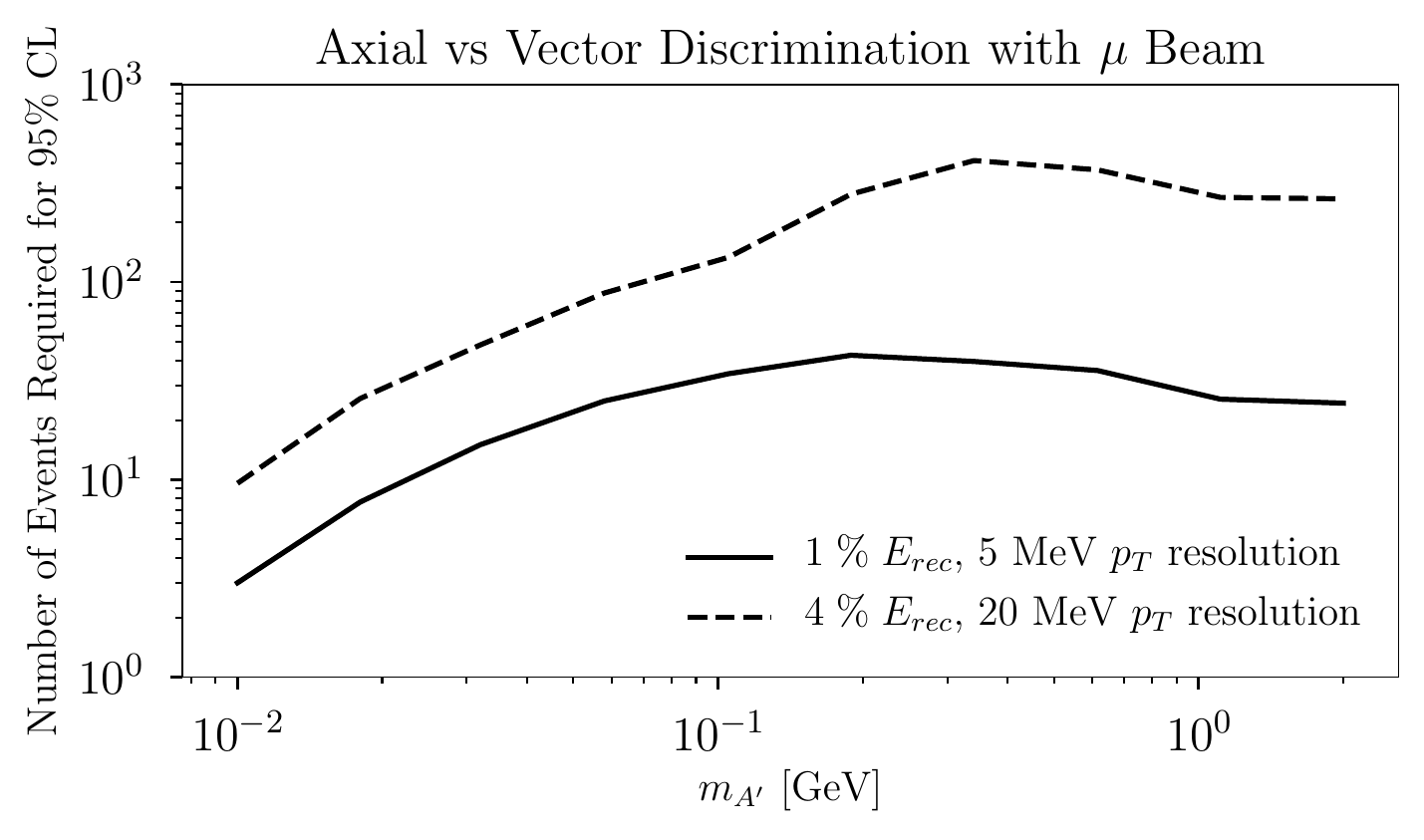}
    \caption{Number of events required to discriminate between axial and vector interactions at a muon beam experiment 
    described in Sec.~\ref{muons}. The solid and dashed lines correspond to two binnings of the kinematic 
  variables (beam muon recoil energy and transverse momentum), representing different experimental resolutions. 
  Higher resolution enables model discrimination with fewer signal events.\label{fig:lorentz_discrimination}}
  \end{figure}

  Using the M$^3$ set up as a benchmark we can estimate the number of signal events 
  required to distinguish between axial and vector interactions using the likelihood ratio method 
  described in Sec.~\ref{kinematic}. 
  As before, we generated mock data from our MC samples for $A'$ of a given mass and the vector interaction, 
  and then studied the distribution of log likelihood ratios for the axial model (with $A'$ of the same mass) and vector 
  model. The result is shown in Fig.~\ref{fig:lorentz_discrimination}.  As the mediator mass becomes 
  larger, the kinematic distributions of axial and vector interactions become more and more similar, requiring 
  larger event samples to disentangle. This figure also illustrates the importance of experimental resolution on 
  the kinematic quantities. The solid and dashed lines correspond to two different binnings in recoil energy and transverse momentum. 
  Higher resolution (i.e., finer binning) enables model discrimination with a fewer number of observed events, as the 
spectral differences between models are spread over a larger number of bins.
  



\section{Conclusion}
\label{conclusion}
Fixed target missing-momentum experiments involving electron or muon beams are powerful
dark matter (and dark sector) discovery tools. 
In this paper we have studied the model discriminating potential of these techniques using various
simplified models categorized according to production topology and Lorentz structure.

In our numerical studies we have varied the masses of BSM particles in the models above and 
developed statistical tests to discriminate between different signal hypotheses using kinematic
variables and beam parameters: lepton recoil energy $E_\ell$, transverse momentum $p_{T,\ell}$, and incident beam energy and polarization. We have also studied the discriminating potential of using electron vs. muon beams. 
Our main conclusions can be summarized as follows:

\begin{enumerate}

\item{ \bf Kinematic Variables:}
For radiative 3-body processes, we find that mediator mass can be well determined using a combination
of recoil energy and transverse momentum variables. Using a likelihood analysis, we compared
$E_\ell$ and $p_{T,\ell}$ distributions against mock data to determine how many
signal events are required to distinguish various mediator masses at given confidence levels. 

We also find that, over the full MeV-GeV mediator mass range, $p_{T,\ell}$ is a superior kinematic variable
and greatly enhances signal discriminating power relative to $E_\ell$-only measurements (e.g., NA64  \cite{NA64:2019imj,Gninenko:2019qiv}).
Indeed, for most test masses in this range, $p_{T,\ell}$  enables clear discrimination between most hypotheses 
with only $\pazocal{O}(10)$ signal events near the nominal discovery threshold for a zero-background experiment. 
In  contrast, the same analysis using only $E_\ell$ distributions typically requires several hundred events or more. 

These conclusions hold independently of the mediator's identity (e.g., scalar vs. vector) as 
the $p_{T,\ell}$ distributions are similar across the different particle variations studied here.
Consequently, $E_\ell$ and $p_{T,\ell}$ are good variables for determining mass assuming
a given BSM scenario, but these alone are not effective at discriminating between different models.
This conclusion also applies to efforts to distinguish between on-shell mediator production  
in $2\to 3$ processes and $2\to 4$ processes that produce DM pairs 
through either light or heavy mediators; for an equivalent invariant mass of 
new states, the kinematic distributions of these models are not easily distinguished.
This highlights the complementarity of different accelerator experiments that 
can potentially resolve some of these ambiguities with higher center-of-mass energies.

\item{\bf Beam Polarization: }
We also find that for different electron-vector Lorentz structures, there are new asymmetry observables
that can discriminate between certain scenarios. In particular, subtracting the energy distributions for 
signal events with incident left and right-handed electrons yields residuals for chiral electron-mediator
couplings of the form $Z^\prime_\mu \bar e \gamma^\mu P_{L,R} e$, where $P_{L,R}$ is a left/right projector; 
if the interaction is a general linear combination of  vector and axial-vector couplings, these observables only 
yield nonzero residuals for the chiral component proportional to $P_{L}$ or $P_R$. Although the feasibility of beam polarization in missing momentum experiments 
has not yet been studied, the discriminating power of this method warrants future work to assess its compatibility with other accelerator requirements (e.g., beam current and structure).

\item{\bf Beam Flavor: } Finally, we find that a combination of electron and muon beam missing-momentum
searches can be used to determine whether a BSM interaction is parametrically
enhanced by the mass of the beam particle. As representative examples, we consider 
pseudoscalar $a \bar \ell \gamma^5 \ell$ and axial-vector $V_\mu \bar \ell \gamma^\mu \gamma^5 \ell$ 
interactions for which $\ell N \to \ell N a/V$ emission requires a lepton chirality flip. Consequently, the
 corresponding cross sections for these processes are proportional to $m^2_\ell$ and the difference between
 electron and muon beam signals can be used to distinguish these models from scenarios that 
 do not require a chiral flip in order to produce the new BSM state.
 
 \end{enumerate}
 
 Our motivation in this paper has been to study model discrimination power of fixed
 target missing-momentum experiments in a variety of well motivated scenarios. 
However, the studies should be interpreted with great care as we only consider simulated
  signal distributions with statistical uncertainties to assess the distinguishability of various BSM scenarios assuming 
  negligible backgrounds from SM processes.
  Since such experiments are expected
 to have negligible irreducible backgrounds and low levels of reducible backgrounds
 from photonuclear reactions \cite{Akesson:2019iul,Kahn:2018cqs} this is a well motivated 
 working assumption. 
However, detailed future studies should include systematic uncertainties, detector level smearing effects, and  
various levels of background contamination to more realistically quantify the model discrimination power of 
the techniques outlined in this paper. 

\begin{acknowledgments}
{\it Acknowledgments} :  
We thank Nhan Tran, Shirley Li, Patrick Draper, Noah Kurinsky, Antonella Palmese, and Andrew Whitbeck for helpful conversations. Fermilab is operated by Fermi Research Alliance, LLC, under Contract No. DE-AC02-07CH11359 with the US Department of Energy. The work of DT was supported in part by the U.S. Department of Energy, Office of Science, Office of Workforce Development for Teachers and Scientists, Office of Science Graduate Student Research (SCGSR) program. The SCGSR program is administered by the Oak Ridge Institute for Science and Education for the DOE under contract number de-sc0014664. 
\end{acknowledgments}

\bibliography{LDMXpol.bib}

\begin{thebibliography}{60}%
\makeatletter
\providecommand \@ifxundefined [1]{%
 \@ifx{#1\undefined}
}%
\providecommand \@ifnum [1]{%
 \ifnum #1\expandafter \@firstoftwo
 \else \expandafter \@secondoftwo
 \fi
}%
\providecommand \@ifx [1]{%
 \ifx #1\expandafter \@firstoftwo
 \else \expandafter \@secondoftwo
 \fi
}%
\providecommand \natexlab [1]{#1}%
\providecommand \enquote  [1]{``#1''}%
\providecommand \bibnamefont  [1]{#1}%
\providecommand \bibfnamefont [1]{#1}%
\providecommand \citenamefont [1]{#1}%
\providecommand \href@noop [0]{\@secondoftwo}%
\providecommand \href [0]{\begingroup \@sanitize@url \@href}%
\providecommand \@href[1]{\@@startlink{#1}\@@href}%
\providecommand \@@href[1]{\endgroup#1\@@endlink}%
\providecommand \@sanitize@url [0]{\catcode `\\12\catcode `\$12\catcode
  `\&12\catcode `\#12\catcode `\^12\catcode `\_12\catcode `\%12\relax}%
\providecommand \@@startlink[1]{}%
\providecommand \@@endlink[0]{}%
\providecommand \url  [0]{\begingroup\@sanitize@url \@url }%
\providecommand \@url [1]{\endgroup\@href {#1}{\urlprefix }}%
\providecommand \urlprefix  [0]{URL }%
\providecommand \Eprint [0]{\href }%
\providecommand \doibase [0]{http://dx.doi.org/}%
\providecommand \selectlanguage [0]{\@gobble}%
\providecommand \bibinfo  [0]{\@secondoftwo}%
\providecommand \bibfield  [0]{\@secondoftwo}%
\providecommand \translation [1]{[#1]}%
\providecommand \BibitemOpen [0]{}%
\providecommand \bibitemStop [0]{}%
\providecommand \bibitemNoStop [0]{.\EOS\space}%
\providecommand \EOS [0]{\spacefactor3000\relax}%
\providecommand \BibitemShut  [1]{\csname bibitem#1\endcsname}%
\let\auto@bib@innerbib\@empty
\bibitem [{\citenamefont {Essig}\ \emph {et~al.}(2012)\citenamefont {Essig},
  \citenamefont {Mardon},\ and\ \citenamefont {Volansky}}]{Essig:2011nj}%
  \BibitemOpen
  \bibfield  {author} {\bibinfo {author} {\bibfnamefont {R.}~\bibnamefont
  {Essig}}, \bibinfo {author} {\bibfnamefont {J.}~\bibnamefont {Mardon}}, \
  and\ \bibinfo {author} {\bibfnamefont {T.}~\bibnamefont {Volansky}},\ }\href
  {\doibase 10.1103/PhysRevD.85.076007} {\bibfield  {journal} {\bibinfo
  {journal} {Phys. Rev.}\ }\textbf {\bibinfo {volume} {D85}},\ \bibinfo {pages}
  {076007} (\bibinfo {year} {2012})},\ \Eprint {http://arxiv.org/abs/1108.5383}
  {arXiv:1108.5383 [hep-ph]} \BibitemShut {NoStop}%
\bibitem [{\citenamefont {Graham}\ \emph {et~al.}(2012)\citenamefont {Graham},
  \citenamefont {Kaplan}, \citenamefont {Rajendran},\ and\ \citenamefont
  {Walters}}]{Graham:2012su}%
  \BibitemOpen
  \bibfield  {author} {\bibinfo {author} {\bibfnamefont {P.~W.}\ \bibnamefont
  {Graham}}, \bibinfo {author} {\bibfnamefont {D.~E.}\ \bibnamefont {Kaplan}},
  \bibinfo {author} {\bibfnamefont {S.}~\bibnamefont {Rajendran}}, \ and\
  \bibinfo {author} {\bibfnamefont {M.~T.}\ \bibnamefont {Walters}},\ }\href
  {\doibase 10.1016/j.dark.2012.09.001} {\bibfield  {journal} {\bibinfo
  {journal} {Phys. Dark Univ.}\ }\textbf {\bibinfo {volume} {1}},\ \bibinfo
  {pages} {32} (\bibinfo {year} {2012})},\ \Eprint
  {http://arxiv.org/abs/1203.2531} {arXiv:1203.2531 [hep-ph]} \BibitemShut
  {NoStop}%
\bibitem [{\citenamefont {Essig}\ \emph
  {et~al.}(2013{\natexlab{a}})\citenamefont {Essig}, \citenamefont {Mardon},
  \citenamefont {Papucci}, \citenamefont {Volansky},\ and\ \citenamefont
  {Zhong}}]{Essig:2013vha}%
  \BibitemOpen
  \bibfield  {author} {\bibinfo {author} {\bibfnamefont {R.}~\bibnamefont
  {Essig}}, \bibinfo {author} {\bibfnamefont {J.}~\bibnamefont {Mardon}},
  \bibinfo {author} {\bibfnamefont {M.}~\bibnamefont {Papucci}}, \bibinfo
  {author} {\bibfnamefont {T.}~\bibnamefont {Volansky}}, \ and\ \bibinfo
  {author} {\bibfnamefont {Y.-M.}\ \bibnamefont {Zhong}},\ }\href {\doibase
  10.1007/JHEP11(2013)167} {\bibfield  {journal} {\bibinfo  {journal} {JHEP}\
  }\textbf {\bibinfo {volume} {11}},\ \bibinfo {pages} {167} (\bibinfo {year}
  {2013}{\natexlab{a}})},\ \Eprint {http://arxiv.org/abs/1309.5084}
  {arXiv:1309.5084 [hep-ph]} \BibitemShut {NoStop}%
\bibitem [{\citenamefont {Hochberg}\ \emph
  {et~al.}(2016{\natexlab{a}})\citenamefont {Hochberg}, \citenamefont {Pyle},
  \citenamefont {Zhao},\ and\ \citenamefont {Zurek}}]{Hochberg:2015fth}%
  \BibitemOpen
  \bibfield  {author} {\bibinfo {author} {\bibfnamefont {Y.}~\bibnamefont
  {Hochberg}}, \bibinfo {author} {\bibfnamefont {M.}~\bibnamefont {Pyle}},
  \bibinfo {author} {\bibfnamefont {Y.}~\bibnamefont {Zhao}}, \ and\ \bibinfo
  {author} {\bibfnamefont {K.~M.}\ \bibnamefont {Zurek}},\ }\href {\doibase
  10.1007/JHEP08(2016)057} {\bibfield  {journal} {\bibinfo  {journal} {JHEP}\
  }\textbf {\bibinfo {volume} {08}},\ \bibinfo {pages} {057} (\bibinfo {year}
  {2016}{\natexlab{a}})},\ \Eprint {http://arxiv.org/abs/1512.04533}
  {arXiv:1512.04533 [hep-ph]} \BibitemShut {NoStop}%
\bibitem [{\citenamefont {Hochberg}\ \emph
  {et~al.}(2016{\natexlab{b}})\citenamefont {Hochberg}, \citenamefont {Zhao},\
  and\ \citenamefont {Zurek}}]{Hochberg:2015pha}%
  \BibitemOpen
  \bibfield  {author} {\bibinfo {author} {\bibfnamefont {Y.}~\bibnamefont
  {Hochberg}}, \bibinfo {author} {\bibfnamefont {Y.}~\bibnamefont {Zhao}}, \
  and\ \bibinfo {author} {\bibfnamefont {K.~M.}\ \bibnamefont {Zurek}},\ }\href
  {\doibase 10.1103/PhysRevLett.116.011301} {\bibfield  {journal} {\bibinfo
  {journal} {Phys. Rev. Lett.}\ }\textbf {\bibinfo {volume} {116}},\ \bibinfo
  {pages} {011301} (\bibinfo {year} {2016}{\natexlab{b}})},\ \Eprint
  {http://arxiv.org/abs/1504.07237} {arXiv:1504.07237 [hep-ph]} \BibitemShut
  {NoStop}%
\bibitem [{\citenamefont {Essig}\ \emph {et~al.}(2017)\citenamefont {Essig},
  \citenamefont {Mardon}, \citenamefont {Slone},\ and\ \citenamefont
  {Volansky}}]{Essig:2016crl}%
  \BibitemOpen
  \bibfield  {author} {\bibinfo {author} {\bibfnamefont {R.}~\bibnamefont
  {Essig}}, \bibinfo {author} {\bibfnamefont {J.}~\bibnamefont {Mardon}},
  \bibinfo {author} {\bibfnamefont {O.}~\bibnamefont {Slone}}, \ and\ \bibinfo
  {author} {\bibfnamefont {T.}~\bibnamefont {Volansky}},\ }\href {\doibase
  10.1103/PhysRevD.95.056011} {\bibfield  {journal} {\bibinfo  {journal} {Phys.
  Rev.}\ }\textbf {\bibinfo {volume} {D95}},\ \bibinfo {pages} {056011}
  (\bibinfo {year} {2017})},\ \Eprint {http://arxiv.org/abs/1608.02940}
  {arXiv:1608.02940 [hep-ph]} \BibitemShut {NoStop}%
\bibitem [{\citenamefont {Knapen}\ \emph {et~al.}(2017)\citenamefont {Knapen},
  \citenamefont {Lin},\ and\ \citenamefont {Zurek}}]{Knapen:2016cue}%
  \BibitemOpen
  \bibfield  {author} {\bibinfo {author} {\bibfnamefont {S.}~\bibnamefont
  {Knapen}}, \bibinfo {author} {\bibfnamefont {T.}~\bibnamefont {Lin}}, \ and\
  \bibinfo {author} {\bibfnamefont {K.~M.}\ \bibnamefont {Zurek}},\ }\href
  {\doibase 10.1103/PhysRevD.95.056019} {\bibfield  {journal} {\bibinfo
  {journal} {Phys. Rev.}\ }\textbf {\bibinfo {volume} {D95}},\ \bibinfo {pages}
  {056019} (\bibinfo {year} {2017})},\ \Eprint
  {http://arxiv.org/abs/1611.06228} {arXiv:1611.06228 [hep-ph]} \BibitemShut
  {NoStop}%
\bibitem [{\citenamefont {Essig}\ \emph {et~al.}(2018)\citenamefont {Essig},
  \citenamefont {Sholapurkar},\ and\ \citenamefont {Yu}}]{Essig:2018tss}%
  \BibitemOpen
  \bibfield  {author} {\bibinfo {author} {\bibfnamefont {R.}~\bibnamefont
  {Essig}}, \bibinfo {author} {\bibfnamefont {M.}~\bibnamefont {Sholapurkar}},
  \ and\ \bibinfo {author} {\bibfnamefont {T.-T.}\ \bibnamefont {Yu}},\ }\href
  {\doibase 10.1103/PhysRevD.97.095029} {\bibfield  {journal} {\bibinfo
  {journal} {Phys. Rev.}\ }\textbf {\bibinfo {volume} {D97}},\ \bibinfo {pages}
  {095029} (\bibinfo {year} {2018})},\ \Eprint
  {http://arxiv.org/abs/1801.10159} {arXiv:1801.10159 [hep-ph]} \BibitemShut
  {NoStop}%
\bibitem [{\citenamefont {Knapen}\ \emph {et~al.}(2018)\citenamefont {Knapen},
  \citenamefont {Lin}, \citenamefont {Pyle},\ and\ \citenamefont
  {Zurek}}]{Knapen:2017ekk}%
  \BibitemOpen
  \bibfield  {author} {\bibinfo {author} {\bibfnamefont {S.}~\bibnamefont
  {Knapen}}, \bibinfo {author} {\bibfnamefont {T.}~\bibnamefont {Lin}},
  \bibinfo {author} {\bibfnamefont {M.}~\bibnamefont {Pyle}}, \ and\ \bibinfo
  {author} {\bibfnamefont {K.~M.}\ \bibnamefont {Zurek}},\ }\href {\doibase
  10.1016/j.physletb.2018.08.064} {\bibfield  {journal} {\bibinfo  {journal}
  {Phys. Lett.}\ }\textbf {\bibinfo {volume} {B785}},\ \bibinfo {pages} {386}
  (\bibinfo {year} {2018})},\ \Eprint {http://arxiv.org/abs/1712.06598}
  {arXiv:1712.06598 [hep-ph]} \BibitemShut {NoStop}%
\bibitem [{\citenamefont {deNiverville}\ \emph {et~al.}(2011)\citenamefont
  {deNiverville}, \citenamefont {Pospelov},\ and\ \citenamefont
  {Ritz}}]{deNiverville:2011it}%
  \BibitemOpen
  \bibfield  {author} {\bibinfo {author} {\bibfnamefont {P.}~\bibnamefont
  {deNiverville}}, \bibinfo {author} {\bibfnamefont {M.}~\bibnamefont
  {Pospelov}}, \ and\ \bibinfo {author} {\bibfnamefont {A.}~\bibnamefont
  {Ritz}},\ }\href {\doibase 10.1103/PhysRevD.84.075020} {\bibfield  {journal}
  {\bibinfo  {journal} {Phys. Rev.}\ }\textbf {\bibinfo {volume} {D84}},\
  \bibinfo {pages} {075020} (\bibinfo {year} {2011})},\ \Eprint
  {http://arxiv.org/abs/1107.4580} {arXiv:1107.4580 [hep-ph]} \BibitemShut
  {NoStop}%
\bibitem [{\citenamefont {deNiverville}\ \emph {et~al.}(2017)\citenamefont
  {deNiverville}, \citenamefont {Chen}, \citenamefont {Pospelov},\ and\
  \citenamefont {Ritz}}]{deNiverville:2016rqh}%
  \BibitemOpen
  \bibfield  {author} {\bibinfo {author} {\bibfnamefont {P.}~\bibnamefont
  {deNiverville}}, \bibinfo {author} {\bibfnamefont {C.-Y.}\ \bibnamefont
  {Chen}}, \bibinfo {author} {\bibfnamefont {M.}~\bibnamefont {Pospelov}}, \
  and\ \bibinfo {author} {\bibfnamefont {A.}~\bibnamefont {Ritz}},\ }\href
  {\doibase 10.1103/PhysRevD.95.035006} {\bibfield  {journal} {\bibinfo
  {journal} {Phys. Rev. D}\ }\textbf {\bibinfo {volume} {95}},\ \bibinfo
  {pages} {035006} (\bibinfo {year} {2017})},\ \Eprint
  {http://arxiv.org/abs/1609.01770} {arXiv:1609.01770 [hep-ph]} \BibitemShut
  {NoStop}%
\bibitem [{\citenamefont {Izaguirre}\ \emph
  {et~al.}(2015{\natexlab{a}})\citenamefont {Izaguirre}, \citenamefont
  {Krnjaic}, \citenamefont {Schuster},\ and\ \citenamefont
  {Toro}}]{Izaguirre:2015yja}%
  \BibitemOpen
  \bibfield  {author} {\bibinfo {author} {\bibfnamefont {E.}~\bibnamefont
  {Izaguirre}}, \bibinfo {author} {\bibfnamefont {G.}~\bibnamefont {Krnjaic}},
  \bibinfo {author} {\bibfnamefont {P.}~\bibnamefont {Schuster}}, \ and\
  \bibinfo {author} {\bibfnamefont {N.}~\bibnamefont {Toro}},\ }\href {\doibase
  10.1103/PhysRevLett.115.251301} {\bibfield  {journal} {\bibinfo  {journal}
  {Phys. Rev. Lett.}\ }\textbf {\bibinfo {volume} {115}},\ \bibinfo {pages}
  {251301} (\bibinfo {year} {2015}{\natexlab{a}})},\ \Eprint
  {http://arxiv.org/abs/1505.00011} {arXiv:1505.00011 [hep-ph]} \BibitemShut
  {NoStop}%
\bibitem [{\citenamefont {Izaguirre}\ \emph {et~al.}(2017)\citenamefont
  {Izaguirre}, \citenamefont {Kahn}, \citenamefont {Krnjaic},\ and\
  \citenamefont {Moschella}}]{Izaguirre:2017bqb}%
  \BibitemOpen
  \bibfield  {author} {\bibinfo {author} {\bibfnamefont {E.}~\bibnamefont
  {Izaguirre}}, \bibinfo {author} {\bibfnamefont {Y.}~\bibnamefont {Kahn}},
  \bibinfo {author} {\bibfnamefont {G.}~\bibnamefont {Krnjaic}}, \ and\
  \bibinfo {author} {\bibfnamefont {M.}~\bibnamefont {Moschella}},\ }\href
  {\doibase 10.1103/PhysRevD.96.055007} {\bibfield  {journal} {\bibinfo
  {journal} {Phys. Rev. D}\ }\textbf {\bibinfo {volume} {96}},\ \bibinfo
  {pages} {055007} (\bibinfo {year} {2017})},\ \Eprint
  {http://arxiv.org/abs/1703.06881} {arXiv:1703.06881 [hep-ph]} \BibitemShut
  {NoStop}%
\bibitem [{\citenamefont {{\AA}kesson}\ \emph {et~al.}(2020)\citenamefont
  {{\AA}kesson} \emph {et~al.}}]{Akesson:2019iul}%
  \BibitemOpen
  \bibfield  {author} {\bibinfo {author} {\bibfnamefont {T.}~\bibnamefont
  {{\AA}kesson}} \emph {et~al.} (\bibinfo {collaboration} {LDMX}),\ }\href
  {\doibase 10.1007/JHEP04(2020)003} {\bibfield  {journal} {\bibinfo  {journal}
  {JHEP}\ }\textbf {\bibinfo {volume} {04}},\ \bibinfo {pages} {003} (\bibinfo
  {year} {2020})},\ \Eprint {http://arxiv.org/abs/1912.05535} {arXiv:1912.05535
  [physics.ins-det]} \BibitemShut {NoStop}%
\bibitem [{\citenamefont {Battaglieri}\ \emph
  {et~al.}(2017{\natexlab{a}})\citenamefont {Battaglieri} \emph
  {et~al.}}]{Battaglieri:2017qen}%
  \BibitemOpen
  \bibfield  {author} {\bibinfo {author} {\bibfnamefont {M.}~\bibnamefont
  {Battaglieri}} \emph {et~al.} (\bibinfo {collaboration} {BDX}),\ }\href@noop
  {} {\  (\bibinfo {year} {2017}{\natexlab{a}})},\ \Eprint
  {http://arxiv.org/abs/1712.01518} {arXiv:1712.01518 [physics.ins-det]}
  \BibitemShut {NoStop}%
\bibitem [{\citenamefont {Kahn}\ \emph {et~al.}(2018)\citenamefont {Kahn},
  \citenamefont {Krnjaic}, \citenamefont {Tran},\ and\ \citenamefont
  {Whitbeck}}]{Kahn:2018cqs}%
  \BibitemOpen
  \bibfield  {author} {\bibinfo {author} {\bibfnamefont {Y.}~\bibnamefont
  {Kahn}}, \bibinfo {author} {\bibfnamefont {G.}~\bibnamefont {Krnjaic}},
  \bibinfo {author} {\bibfnamefont {N.}~\bibnamefont {Tran}}, \ and\ \bibinfo
  {author} {\bibfnamefont {A.}~\bibnamefont {Whitbeck}},\ }\href {\doibase
  10.1007/JHEP09(2018)153} {\bibfield  {journal} {\bibinfo  {journal} {JHEP}\
  }\textbf {\bibinfo {volume} {09}},\ \bibinfo {pages} {153} (\bibinfo {year}
  {2018})},\ \Eprint {http://arxiv.org/abs/1804.03144} {arXiv:1804.03144
  [hep-ph]} \BibitemShut {NoStop}%
\bibitem [{\citenamefont {Berlin}\ \emph {et~al.}(2019)\citenamefont {Berlin},
  \citenamefont {Blinov}, \citenamefont {Krnjaic}, \citenamefont {Schuster},\
  and\ \citenamefont {Toro}}]{Berlin:2018bsc}%
  \BibitemOpen
  \bibfield  {author} {\bibinfo {author} {\bibfnamefont {A.}~\bibnamefont
  {Berlin}}, \bibinfo {author} {\bibfnamefont {N.}~\bibnamefont {Blinov}},
  \bibinfo {author} {\bibfnamefont {G.}~\bibnamefont {Krnjaic}}, \bibinfo
  {author} {\bibfnamefont {P.}~\bibnamefont {Schuster}}, \ and\ \bibinfo
  {author} {\bibfnamefont {N.}~\bibnamefont {Toro}},\ }\href {\doibase
  10.1103/PhysRevD.99.075001} {\bibfield  {journal} {\bibinfo  {journal} {Phys.
  Rev.}\ }\textbf {\bibinfo {volume} {D99}},\ \bibinfo {pages} {075001}
  (\bibinfo {year} {2019})},\ \Eprint {http://arxiv.org/abs/1807.01730}
  {arXiv:1807.01730 [hep-ph]} \BibitemShut {NoStop}%
\bibitem [{\citenamefont {Izaguirre}\ \emph
  {et~al.}(2015{\natexlab{b}})\citenamefont {Izaguirre}, \citenamefont
  {Krnjaic}, \citenamefont {Schuster},\ and\ \citenamefont
  {Toro}}]{Izaguirre:2014bca}%
  \BibitemOpen
  \bibfield  {author} {\bibinfo {author} {\bibfnamefont {E.}~\bibnamefont
  {Izaguirre}}, \bibinfo {author} {\bibfnamefont {G.}~\bibnamefont {Krnjaic}},
  \bibinfo {author} {\bibfnamefont {P.}~\bibnamefont {Schuster}}, \ and\
  \bibinfo {author} {\bibfnamefont {N.}~\bibnamefont {Toro}},\ }\href {\doibase
  10.1103/PhysRevD.91.094026} {\bibfield  {journal} {\bibinfo  {journal} {Phys.
  Rev. D}\ }\textbf {\bibinfo {volume} {91}},\ \bibinfo {pages} {094026}
  (\bibinfo {year} {2015}{\natexlab{b}})},\ \Eprint
  {http://arxiv.org/abs/1411.1404} {arXiv:1411.1404 [hep-ph]} \BibitemShut
  {NoStop}%
\bibitem [{\citenamefont {Kahn}\ \emph {et~al.}(2015)\citenamefont {Kahn},
  \citenamefont {Krnjaic}, \citenamefont {Thaler},\ and\ \citenamefont
  {Toups}}]{Kahn:2014sra}%
  \BibitemOpen
  \bibfield  {author} {\bibinfo {author} {\bibfnamefont {Y.}~\bibnamefont
  {Kahn}}, \bibinfo {author} {\bibfnamefont {G.}~\bibnamefont {Krnjaic}},
  \bibinfo {author} {\bibfnamefont {J.}~\bibnamefont {Thaler}}, \ and\ \bibinfo
  {author} {\bibfnamefont {M.}~\bibnamefont {Toups}},\ }\href {\doibase
  10.1103/PhysRevD.91.055006} {\bibfield  {journal} {\bibinfo  {journal} {Phys.
  Rev. D}\ }\textbf {\bibinfo {volume} {91}},\ \bibinfo {pages} {055006}
  (\bibinfo {year} {2015})},\ \Eprint {http://arxiv.org/abs/1411.1055}
  {arXiv:1411.1055 [hep-ph]} \BibitemShut {NoStop}%
\bibitem [{\citenamefont {Izaguirre}\ \emph
  {et~al.}(2015{\natexlab{c}})\citenamefont {Izaguirre}, \citenamefont
  {Krnjaic},\ and\ \citenamefont {Pospelov}}]{Izaguirre:2014cza}%
  \BibitemOpen
  \bibfield  {author} {\bibinfo {author} {\bibfnamefont {E.}~\bibnamefont
  {Izaguirre}}, \bibinfo {author} {\bibfnamefont {G.}~\bibnamefont {Krnjaic}},
  \ and\ \bibinfo {author} {\bibfnamefont {M.}~\bibnamefont {Pospelov}},\
  }\href {\doibase 10.1016/j.physletb.2014.11.037} {\bibfield  {journal}
  {\bibinfo  {journal} {Phys. Lett. B}\ }\textbf {\bibinfo {volume} {740}},\
  \bibinfo {pages} {61} (\bibinfo {year} {2015}{\natexlab{c}})},\ \Eprint
  {http://arxiv.org/abs/1405.4864} {arXiv:1405.4864 [hep-ph]} \BibitemShut
  {NoStop}%
\bibitem [{\citenamefont {Izaguirre}\ \emph {et~al.}(2014)\citenamefont
  {Izaguirre}, \citenamefont {Krnjaic}, \citenamefont {Schuster},\ and\
  \citenamefont {Toro}}]{Izaguirre:2014dua}%
  \BibitemOpen
  \bibfield  {author} {\bibinfo {author} {\bibfnamefont {E.}~\bibnamefont
  {Izaguirre}}, \bibinfo {author} {\bibfnamefont {G.}~\bibnamefont {Krnjaic}},
  \bibinfo {author} {\bibfnamefont {P.}~\bibnamefont {Schuster}}, \ and\
  \bibinfo {author} {\bibfnamefont {N.}~\bibnamefont {Toro}},\ }\href {\doibase
  10.1103/PhysRevD.90.014052} {\bibfield  {journal} {\bibinfo  {journal} {Phys.
  Rev. D}\ }\textbf {\bibinfo {volume} {90}},\ \bibinfo {pages} {014052}
  (\bibinfo {year} {2014})},\ \Eprint {http://arxiv.org/abs/1403.6826}
  {arXiv:1403.6826 [hep-ph]} \BibitemShut {NoStop}%
\bibitem [{\citenamefont {Izaguirre}\ \emph {et~al.}(2013)\citenamefont
  {Izaguirre}, \citenamefont {Krnjaic}, \citenamefont {Schuster},\ and\
  \citenamefont {Toro}}]{Izaguirre:2013uxa}%
  \BibitemOpen
  \bibfield  {author} {\bibinfo {author} {\bibfnamefont {E.}~\bibnamefont
  {Izaguirre}}, \bibinfo {author} {\bibfnamefont {G.}~\bibnamefont {Krnjaic}},
  \bibinfo {author} {\bibfnamefont {P.}~\bibnamefont {Schuster}}, \ and\
  \bibinfo {author} {\bibfnamefont {N.}~\bibnamefont {Toro}},\ }\href {\doibase
  10.1103/PhysRevD.88.114015} {\bibfield  {journal} {\bibinfo  {journal} {Phys.
  Rev. D}\ }\textbf {\bibinfo {volume} {88}},\ \bibinfo {pages} {114015}
  (\bibinfo {year} {2013})},\ \Eprint {http://arxiv.org/abs/1307.6554}
  {arXiv:1307.6554 [hep-ph]} \BibitemShut {NoStop}%
\bibitem [{\citenamefont {Berlin}\ \emph {et~al.}(2020)\citenamefont {Berlin},
  \citenamefont {deNiverville}, \citenamefont {Ritz}, \citenamefont
  {Schuster},\ and\ \citenamefont {Toro}}]{Berlin:2020uwy}%
  \BibitemOpen
  \bibfield  {author} {\bibinfo {author} {\bibfnamefont {A.}~\bibnamefont
  {Berlin}}, \bibinfo {author} {\bibfnamefont {P.}~\bibnamefont
  {deNiverville}}, \bibinfo {author} {\bibfnamefont {A.}~\bibnamefont {Ritz}},
  \bibinfo {author} {\bibfnamefont {P.}~\bibnamefont {Schuster}}, \ and\
  \bibinfo {author} {\bibfnamefont {N.}~\bibnamefont {Toro}},\ }\href@noop {}
  {\  (\bibinfo {year} {2020})},\ \Eprint {http://arxiv.org/abs/2003.03379}
  {arXiv:2003.03379 [hep-ph]} \BibitemShut {NoStop}%
\bibitem [{\citenamefont {Buonocore}\ \emph {et~al.}(2019)\citenamefont
  {Buonocore}, \citenamefont {deNiverville},\ and\ \citenamefont
  {Frugiuele}}]{Buonocore:2019esg}%
  \BibitemOpen
  \bibfield  {author} {\bibinfo {author} {\bibfnamefont {L.}~\bibnamefont
  {Buonocore}}, \bibinfo {author} {\bibfnamefont {P.}~\bibnamefont
  {deNiverville}}, \ and\ \bibinfo {author} {\bibfnamefont {C.}~\bibnamefont
  {Frugiuele}},\ }\href@noop {} {\  (\bibinfo {year} {2019})},\ \Eprint
  {http://arxiv.org/abs/1912.09346} {arXiv:1912.09346 [hep-ph]} \BibitemShut
  {NoStop}%
\bibitem [{\citenamefont {Tsai}\ \emph {et~al.}(2019)\citenamefont {Tsai},
  \citenamefont {deNiverville},\ and\ \citenamefont {Liu}}]{Tsai:2019mtm}%
  \BibitemOpen
  \bibfield  {author} {\bibinfo {author} {\bibfnamefont {Y.-D.}\ \bibnamefont
  {Tsai}}, \bibinfo {author} {\bibfnamefont {P.}~\bibnamefont {deNiverville}},
  \ and\ \bibinfo {author} {\bibfnamefont {M.~X.}\ \bibnamefont {Liu}},\
  }\href@noop {} {\  (\bibinfo {year} {2019})},\ \Eprint
  {http://arxiv.org/abs/1908.07525} {arXiv:1908.07525 [hep-ph]} \BibitemShut
  {NoStop}%
\bibitem [{\citenamefont {deNiverville}\ and\ \citenamefont
  {Frugiuele}(2019)}]{deNiverville:2018dbu}%
  \BibitemOpen
  \bibfield  {author} {\bibinfo {author} {\bibfnamefont {P.}~\bibnamefont
  {deNiverville}}\ and\ \bibinfo {author} {\bibfnamefont {C.}~\bibnamefont
  {Frugiuele}},\ }\href {\doibase 10.1103/PhysRevD.99.051701} {\bibfield
  {journal} {\bibinfo  {journal} {Phys. Rev. D}\ }\textbf {\bibinfo {volume}
  {99}},\ \bibinfo {pages} {051701} (\bibinfo {year} {2019})},\ \Eprint
  {http://arxiv.org/abs/1807.06501} {arXiv:1807.06501 [hep-ph]} \BibitemShut
  {NoStop}%
\bibitem [{\citenamefont {Aguilar-Arevalo}\ \emph {et~al.}(2018)\citenamefont
  {Aguilar-Arevalo} \emph {et~al.}}]{Aguilar-Arevalo:2018wea}%
  \BibitemOpen
  \bibfield  {author} {\bibinfo {author} {\bibfnamefont {A.}~\bibnamefont
  {Aguilar-Arevalo}} \emph {et~al.} (\bibinfo {collaboration} {MiniBooNE DM}),\
  }\href {\doibase 10.1103/PhysRevD.98.112004} {\bibfield  {journal} {\bibinfo
  {journal} {Phys. Rev. D}\ }\textbf {\bibinfo {volume} {98}},\ \bibinfo
  {pages} {112004} (\bibinfo {year} {2018})},\ \Eprint
  {http://arxiv.org/abs/1807.06137} {arXiv:1807.06137 [hep-ex]} \BibitemShut
  {NoStop}%
\bibitem [{\citenamefont {Berlin}\ \emph {et~al.}(2018)\citenamefont {Berlin},
  \citenamefont {Gori}, \citenamefont {Schuster},\ and\ \citenamefont
  {Toro}}]{Berlin:2018pwi}%
  \BibitemOpen
  \bibfield  {author} {\bibinfo {author} {\bibfnamefont {A.}~\bibnamefont
  {Berlin}}, \bibinfo {author} {\bibfnamefont {S.}~\bibnamefont {Gori}},
  \bibinfo {author} {\bibfnamefont {P.}~\bibnamefont {Schuster}}, \ and\
  \bibinfo {author} {\bibfnamefont {N.}~\bibnamefont {Toro}},\ }\href {\doibase
  10.1103/PhysRevD.98.035011} {\bibfield  {journal} {\bibinfo  {journal} {Phys.
  Rev. D}\ }\textbf {\bibinfo {volume} {98}},\ \bibinfo {pages} {035011}
  (\bibinfo {year} {2018})},\ \Eprint {http://arxiv.org/abs/1804.00661}
  {arXiv:1804.00661 [hep-ph]} \BibitemShut {NoStop}%
\bibitem [{\citenamefont {Aguilar-Arevalo}\ \emph {et~al.}(2017)\citenamefont
  {Aguilar-Arevalo} \emph {et~al.}}]{Aguilar-Arevalo:2017mqx}%
  \BibitemOpen
  \bibfield  {author} {\bibinfo {author} {\bibfnamefont {A.}~\bibnamefont
  {Aguilar-Arevalo}} \emph {et~al.} (\bibinfo {collaboration} {MiniBooNE}),\
  }\href {\doibase 10.1103/PhysRevLett.118.221803} {\bibfield  {journal}
  {\bibinfo  {journal} {Phys. Rev. Lett.}\ }\textbf {\bibinfo {volume} {118}},\
  \bibinfo {pages} {221803} (\bibinfo {year} {2017})},\ \Eprint
  {http://arxiv.org/abs/1702.02688} {arXiv:1702.02688 [hep-ex]} \BibitemShut
  {NoStop}%
\bibitem [{\citenamefont {Essig}\ \emph
  {et~al.}(2013{\natexlab{b}})\citenamefont {Essig} \emph
  {et~al.}}]{Essig:2013lka}%
  \BibitemOpen
  \bibfield  {author} {\bibinfo {author} {\bibfnamefont {R.}~\bibnamefont
  {Essig}} \emph {et~al.},\ }in\ \href@noop {} {\emph {\bibinfo {booktitle}
  {{Proceedings, 2013 Community Summer Study on the Future of U.S. Particle
  Physics: Snowmass on the Mississippi (CSS2013)}: {Minneapolis, MN, USA, July
  29-August 6, 2013}}}}\ (\bibinfo {year} {2013})\ \Eprint
  {http://arxiv.org/abs/1311.0029} {arXiv:1311.0029 [hep-ph]} \BibitemShut
  {NoStop}%
\bibitem [{\citenamefont {Alexander}\ \emph {et~al.}(2016)\citenamefont
  {Alexander} \emph {et~al.}}]{Alexander:2016aln}%
  \BibitemOpen
  \bibfield  {author} {\bibinfo {author} {\bibfnamefont {J.}~\bibnamefont
  {Alexander}} \emph {et~al.}\ }(\bibinfo {year} {2016})\ \Eprint
  {http://arxiv.org/abs/1608.08632} {arXiv:1608.08632 [hep-ph]} \BibitemShut
  {NoStop}%
\bibitem [{\citenamefont {Battaglieri}\ \emph
  {et~al.}(2017{\natexlab{b}})\citenamefont {Battaglieri} \emph
  {et~al.}}]{Battaglieri:2017aum}%
  \BibitemOpen
  \bibfield  {author} {\bibinfo {author} {\bibfnamefont {M.}~\bibnamefont
  {Battaglieri}} \emph {et~al.},\ }\href@noop {} {\  (\bibinfo {year}
  {2017}{\natexlab{b}})},\ \Eprint {http://arxiv.org/abs/1707.04591}
  {arXiv:1707.04591 [hep-ph]} \BibitemShut {NoStop}%
\bibitem [{\citenamefont {Ellis}\ \emph {et~al.}(2019)\citenamefont {Ellis}
  \emph {et~al.}}]{Strategy:2019vxc}%
  \BibitemOpen
  \bibfield  {author} {\bibinfo {author} {\bibfnamefont {R.~K.}\ \bibnamefont
  {Ellis}} \emph {et~al.},\ }\href@noop {} {\  (\bibinfo {year} {2019})},\
  \Eprint {http://arxiv.org/abs/1910.11775} {arXiv:1910.11775 [hep-ex]}
  \BibitemShut {NoStop}%
\bibitem [{\citenamefont {{\AA}kesson}\ \emph {et~al.}(2018)\citenamefont
  {{\AA}kesson} \emph {et~al.}}]{Akesson:2018vlm}%
  \BibitemOpen
  \bibfield  {author} {\bibinfo {author} {\bibfnamefont {T.}~\bibnamefont
  {{\AA}kesson}} \emph {et~al.} (\bibinfo {collaboration} {LDMX}),\ }\href@noop
  {} {\  (\bibinfo {year} {2018})},\ \Eprint {http://arxiv.org/abs/1808.05219}
  {arXiv:1808.05219 [hep-ex]} \BibitemShut {NoStop}%
\bibitem [{\citenamefont {Feng}\ and\ \citenamefont
  {Smolinsky}(2017)}]{Feng:2017drg}%
  \BibitemOpen
  \bibfield  {author} {\bibinfo {author} {\bibfnamefont {J.~L.}\ \bibnamefont
  {Feng}}\ and\ \bibinfo {author} {\bibfnamefont {J.}~\bibnamefont
  {Smolinsky}},\ }\href {\doibase 10.1103/PhysRevD.96.095022} {\bibfield
  {journal} {\bibinfo  {journal} {Phys. Rev.}\ }\textbf {\bibinfo {volume}
  {D96}},\ \bibinfo {pages} {095022} (\bibinfo {year} {2017})},\ \Eprint
  {http://arxiv.org/abs/1707.03835} {arXiv:1707.03835 [hep-ph]} \BibitemShut
  {NoStop}%
\bibitem [{\citenamefont {Ilten}\ \emph {et~al.}(2018)\citenamefont {Ilten},
  \citenamefont {Soreq}, \citenamefont {Williams},\ and\ \citenamefont
  {Xue}}]{Ilten:2018crw}%
  \BibitemOpen
  \bibfield  {author} {\bibinfo {author} {\bibfnamefont {P.}~\bibnamefont
  {Ilten}}, \bibinfo {author} {\bibfnamefont {Y.}~\bibnamefont {Soreq}},
  \bibinfo {author} {\bibfnamefont {M.}~\bibnamefont {Williams}}, \ and\
  \bibinfo {author} {\bibfnamefont {W.}~\bibnamefont {Xue}},\ }\href {\doibase
  10.1007/JHEP06(2018)004} {\bibfield  {journal} {\bibinfo  {journal} {JHEP}\
  }\textbf {\bibinfo {volume} {06}},\ \bibinfo {pages} {004} (\bibinfo {year}
  {2018})},\ \Eprint {http://arxiv.org/abs/1801.04847} {arXiv:1801.04847
  [hep-ph]} \BibitemShut {NoStop}%
\bibitem [{\citenamefont {Bauer}\ \emph {et~al.}(2018)\citenamefont {Bauer},
  \citenamefont {Foldenauer},\ and\ \citenamefont {Jaeckel}}]{Bauer:2018onh}%
  \BibitemOpen
  \bibfield  {author} {\bibinfo {author} {\bibfnamefont {M.}~\bibnamefont
  {Bauer}}, \bibinfo {author} {\bibfnamefont {P.}~\bibnamefont {Foldenauer}}, \
  and\ \bibinfo {author} {\bibfnamefont {J.}~\bibnamefont {Jaeckel}},\
  }\href@noop {} {\  (\bibinfo {year} {2018})},\ \Eprint
  {http://arxiv.org/abs/1803.05466} {arXiv:1803.05466 [hep-ph]} \BibitemShut
  {NoStop}%
\bibitem [{\citenamefont {Kahn}\ \emph {et~al.}(2017)\citenamefont {Kahn},
  \citenamefont {Krnjaic}, \citenamefont {Mishra-Sharma},\ and\ \citenamefont
  {Tait}}]{Kahn:2016vjr}%
  \BibitemOpen
  \bibfield  {author} {\bibinfo {author} {\bibfnamefont {Y.}~\bibnamefont
  {Kahn}}, \bibinfo {author} {\bibfnamefont {G.}~\bibnamefont {Krnjaic}},
  \bibinfo {author} {\bibfnamefont {S.}~\bibnamefont {Mishra-Sharma}}, \ and\
  \bibinfo {author} {\bibfnamefont {T.~M.~P.}\ \bibnamefont {Tait}},\ }\href
  {\doibase 10.1007/JHEP05(2017)002} {\bibfield  {journal} {\bibinfo  {journal}
  {JHEP}\ }\textbf {\bibinfo {volume} {05}},\ \bibinfo {pages} {002} (\bibinfo
  {year} {2017})},\ \Eprint {http://arxiv.org/abs/1609.09072} {arXiv:1609.09072
  [hep-ph]} \BibitemShut {NoStop}%
\bibitem [{\citenamefont {Dror}\ \emph {et~al.}(2017)\citenamefont {Dror},
  \citenamefont {Lasenby},\ and\ \citenamefont {Pospelov}}]{Dror_2017dark}%
  \BibitemOpen
  \bibfield  {author} {\bibinfo {author} {\bibfnamefont {J.~A.}\ \bibnamefont
  {Dror}}, \bibinfo {author} {\bibfnamefont {R.}~\bibnamefont {Lasenby}}, \
  and\ \bibinfo {author} {\bibfnamefont {M.}~\bibnamefont {Pospelov}},\ }\href
  {\doibase 10.1103/physrevd.96.075036} {\bibfield  {journal} {\bibinfo
  {journal} {Physical Review D}\ }\textbf {\bibinfo {volume} {96}} (\bibinfo
  {year} {2017}),\ 10.1103/physrevd.96.075036}\BibitemShut {NoStop}%
\bibitem [{\citenamefont {Essig}\ \emph {et~al.}(2010)\citenamefont {Essig},
  \citenamefont {Harnik}, \citenamefont {Kaplan},\ and\ \citenamefont
  {Toro}}]{Essig:2010gu}%
  \BibitemOpen
  \bibfield  {author} {\bibinfo {author} {\bibfnamefont {R.}~\bibnamefont
  {Essig}}, \bibinfo {author} {\bibfnamefont {R.}~\bibnamefont {Harnik}},
  \bibinfo {author} {\bibfnamefont {J.}~\bibnamefont {Kaplan}}, \ and\ \bibinfo
  {author} {\bibfnamefont {N.}~\bibnamefont {Toro}},\ }\href {\doibase
  10.1103/PhysRevD.82.113008} {\bibfield  {journal} {\bibinfo  {journal} {Phys.
  Rev. D}\ }\textbf {\bibinfo {volume} {82}},\ \bibinfo {pages} {113008}
  (\bibinfo {year} {2010})},\ \Eprint {http://arxiv.org/abs/1008.0636}
  {arXiv:1008.0636 [hep-ph]} \BibitemShut {NoStop}%
\bibitem [{\citenamefont {Dolan}\ \emph {et~al.}(2015)\citenamefont {Dolan},
  \citenamefont {Kahlhoefer}, \citenamefont {McCabe},\ and\ \citenamefont
  {Schmidt-Hoberg}}]{Dolan:2014ska}%
  \BibitemOpen
  \bibfield  {author} {\bibinfo {author} {\bibfnamefont {M.~J.}\ \bibnamefont
  {Dolan}}, \bibinfo {author} {\bibfnamefont {F.}~\bibnamefont {Kahlhoefer}},
  \bibinfo {author} {\bibfnamefont {C.}~\bibnamefont {McCabe}}, \ and\ \bibinfo
  {author} {\bibfnamefont {K.}~\bibnamefont {Schmidt-Hoberg}},\ }\href
  {\doibase 10.1007/JHEP03(2015)171} {\bibfield  {journal} {\bibinfo  {journal}
  {JHEP}\ }\textbf {\bibinfo {volume} {03}},\ \bibinfo {pages} {171} (\bibinfo
  {year} {2015})},\ \bibinfo {note} {[Erratum: JHEP 07, 103 (2015)]},\ \Eprint
  {http://arxiv.org/abs/1412.5174} {arXiv:1412.5174 [hep-ph]} \BibitemShut
  {NoStop}%
\bibitem [{\citenamefont {Krnjaic}(2016)}]{Krnjaic:2015mbs}%
  \BibitemOpen
  \bibfield  {author} {\bibinfo {author} {\bibfnamefont {G.}~\bibnamefont
  {Krnjaic}},\ }\href {\doibase 10.1103/PhysRevD.94.073009} {\bibfield
  {journal} {\bibinfo  {journal} {Phys. Rev. D}\ }\textbf {\bibinfo {volume}
  {94}},\ \bibinfo {pages} {073009} (\bibinfo {year} {2016})},\ \Eprint
  {http://arxiv.org/abs/1512.04119} {arXiv:1512.04119 [hep-ph]} \BibitemShut
  {NoStop}%
\bibitem [{\citenamefont {Batell}\ \emph {et~al.}(2018)\citenamefont {Batell},
  \citenamefont {Freitas}, \citenamefont {Ismail},\ and\ \citenamefont
  {Mckeen}}]{Batell:2017kty}%
  \BibitemOpen
  \bibfield  {author} {\bibinfo {author} {\bibfnamefont {B.}~\bibnamefont
  {Batell}}, \bibinfo {author} {\bibfnamefont {A.}~\bibnamefont {Freitas}},
  \bibinfo {author} {\bibfnamefont {A.}~\bibnamefont {Ismail}}, \ and\ \bibinfo
  {author} {\bibfnamefont {D.}~\bibnamefont {Mckeen}},\ }\href {\doibase
  10.1103/PhysRevD.98.055026} {\bibfield  {journal} {\bibinfo  {journal} {Phys.
  Rev. D}\ }\textbf {\bibinfo {volume} {98}},\ \bibinfo {pages} {055026}
  (\bibinfo {year} {2018})},\ \Eprint {http://arxiv.org/abs/1712.10022}
  {arXiv:1712.10022 [hep-ph]} \BibitemShut {NoStop}%
\bibitem [{\citenamefont {Egana-Ugrinovic}\ \emph {et~al.}(2020)\citenamefont
  {Egana-Ugrinovic}, \citenamefont {Homiller},\ and\ \citenamefont
  {Meade}}]{Egana_Ugrinovic_2020}%
  \BibitemOpen
  \bibfield  {author} {\bibinfo {author} {\bibfnamefont {D.}~\bibnamefont
  {Egana-Ugrinovic}}, \bibinfo {author} {\bibfnamefont {S.}~\bibnamefont
  {Homiller}}, \ and\ \bibinfo {author} {\bibfnamefont {P.}~\bibnamefont
  {Meade}},\ }\href {\doibase 10.1103/physrevlett.124.191801} {\bibfield
  {journal} {\bibinfo  {journal} {Physical Review Letters}\ }\textbf {\bibinfo
  {volume} {124}} (\bibinfo {year} {2020}),\
  10.1103/physrevlett.124.191801}\BibitemShut {NoStop}%
\bibitem [{\citenamefont {Liu}\ \emph {et~al.}(2020)\citenamefont {Liu},
  \citenamefont {McGinnis}, \citenamefont {Wagner},\ and\ \citenamefont
  {Wang}}]{Liu:2020qgx}%
  \BibitemOpen
  \bibfield  {author} {\bibinfo {author} {\bibfnamefont {J.}~\bibnamefont
  {Liu}}, \bibinfo {author} {\bibfnamefont {N.}~\bibnamefont {McGinnis}},
  \bibinfo {author} {\bibfnamefont {C.~E.}\ \bibnamefont {Wagner}}, \ and\
  \bibinfo {author} {\bibfnamefont {X.-P.}\ \bibnamefont {Wang}},\ }\href
  {\doibase 10.1007/JHEP04(2020)197} {\bibfield  {journal} {\bibinfo  {journal}
  {JHEP}\ }\textbf {\bibinfo {volume} {04}},\ \bibinfo {pages} {197} (\bibinfo
  {year} {2020})},\ \Eprint {http://arxiv.org/abs/2001.06522} {arXiv:2001.06522
  [hep-ph]} \BibitemShut {NoStop}%
\bibitem [{\citenamefont {Alwall}\ \emph {et~al.}(2014)\citenamefont {Alwall},
  \citenamefont {Frederix}, \citenamefont {Frixione}, \citenamefont {Hirschi},
  \citenamefont {Maltoni}, \citenamefont {Mattelaer}, \citenamefont {Shao},
  \citenamefont {Stelzer}, \citenamefont {Torrielli},\ and\ \citenamefont
  {Zaro}}]{Alwall:2014hca}%
  \BibitemOpen
  \bibfield  {author} {\bibinfo {author} {\bibfnamefont {J.}~\bibnamefont
  {Alwall}}, \bibinfo {author} {\bibfnamefont {R.}~\bibnamefont {Frederix}},
  \bibinfo {author} {\bibfnamefont {S.}~\bibnamefont {Frixione}}, \bibinfo
  {author} {\bibfnamefont {V.}~\bibnamefont {Hirschi}}, \bibinfo {author}
  {\bibfnamefont {F.}~\bibnamefont {Maltoni}}, \bibinfo {author} {\bibfnamefont
  {O.}~\bibnamefont {Mattelaer}}, \bibinfo {author} {\bibfnamefont {H.~S.}\
  \bibnamefont {Shao}}, \bibinfo {author} {\bibfnamefont {T.}~\bibnamefont
  {Stelzer}}, \bibinfo {author} {\bibfnamefont {P.}~\bibnamefont {Torrielli}},
  \ and\ \bibinfo {author} {\bibfnamefont {M.}~\bibnamefont {Zaro}},\ }\href
  {\doibase 10.1007/JHEP07(2014)079} {\bibfield  {journal} {\bibinfo  {journal}
  {JHEP}\ }\textbf {\bibinfo {volume} {07}},\ \bibinfo {pages} {079} (\bibinfo
  {year} {2014})},\ \Eprint {http://arxiv.org/abs/1405.0301} {arXiv:1405.0301
  [hep-ph]} \BibitemShut {NoStop}%
\bibitem [{\citenamefont {Kim}\ and\ \citenamefont {Tsai}(1973)}]{Kim:1973he}%
  \BibitemOpen
  \bibfield  {author} {\bibinfo {author} {\bibfnamefont {K.~J.}\ \bibnamefont
  {Kim}}\ and\ \bibinfo {author} {\bibfnamefont {Y.-S.}\ \bibnamefont {Tsai}},\
  }\href {\doibase 10.1103/PhysRevD.8.3109} {\bibfield  {journal} {\bibinfo
  {journal} {Phys. Rev. D}\ }\textbf {\bibinfo {volume} {8}},\ \bibinfo {pages}
  {3109} (\bibinfo {year} {1973})}\BibitemShut {NoStop}%
\bibitem [{\citenamefont {Tsai}(1974)}]{Tsai:1973py}%
  \BibitemOpen
  \bibfield  {author} {\bibinfo {author} {\bibfnamefont {Y.-S.}\ \bibnamefont
  {Tsai}},\ }\href {\doibase 10.1103/RevModPhys.46.815} {\bibfield  {journal}
  {\bibinfo  {journal} {Rev. Mod. Phys.}\ }\textbf {\bibinfo {volume} {46}},\
  \bibinfo {pages} {815} (\bibinfo {year} {1974})},\ \bibinfo {note} {[Erratum:
  Rev.Mod.Phys. 49, 521--423 (1977)]}\BibitemShut {NoStop}%
\bibitem [{\citenamefont {Cowan}(1998)}]{Cowan:1998ji}%
  \BibitemOpen
  \bibfield  {author} {\bibinfo {author} {\bibfnamefont {G.}~\bibnamefont
  {Cowan}},\ }\href@noop {} {\emph {\bibinfo {title} {{Statistical data
  analysis}}}}\ (\bibinfo {year} {1998})\BibitemShut {NoStop}%
\bibitem [{\citenamefont {Cowan}(2013)}]{Cowan:2013pha}%
  \BibitemOpen
  \bibfield  {author} {\bibinfo {author} {\bibfnamefont {G.}~\bibnamefont
  {Cowan}},\ }in\ \href {\doibase 10.1007/978-3-319-05362-2\_9} {\emph
  {\bibinfo {booktitle} {{69th Scottish Universities Summer School in Physics}:
  {LHC Physics}}}}\ (\bibinfo {year} {2013})\ pp.\ \bibinfo {pages}
  {321--355},\ \Eprint {http://arxiv.org/abs/1307.2487} {arXiv:1307.2487
  [hep-ex]} \BibitemShut {NoStop}%
\bibitem [{\citenamefont {Barlow}\ and\ \citenamefont
  {Beeston}(1993)}]{Barlow:1993dm}%
  \BibitemOpen
  \bibfield  {author} {\bibinfo {author} {\bibfnamefont {R.~J.}\ \bibnamefont
  {Barlow}}\ and\ \bibinfo {author} {\bibfnamefont {C.}~\bibnamefont
  {Beeston}},\ }\href {\doibase 10.1016/0010-4655(93)90005-W} {\bibfield
  {journal} {\bibinfo  {journal} {Comput. Phys. Commun.}\ }\textbf {\bibinfo
  {volume} {77}},\ \bibinfo {pages} {219} (\bibinfo {year} {1993})}\BibitemShut
  {NoStop}%
\bibitem [{\citenamefont {Banerjee}\ \emph {et~al.}(2019)\citenamefont
  {Banerjee} \emph {et~al.}}]{NA64:2019imj}%
  \BibitemOpen
  \bibfield  {author} {\bibinfo {author} {\bibfnamefont {D.}~\bibnamefont
  {Banerjee}} \emph {et~al.},\ }\href {\doibase 10.1103/PhysRevLett.123.121801}
  {\bibfield  {journal} {\bibinfo  {journal} {Phys. Rev. Lett.}\ }\textbf
  {\bibinfo {volume} {123}},\ \bibinfo {pages} {121801} (\bibinfo {year}
  {2019})},\ \Eprint {http://arxiv.org/abs/1906.00176} {arXiv:1906.00176
  [hep-ex]} \BibitemShut {NoStop}%
\bibitem [{\citenamefont {Gninenko}\ \emph {et~al.}(2019)\citenamefont
  {Gninenko}, \citenamefont {Kirpichnikov}, \citenamefont {Kirsanov},\ and\
  \citenamefont {Krasnikov}}]{Gninenko:2019qiv}%
  \BibitemOpen
  \bibfield  {author} {\bibinfo {author} {\bibfnamefont {S.}~\bibnamefont
  {Gninenko}}, \bibinfo {author} {\bibfnamefont {D.}~\bibnamefont
  {Kirpichnikov}}, \bibinfo {author} {\bibfnamefont {M.}~\bibnamefont
  {Kirsanov}}, \ and\ \bibinfo {author} {\bibfnamefont {N.}~\bibnamefont
  {Krasnikov}},\ }\href {\doibase 10.1016/j.physletb.2019.07.015} {\bibfield
  {journal} {\bibinfo  {journal} {Phys. Lett. B}\ }\textbf {\bibinfo {volume}
  {796}},\ \bibinfo {pages} {117} (\bibinfo {year} {2019})},\ \Eprint
  {http://arxiv.org/abs/1903.07899} {arXiv:1903.07899 [hep-ph]} \BibitemShut
  {NoStop}%
\bibitem [{\citenamefont {Bevan}\ \emph {et~al.}(2014)\citenamefont {Bevan}
  \emph {et~al.}}]{Bevan:2014iga}%
  \BibitemOpen
  \bibfield  {author} {\bibinfo {author} {\bibfnamefont {A.}~\bibnamefont
  {Bevan}} \emph {et~al.} (\bibinfo {collaboration} {BaBar, Belle}),\ }\href
  {\doibase 10.1140/epjc/s10052-014-3026-9} {\bibfield  {journal} {\bibinfo
  {journal} {Eur. Phys. J. C}\ }\textbf {\bibinfo {volume} {74}},\ \bibinfo
  {pages} {3026} (\bibinfo {year} {2014})},\ \Eprint
  {http://arxiv.org/abs/1406.6311} {arXiv:1406.6311 [hep-ex]} \BibitemShut
  {NoStop}%
\bibitem [{\citenamefont {Lees}\ \emph {et~al.}(2017)\citenamefont {Lees} \emph
  {et~al.}}]{Lees:2017lec}%
  \BibitemOpen
  \bibfield  {author} {\bibinfo {author} {\bibfnamefont {J.}~\bibnamefont
  {Lees}} \emph {et~al.} (\bibinfo {collaboration} {BaBar}),\ }\href {\doibase
  10.1103/PhysRevLett.119.131804} {\bibfield  {journal} {\bibinfo  {journal}
  {Phys. Rev. Lett.}\ }\textbf {\bibinfo {volume} {119}},\ \bibinfo {pages}
  {131804} (\bibinfo {year} {2017})},\ \Eprint
  {http://arxiv.org/abs/1702.03327} {arXiv:1702.03327 [hep-ex]} \BibitemShut
  {NoStop}%
\bibitem [{\citenamefont {Altmannshofer}\ \emph {et~al.}(2019)\citenamefont
  {Altmannshofer} \emph {et~al.}}]{Kou:2018nap}%
  \BibitemOpen
  \bibfield  {author} {\bibinfo {author} {\bibfnamefont {W.}~\bibnamefont
  {Altmannshofer}} \emph {et~al.} (\bibinfo {collaboration} {Belle-II}),\
  }\href {\doibase 10.1093/ptep/ptz106} {\bibfield  {journal} {\bibinfo
  {journal} {PTEP}\ }\textbf {\bibinfo {volume} {2019}},\ \bibinfo {pages}
  {123C01} (\bibinfo {year} {2019})},\ \bibinfo {note} {[Erratum: PTEP 2020,
  029201 (2020)]},\ \Eprint {http://arxiv.org/abs/1808.10567} {arXiv:1808.10567
  [hep-ex]} \BibitemShut {NoStop}%
\bibitem [{\citenamefont {Liang}\ \emph {et~al.}(2020)\citenamefont {Liang},
  \citenamefont {Liu}, \citenamefont {Ma},\ and\ \citenamefont
  {Zhang}}]{Liang:2019zkb}%
  \BibitemOpen
  \bibfield  {author} {\bibinfo {author} {\bibfnamefont {J.}~\bibnamefont
  {Liang}}, \bibinfo {author} {\bibfnamefont {Z.}~\bibnamefont {Liu}}, \bibinfo
  {author} {\bibfnamefont {Y.}~\bibnamefont {Ma}}, \ and\ \bibinfo {author}
  {\bibfnamefont {Y.}~\bibnamefont {Zhang}},\ }\href {\doibase
  10.1103/PhysRevD.102.015002} {\bibfield  {journal} {\bibinfo  {journal}
  {Phys. Rev. D}\ }\textbf {\bibinfo {volume} {102}},\ \bibinfo {pages}
  {015002} (\bibinfo {year} {2020})},\ \Eprint
  {http://arxiv.org/abs/1909.06847} {arXiv:1909.06847 [hep-ph]} \BibitemShut
  {NoStop}%
\bibitem [{\citenamefont {Benesch}\ \emph {et~al.}(2014)\citenamefont {Benesch}
  \emph {et~al.}}]{Benesch:2014bas}%
  \BibitemOpen
  \bibfield  {author} {\bibinfo {author} {\bibfnamefont {J.}~\bibnamefont
  {Benesch}} \emph {et~al.} (\bibinfo {collaboration} {MOLLER}),\ }\href@noop
  {} {\  (\bibinfo {year} {2014})},\ \Eprint {http://arxiv.org/abs/1411.4088}
  {arXiv:1411.4088 [nucl-ex]} \BibitemShut {NoStop}%
\bibitem [{\citenamefont {Zhao}(2017)}]{Zhao:2017xej}%
  \BibitemOpen
  \bibfield  {author} {\bibinfo {author} {\bibfnamefont {Y.}~\bibnamefont
  {Zhao}} (\bibinfo {collaboration} {SoLID}),\ }in\ \href@noop {} {\emph
  {\bibinfo {booktitle} {{22nd International Symposium on Spin Physics}}}}\
  (\bibinfo {year} {2017})\ \Eprint {http://arxiv.org/abs/1701.02780}
  {arXiv:1701.02780 [nucl-ex]} \BibitemShut {NoStop}%
\bibitem [{\citenamefont {Arg{\"u}elles}\ \emph {et~al.}(2019)\citenamefont
  {Arg{\"u}elles}, \citenamefont {Schneider},\ and\ \citenamefont
  {Yuan}}]{Arguelles:2019izp}%
  \BibitemOpen
  \bibfield  {author} {\bibinfo {author} {\bibfnamefont {C.~A.}\ \bibnamefont
  {Arg{\"u}elles}}, \bibinfo {author} {\bibfnamefont {A.}~\bibnamefont
  {Schneider}}, \ and\ \bibinfo {author} {\bibfnamefont {T.}~\bibnamefont
  {Yuan}},\ }\href {\doibase 10.1007/JHEP06(2019)030} {\bibfield  {journal}
  {\bibinfo  {journal} {JHEP}\ }\textbf {\bibinfo {volume} {06}},\ \bibinfo
  {pages} {030} (\bibinfo {year} {2019})},\ \Eprint
  {http://arxiv.org/abs/1901.04645} {arXiv:1901.04645 [physics.data-an]}
  \BibitemShut {NoStop}%
\end{thebibliography}%


%

\begin{appendix}

\section{Likelihood for Finite Monte Carlo}
\label{sec:stat}
In this appendix we describe the likelihood analysis
used in Sec \ref{kinematic} to distinguish between particle masses 
using final state beam recoil energy $E_e$ and transverse
momentum $p_{T,e}$. The simple likelihood given in Eq.~(\ref{eq:simple_log_lkl}) correctly 
accounts for statistical fluctuations in the observations. However, it assumed that 
each bin contains a large number of simulated events, i.e., the relative fluctuations in the bin counts are small.
The predictions for missing recoil energy and transverse momentum of the 
electrons are obtained using finitely-sized MC samples. Even if the sample is large, 
if we bin it finely enough (and especially if we bin in multiple kinematic variables at the same time) 
each bin will contain a small number of events which are subject to Poisson fluctuations due to the 
finite size of the MC sample. If we are comparing two distributions that only differ in such bins 
where fluctuations are important, we miss the theoretical uncertainty due to MC statistics. 
This uncertainty can be incorporated following Ref.~\cite{Barlow:1993dm} (see also Ref.~\cite{Arguelles:2019izp} 
for a concise description of the problem). 
The idea is to introduce nuisance parameters that encode the \emph{true} expected values in each bin; 
the count of MC events is then a specific random Poisson realization of this expected value. Since we cannot 
evaluate the cross-section exactly in that kinematic bin, we do not know what this true value is, and therefore 
we must marginalize over it. The likelihood function that takes this effect into account is
\begin{align}
\widetilde{L}(\mathbf{d}, \mathbf{a}, \mathbf{A}, p) &  = \prod_i f_P(d_i, pA_i) f_P(a_i, A_i)  \\
& = \prod_i \frac{(pA_i)^{d_i}}{d_i!} e^{-pA_i} \times \frac{A_i^{a_i}}{a_i!} e^{-A_i},
\label{eq:finite_mc_likelihood}
\end{align}
where $a_i$ are the ``raw'' (unnormalized) MC bin counts (i.e., the number of events in each kinematic bin as it 
comes out of the simulation -- these numbers grow as the MC sample gets larger); $A_i$ are the true expected values  in each bin 
(equal to $a_i$ in the limit of an infinitely large MC sample, and therefore not known); $p$ is the signal strength, such that 
$pA_i$ corresponds to the theory prediction of the expected count in bin $i$.
$A_i$ and $p$ are nuisance parameters over which we must maximize the likelihood -- luckily we will be able to do this analytically. 
The notation used here corresponds to that of Ref.~\cite{Barlow:1993dm}. 
In other words, Eq.~(\ref{eq:finite_mc_likelihood}) treats the MC sample as another dataset; both the 
``real'' data $\mathbf{d}$ and the MC data $\mathbf{a}$ constrain $p$ and $\mathbf{A}$. 
The logarithm of this likelihood is
\be
\ln \widetilde{L}(\mathbf{d}, \mathbf{a}, \mathbf{A}, p) 
&=& \sum_i d_i \ln p A_i - p A_i - \ln d_i!
 \nonumber \\
&&
~~~+  a_i \ln A_i - A_i - \ln a_i!~~. 
\label{eq:finite_mc_loglikelihood}
\ee
Next we maximize this with respect to $p$ and $A_i$:
\begin{align}
  \frac{\partial \ln \widetilde{L}}{\partial p} & = \sum_i \frac{d_i}{p} - A_i = 0\\
  \frac{\partial \ln \widetilde{L}}{\partial A_i} & = \frac{d_i + a_i}{A_i} - p - 1= 0
\end{align}
The solutions of these equations are 
\begin{align}
  p & = \frac{\sum_i d_i}{\sum_i a_i} = \frac{N_d}{N_{mc}}\\
  A_i & = \frac{a_i + d_i}{1 + p}.
\end{align} 
Note that the signal strength $p$ takes the natural value which ensures that the predicted 
number of events $\sum_i p A_i$ matches the number of observed data events $N_d$. At the maximum 
of the likelihood, $A_i \neq a_i$, unless $a_i \gg d_i$; this encodes the fact that for a finite MC 
sample we do not quite know what the actual expected value of the bin counts is.
Plugging these solutions into Eq.~(\ref{eq:finite_mc_loglikelihood}) we find 
$\ln L \equiv \max_{p,A_i} \ln \widetilde{L}$
\be
\ln L(\mathbf{d}, \mathbf{a}) 
&=& 
\sum_i (d_i + a_i) \ln \frac{N_d}{N_d + N_{mc}}(d_i + a_i) 
\nonumber \\
&& 
\hspace{-0.75cm}
- (d_i + a_i)
 + a_i \ln \frac{N_{mc}}{N_d} 
 - \ln d_i! - \ln a_i!~~~~~~
\label{eq:finite_mc_loglikelihood_max}
\ee
The last line contains factors that are commonly dropped when discussing real data (i.e., $\mathbf{d}$); however 
since $a_i$ depends on the model, we must keep this factor such that we can meaningfully 
compare the likelihoods in different models (which generically have differently-sized MC samples). $d_i!$, however, 
\emph{is} the same in each model, so we can drop it.

The likelihood in Eq.~(\ref{eq:finite_mc_loglikelihood_max}) is not very intuitive so it is useful to consider some illustrative limits:
\begin{itemize}
  \item $a_i \gg d_i$ $\forall i$, and therefore $N_{mc}\gg N_d$. Expanding in small quantities one 
    finds
    \beq
    \ln L(\mathbf{d}, \mathbf{a}) \approx \sum_i d_i \ln \frac{N_d}{N_{mc}} a_i - \frac{N_d}{N_{mc}} a_i,
    \eeq
    which is just the standard Poisson likelihood of Eq.~(\ref{eq:simple_log_lkl}) with $\nu_i = (N_d/N_{mc})a_i$ 
    (there are corrections that go like $\ln a_i$ from approximating $\ln a_i!$). 
    Thus in the limit in which theory (MC statistical) uncertainties are not important we recover the naive result.
  \item Without taking theory uncertainties into account, if MC predicts $a_i = 0$ events in a bin, but the data $d_i$ is not $0$ there, 
    the standard log likelihood for that bin is $d_i \ln 0 - d_i = -\infty$, i.e., the model is immediately ruled out. 
    Using the above log likelihood instead, we find that such a bin would instead contribute
    \beq
    d_i \ln \frac{N_d}{N_d + N_{mc}} d_i - d_i
    \eeq
    \newline
    to the log likelihood. The would-be infinity is regularized by the fact that $N_d/N_{mc}$ is not 0. 
    This is the desired behavior, since it prevents us from overstating the discriminating power of certain bins if 
    we have insufficient MC statistics.
\end{itemize}
The full likelihood of Eq.~(\ref{eq:finite_mc_loglikelihood_max}) is used to define the test 
statistic in Eq.~(\ref{eq:ts}), which is then studied in the various model discrimination examples in Sections~\ref{kinematic} and~\ref{muons}.

\end{appendix}

\end{document}